\PassOptionsToPackage{dvipsnames}{xcolor}
\documentclass[acmtog]{acmart}
\acmSubmissionID{1355}
\usepackage{booktabs} 
\usepackage{xcolor}
\usepackage[dvipsnames]{xcolor}
\usepackage{graphicx}
\usepackage{caption}
\usepackage{subcaption}
\usepackage{tikz}
\usetikzlibrary{positioning}

\usepackage[ruled]{algorithm2e} 

\SetAlFnt{\small}
\SetAlCapFnt{\small}
\SetAlCapNameFnt{\small}
\SetAlCapHSkip{0pt}

\newcommand{\videoInput}{\mathbf{I}}
\newcommand{\matmap}{\mathbf{mat}}
\newcommand{\diffusionModel}{\mathbf{f}}
\newcommand{\diffusionModelParams}{\theta}
\newcommand{\diffusionModelFn}{\diffusionModel_{\diffusionModelParams}}
\newcommand{\vaeEncoder}{\mathcal{E}}
\newcommand{\vaeDecoder}{\mathcal{D}}
\newcommand{\typeEmb}{\mathbf{c}_{\text{prompt}}}
\newcommand{\dataDistribution}{p_{\text{data}}}
\newcommand{\diffusionNoise}{\mathbb{\epsilon}}

\newcommand{\FLIP}{\protect\reflectbox{F}LIP\xspace}

\definecolor{myorange}{rgb}{1, 0.85, 0.7}
\definecolor{myred}{rgb}{1, 0.7, 0.7}

\newcommand{\reducedstrut}{\rule{0pt}{2.2ex}}
\newcommand{\sota}[1]{%
  \begingroup
  \setlength{\fboxsep}{0pt}%
  \colorbox{myred}{\reducedstrut#1}%
  \endgroup
}

\copyrightyear{2026}
\acmYear{2026}
\setcopyright{cc}
\setcctype{by}
\acmConference[SIGGRAPH Conference Papers '26]{Special Interest Group on Computer Graphics and Interactive Techniques Conference Conference Papers}{July 19--23, 2026}{Los Angeles, CA, USA}
\acmBooktitle{Special Interest Group on Computer Graphics and Interactive Techniques Conference Conference Papers (SIGGRAPH Conference Papers '26), July 19--23, 2026, Los Angeles, CA, USA}
\acmDOI{10.1145/3799902.3811210}
\acmISBN{979-8-4007-2554-8/2026/07}

\begin{document}
\title{Toward Richer Material Generation via Procedural Data Enhancement}

\author{Yunchen Yu}
\email{yy735@cornell.edu}
\orcid{0009-0009-3431-8758}
\affiliation{
\institution{Cornell University \& NVIDIA}
\country{USA}
}

\author{Jacob Munkberg}
\email{jmunkberg@nvidia.com}
\orcid{0009-0004-0451-7442}
\affiliation{
\institution{NVIDIA}
\country{Sweden}
}

\author{Jon Hasselgren}
\email{jhasselgren@nvidia.com}
\orcid{0009-0002-3423-190X}
\affiliation{
\institution{NVIDIA}
\country{Sweden}
}

\author{Chris Cummings}
\email{ccummings@nvidia.com}
\orcid{0009-0006-9281-8644}
\affiliation{
\institution{NVIDIA}
\country{United Kingdom}
}

\author{Steve Marschner}
\email{srm@cs.cornell.edu}
\orcid{0000-0001-9810-0306}
\affiliation{
\institution{Cornell University \& NVIDIA}
\country{USA}
}

\author{Andrea Weidlich}
\email{aweidlich@nvidia.com}
\orcid{0000-0002-4146-187X}
\affiliation{
\institution{NVIDIA}
\country{Canada}
}

\renewcommand{\shortauthors}{Yunchen Yu, Jacob Munkberg, Jon Hasselgren, Chris Cummings, Steve Marschner, and Andrea Weidlich}

\begin{abstract}
Generative models for material creation are fundamentally limited by the quality and expressivity of available training data. Simple physically based rendering (PBR) materials, which combine a diffuse term with a single-lobe specular component, are commonly used for training but are insufficient to capture many important visual effects present in real materials.

We present a method that enhances such simple PBR materials to more expressive ones, by augmenting the single GGX specular lobe into a layered model that captures a broader range of non-diffuse effects. Starting from a simple material, we procedurally construct a corresponding multi-lobe non-diffuse component guided by physical priors, enabling effects such as dust, clearcoat, and layered scattering. To provide a compact representation for downstream applications, we encode this non-diffuse component as a neural material with a shared 6D latent space, where each material instance is represented by two latent textures and decoded by a pretrained universal MLP. We further regularize the latent space to support material generation.

The resulting neural material dataset enables training generative models for richer material creation. To demonstrate this application, we finetune a video diffusion model to produce neural latent textures that encode our multi-lobe material, and present generative results as proof of feasibility. Our procedural data enhancement approach is an important step toward improving expressivity in material generation.
\end{abstract}



\begin{teaserfigure}
    \centering
    {\includegraphics [width=\textwidth] {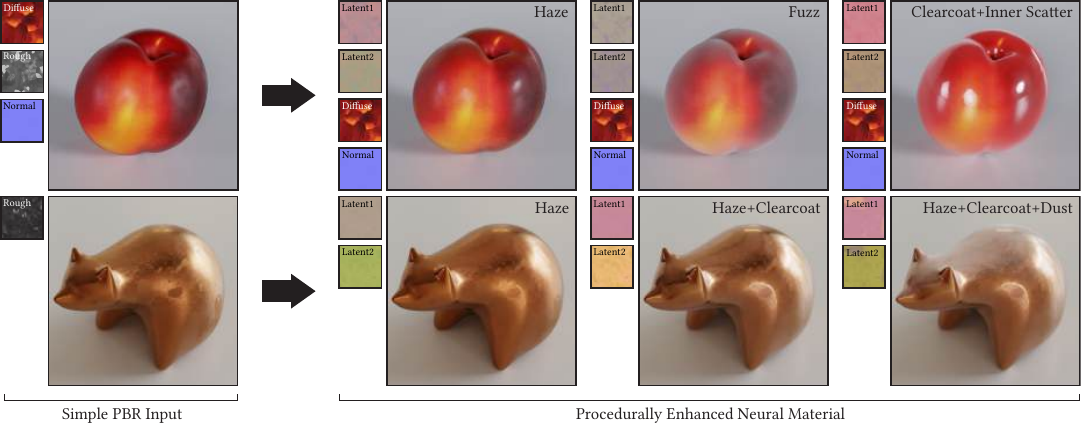}}
    \caption{Creating complex materials at scale is challenging. We introduce a material enhancement procedure that uplifts PBR materials into visually rich appearances that can be compactly represented as neural materials. By adding surface effects, we can make an object appear waxy, peach‑fuzzed, or sugar‑coated (top row). Similarly, we can introduce subtle spatial specular breakup, as well as large-scale changes like clearcoat or dust layers (bottom row). The small insets show the textures used by each model. Normal maps are applied unchanged to the neural material.}
    \label{fig:teaser}
\end{teaserfigure}


\newcommand{\figMasks}{
\begin{figure}
    \centering
    \includegraphics[width=0.24\linewidth]{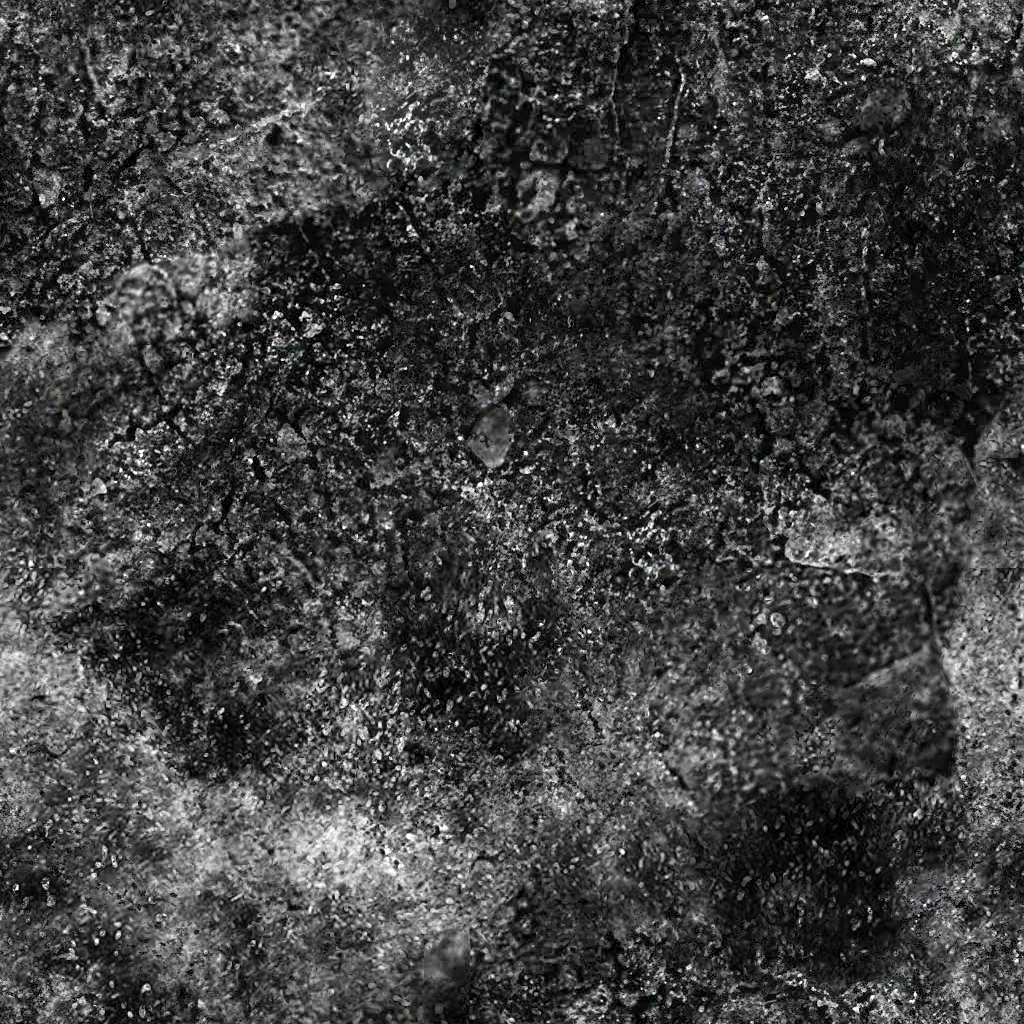}
    \includegraphics[width=0.24\linewidth]{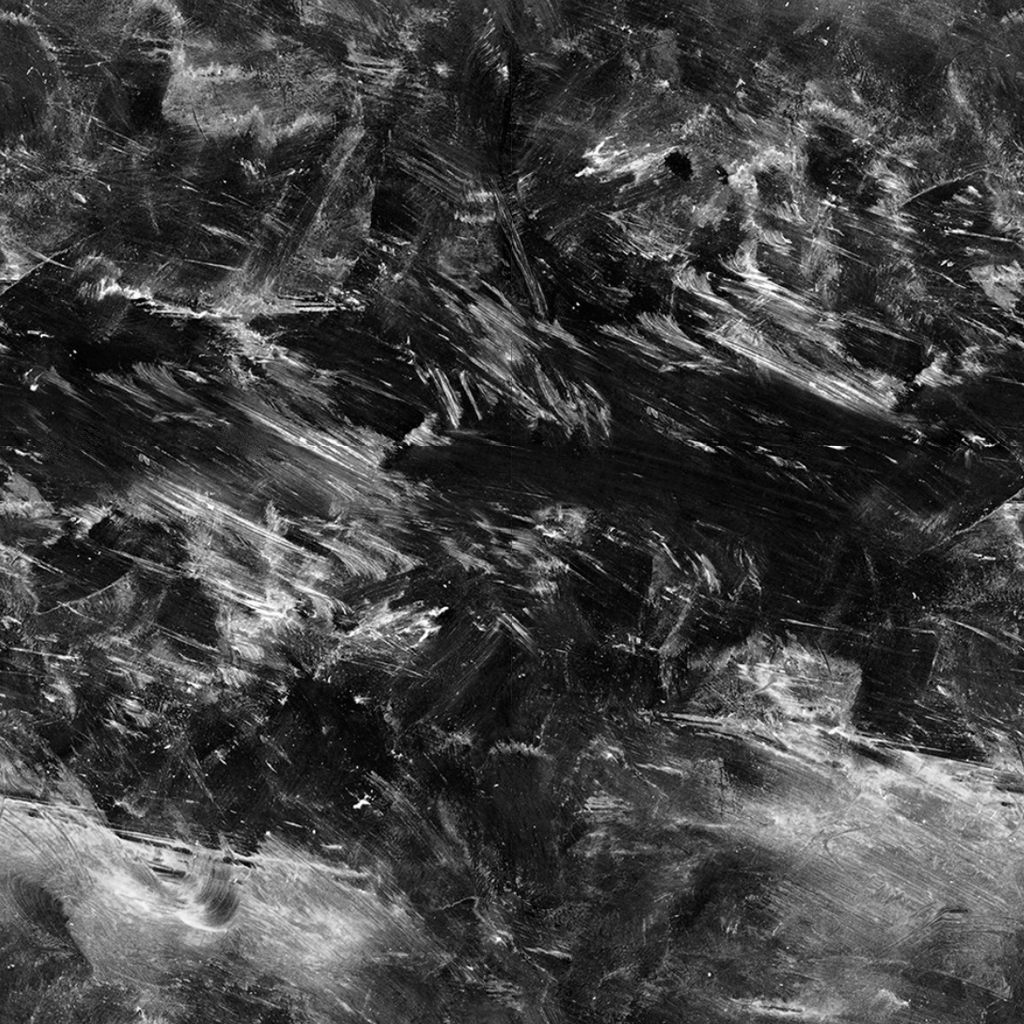}
    \includegraphics[width=0.24\linewidth]{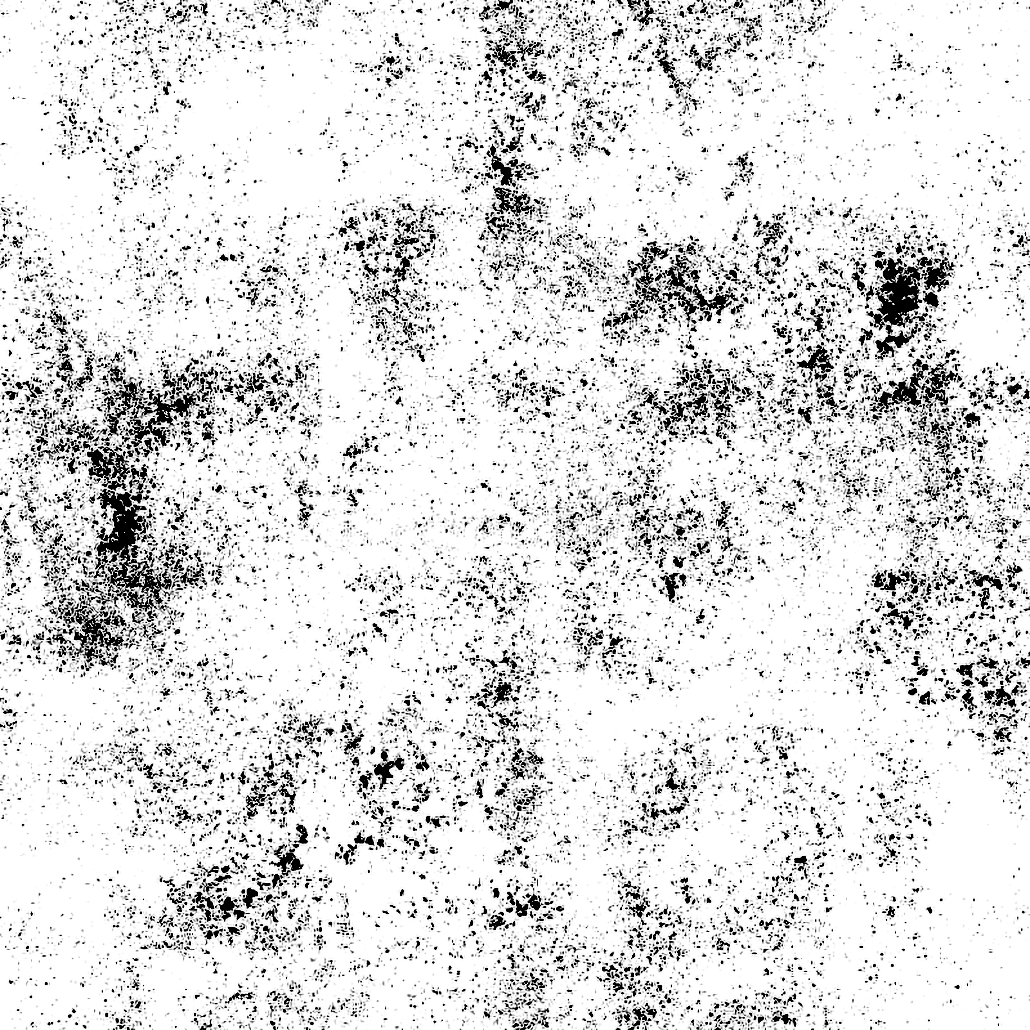}
    \includegraphics[width=0.24\linewidth]{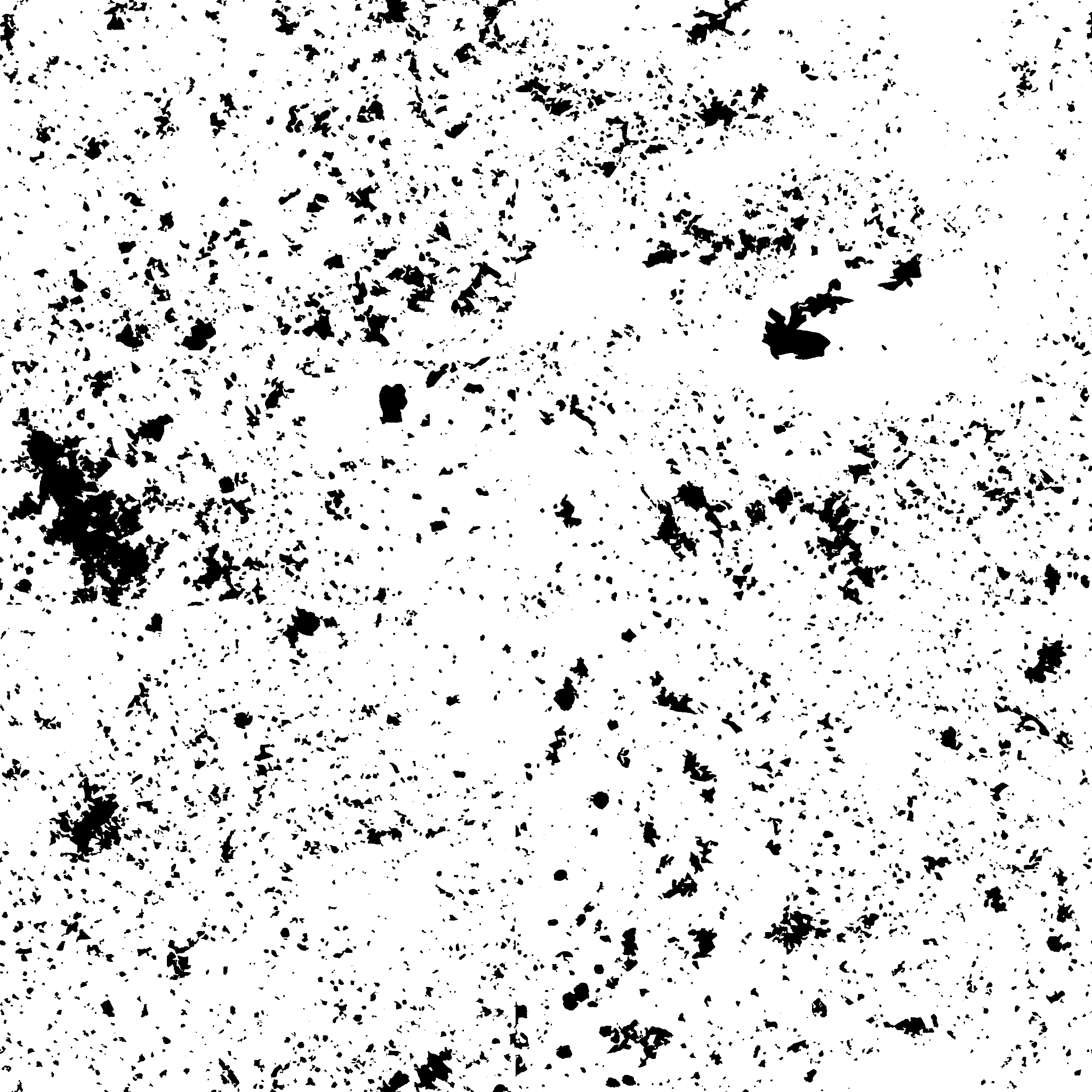}
    \caption{Example masks that modulate spatial variations in dust (left images) and clearcoat (right images), simulating fingerprints or ripple-like patterns on dust layers and dents or scratches on shiny coatings.}
    \label{fig:masks}
\end{figure}
}

\newcommand{\figTrainingData}{
\begin{figure}
    \centering
    \includegraphics[width=0.99\linewidth]{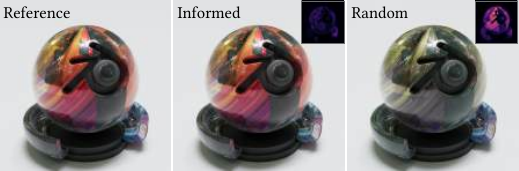}
    \caption{Comparison of informed vs. random training data and their \FLIP error images \cite{Andersson2020}. We reconstruct the same reference material with both approaches. The network built with informed data captures the reference material much better than the one with random data. 
    }
    \label{fig:trainingdata}
\end{figure}
}

\newcommand{\figSystem}{
\begin{figure*}[t]
    \centering
        \includegraphics[width=0.99\textwidth]{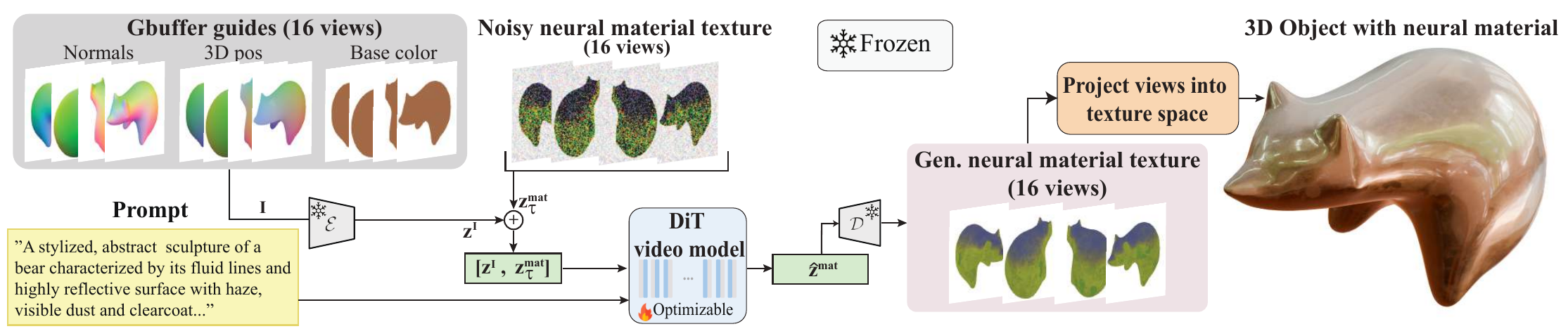}
    \vspace*{-2mm}
    \caption{Our generative pipeline starts from a known 3D model and a text prompt. We first render views of normals, 
    world space positions, and base color. These conditions are encoded using a frozen VAE encoder, $\mathcal{E}$, to produce latent conditions, $\textbf{z}^{\videoInput}$. These are concatenated with noisy latents, $\textbf{z}_\tau^{\matmap}$, representing neural material textures.
    The latents and text prompt are passed to a finetuned video model, which generates a denoised latent, $\hat{\textbf{z}}^{\matmap}$, which is decoded into 
    views of neural material textures, using a frozen VAE decoder $\mathcal{D}$. Finally, we project the generated views into texture space to extract a neural material.}
    \label{fig:system}
\end{figure*}
}


\newcommand{\figGenMat}{
\begin{figure*}
\centering
\begin{tikzpicture}
    \node at (0.0, 0.0) 
    {\includegraphics[width=0.19\linewidth]{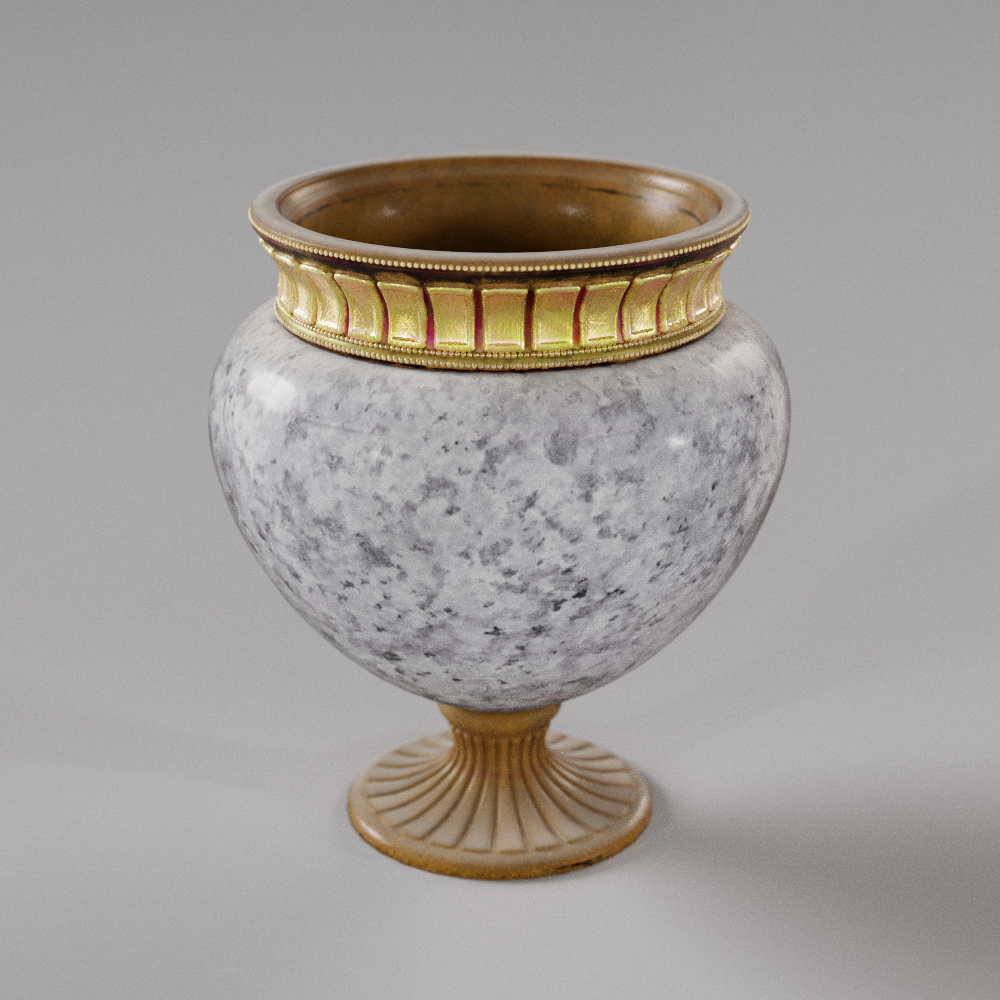}};
    \node at (-1.45, 1.45) {(a)};
    \node at (3.5, 0.0) 
    {\includegraphics[width=0.19\linewidth]{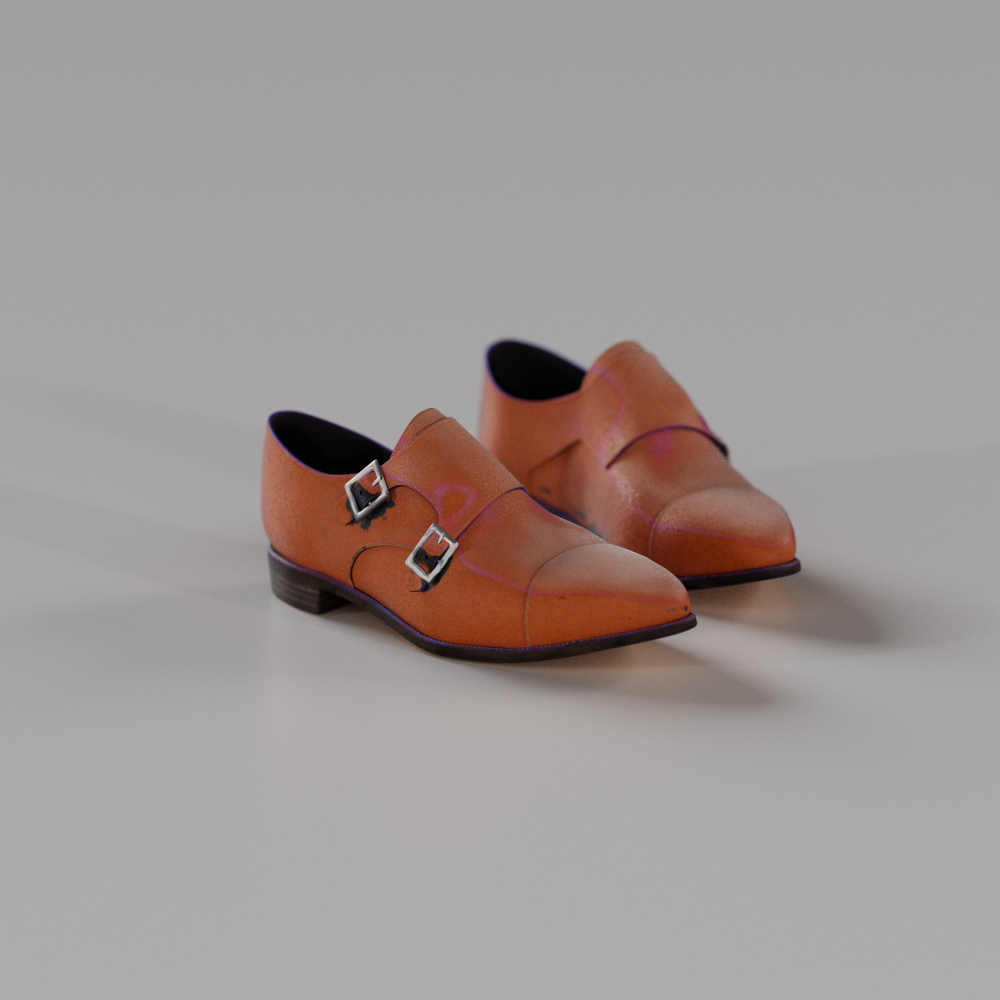}};
    \node at (2.05, 1.45) {(b)};    
    \node at (7.0, 0.0) 
    {\includegraphics[width=0.19\linewidth]{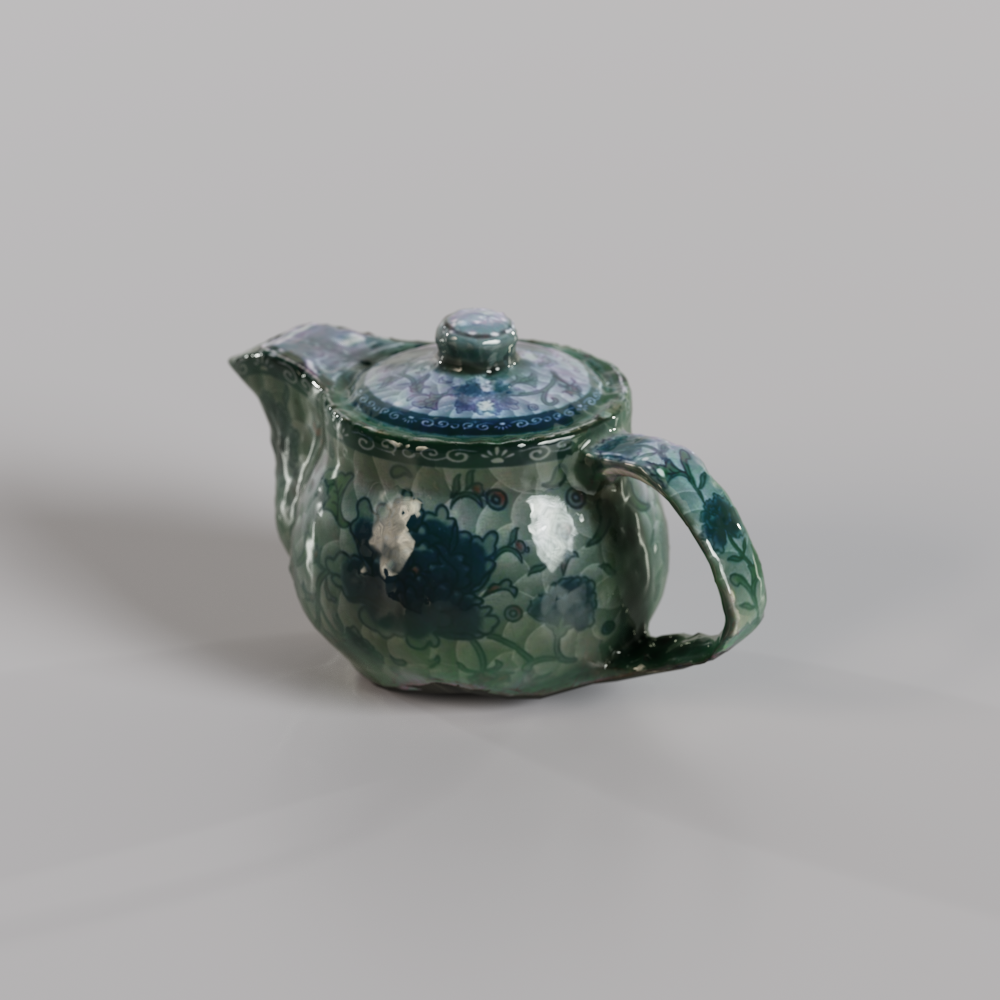}};
    \node at (5.55, 1.45) {(c)};    
    \node at (10.5, 0.0) 
    {\includegraphics[width=0.19\linewidth]{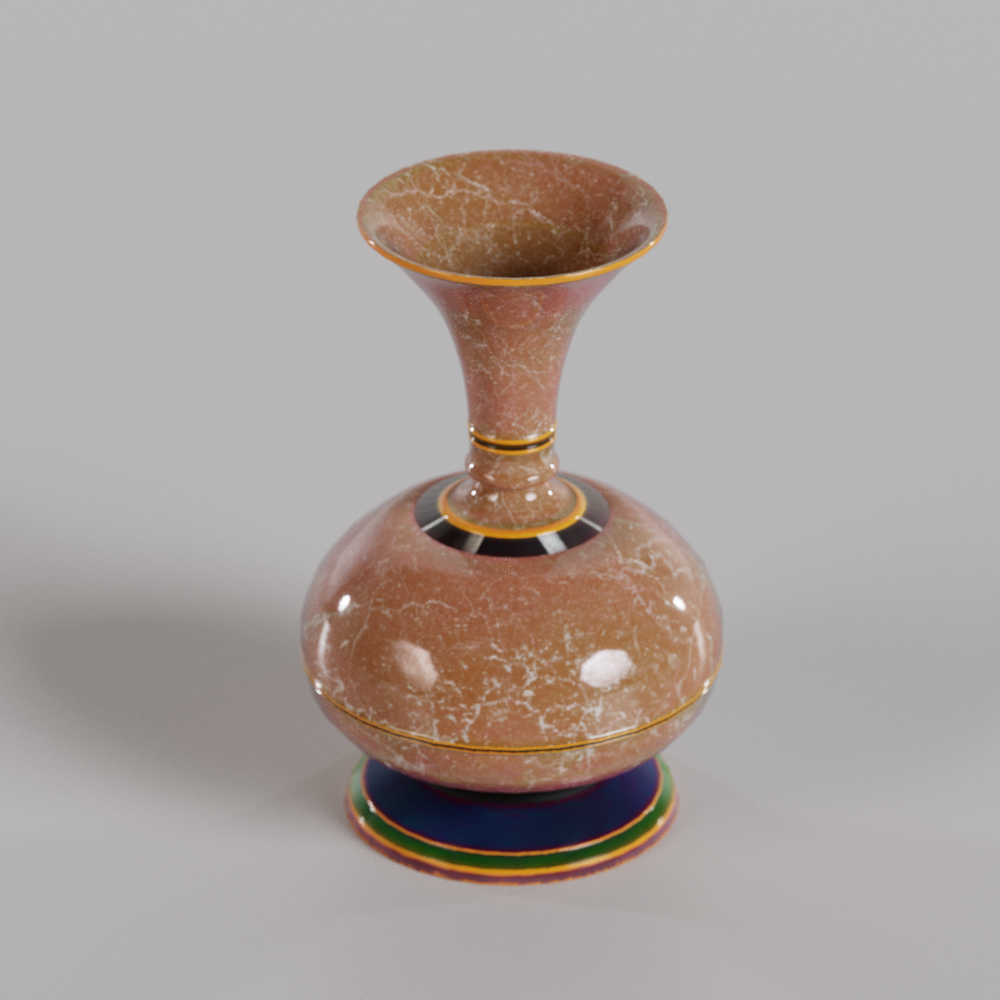}};
    \node at (9.05, 1.45) {(d)};    
    \node at (14.0, 0.0) 
    {\includegraphics[width=0.19\linewidth]{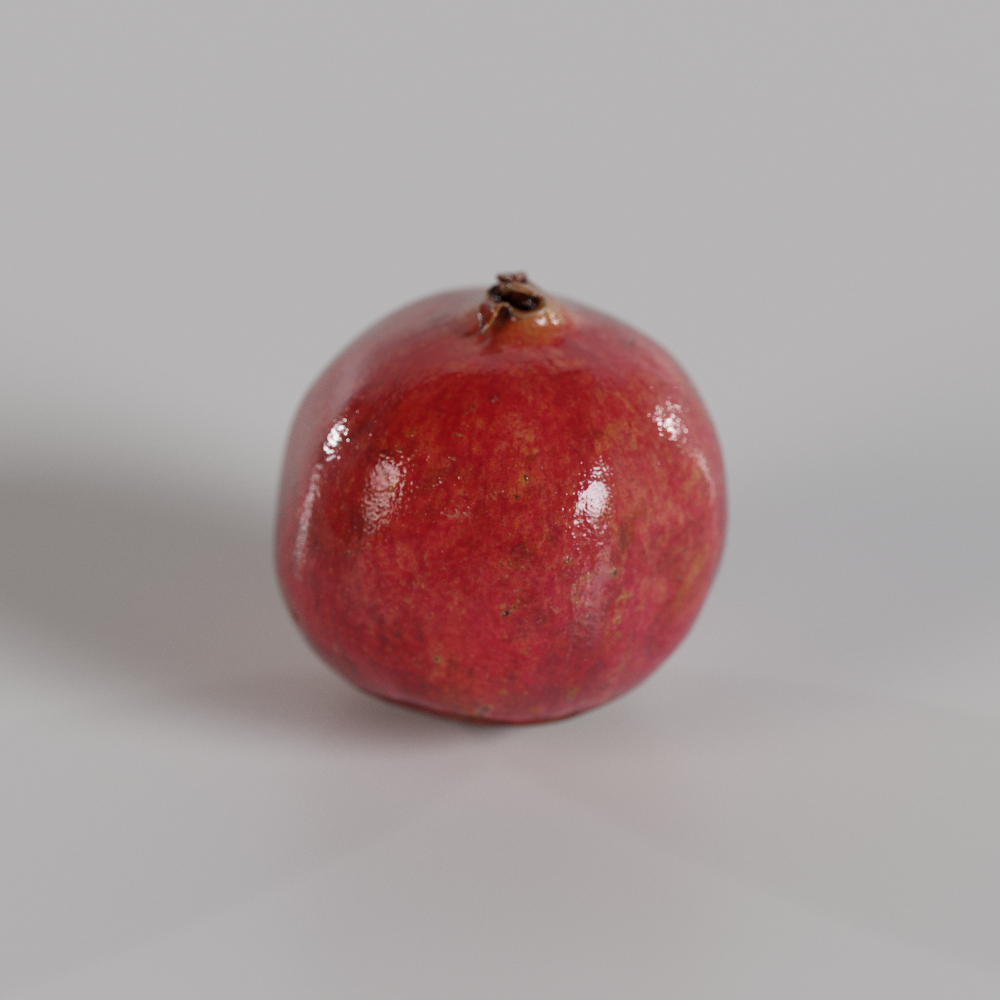}};
    \node at (12.55, 1.45) {(e)};
    \node at (0.0, -3.5) 
    {\includegraphics[width=0.19\linewidth]{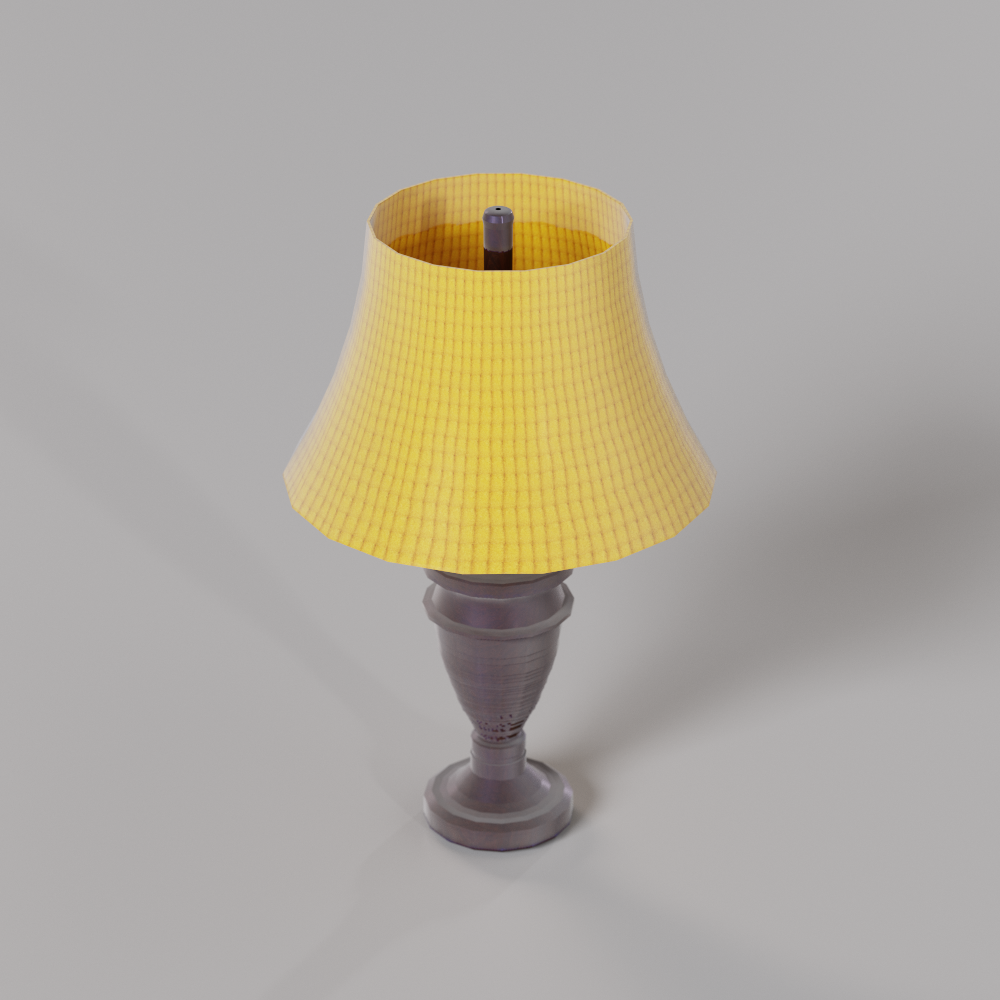}};
    \node at (-1.45, -2.05) {(f)};
    \node at (3.5, -3.5) 
    {\includegraphics[width=0.19\linewidth]{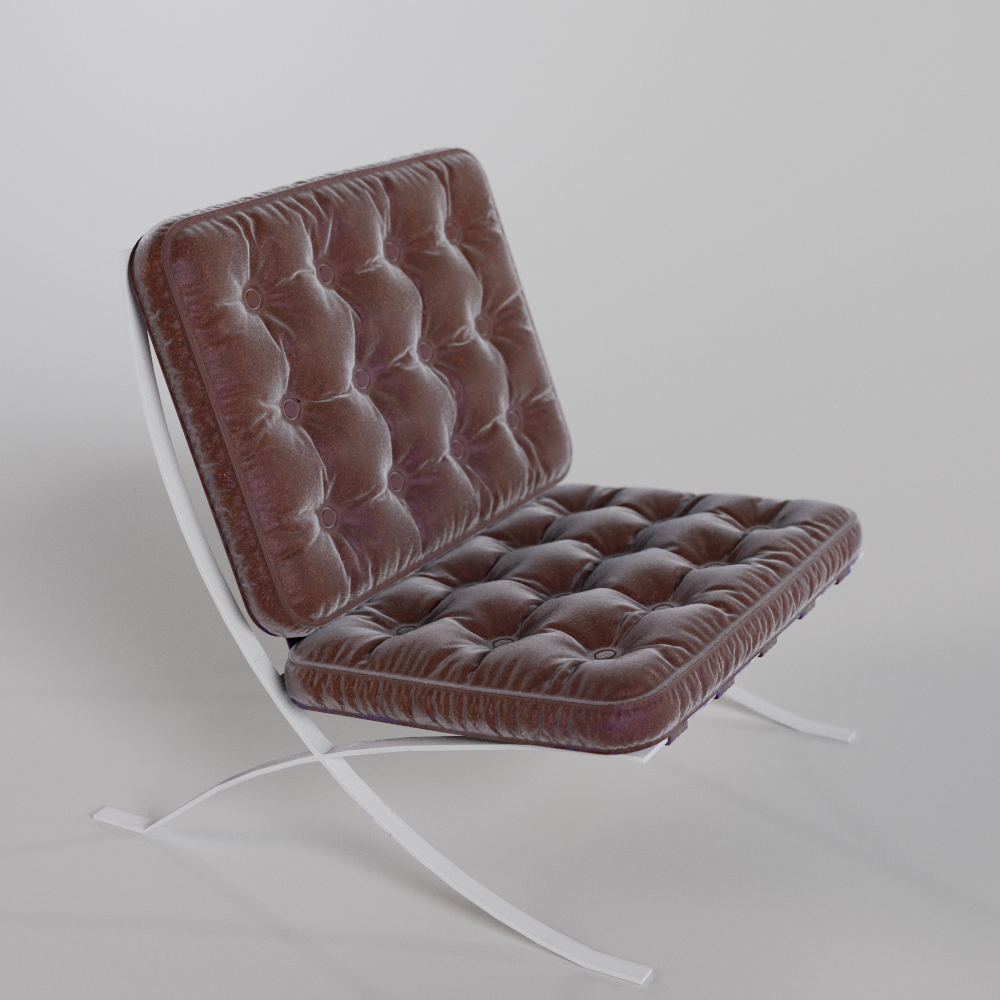}};
    \node at (2.05, -2.05) {(g)};    
    \node at (7.0, -3.5) 
    {\includegraphics[width=0.19\linewidth]{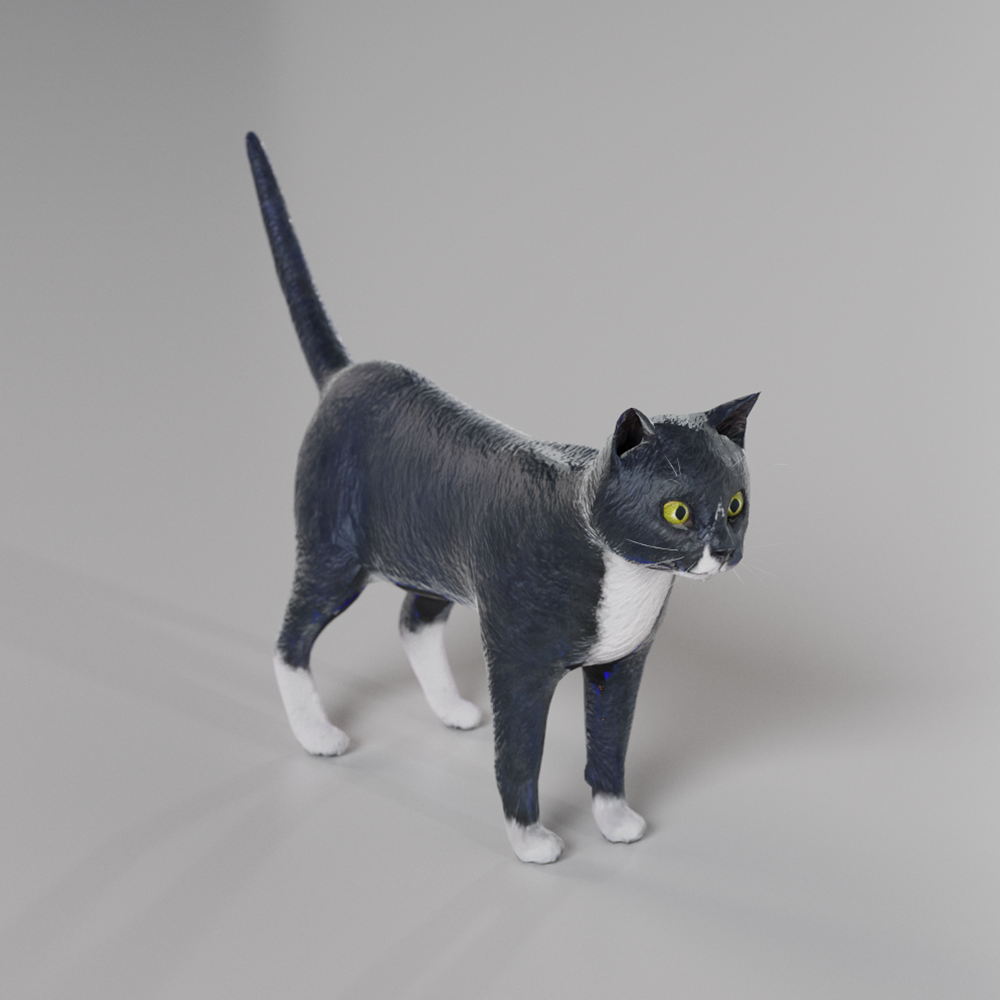}};
    \node at (5.55, -2.05) {(h)};    
    \node at (10.5, -3.5) 
    {\includegraphics[width=0.19\linewidth]{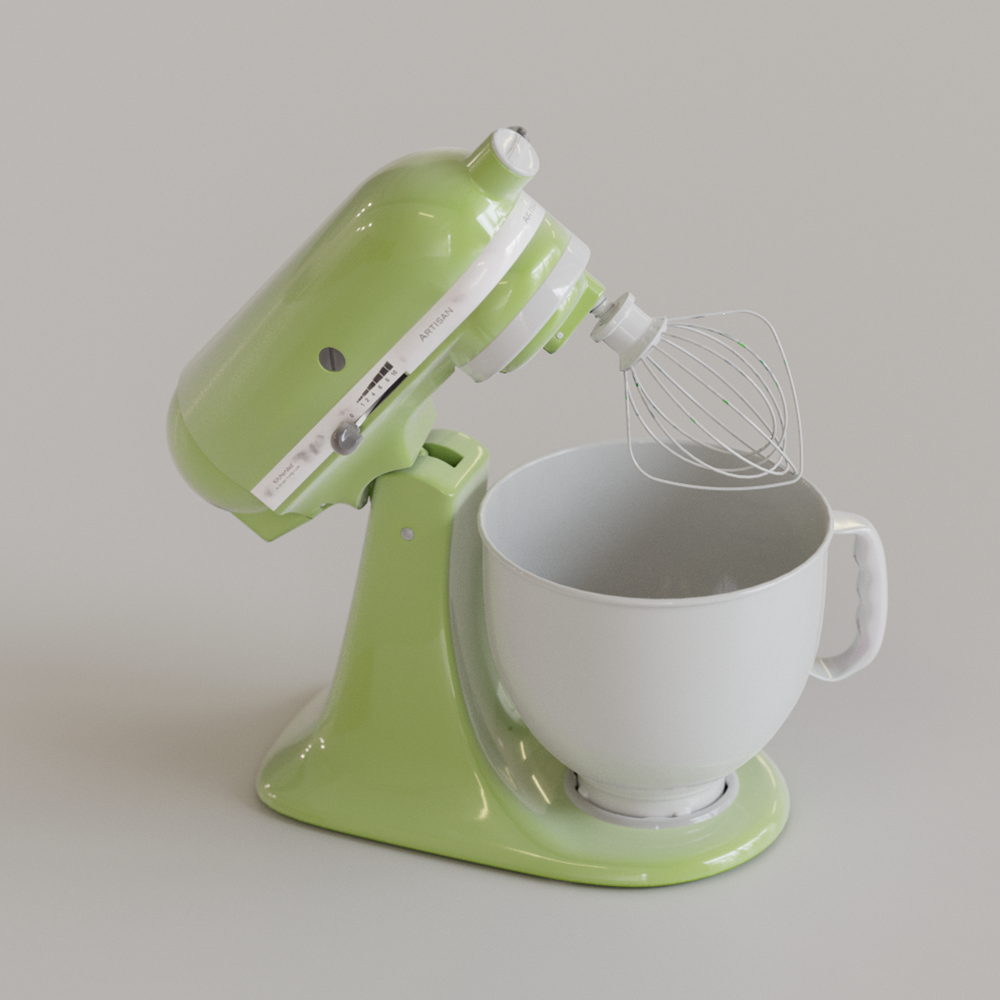}};
    \node at (9.05, -2.05) {(i)};    
    \node at (14.0, -3.5) 
    {\includegraphics[width=0.19\linewidth]{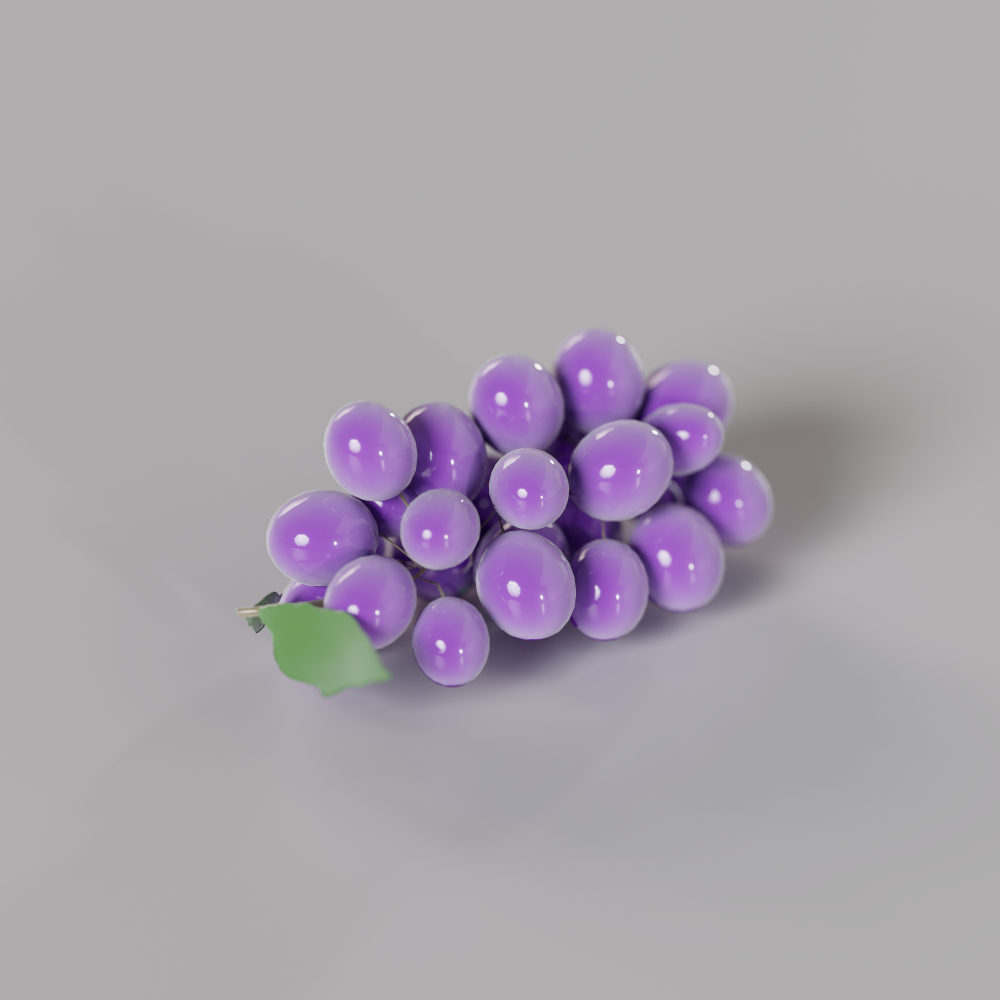}};
    \node at (12.55, -2.05) {(j)};    
\end{tikzpicture}
\vspace*{-2mm}
\caption{Examples of our generated neural materials for a selection of 3D models from our test set. Intended enhancement types of these assets' reference neural materials are dust (a, b), clearcoat (c, d), dust + clearcoat (e), fuzz + scatter (f, g, h), all lobes (i, j). Text prompts are provided in the supplemental material. The material in (b) contains some artifacts from projecting views of latents into the texture space, and (e) is a near-failure case as the dust is not pronounced. }
\label{fig:genmat}
\end{figure*}
}


\newcommand{\figDataset}{
\begin{figure*}
\centering
\small
\setlength{\tabcolsep}{1pt}
\begin{tabular}{lccccc}

    \vspace{0.025in}

    \rotatebox[origin=c]{90}{Haze} &
	\raisebox{-0.5\height}{\includegraphics[width=0.19\textwidth]{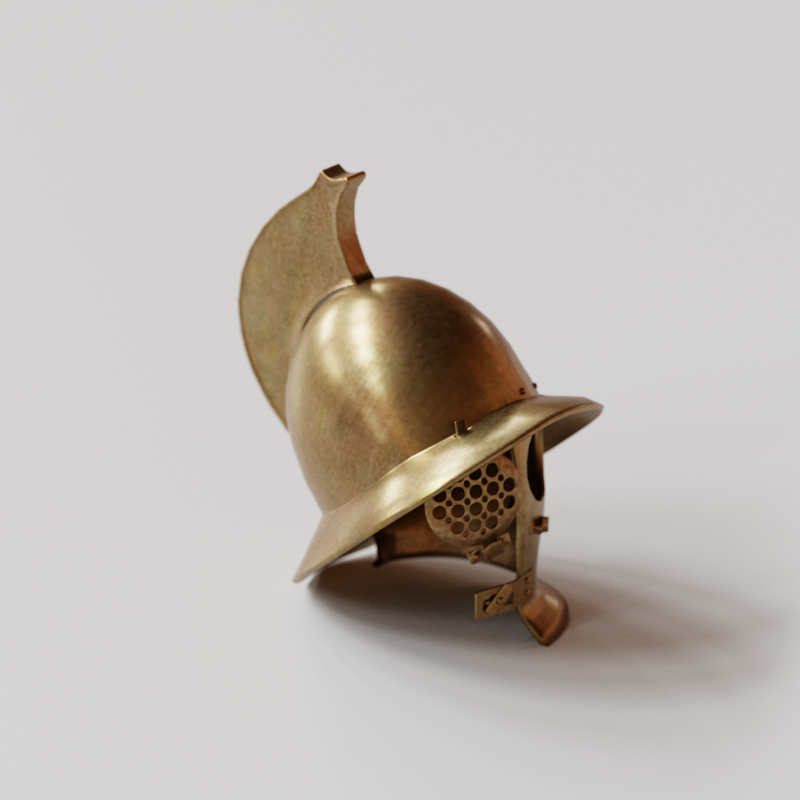}} &
	\raisebox{-0.5\height}{\includegraphics[width=0.19\textwidth]{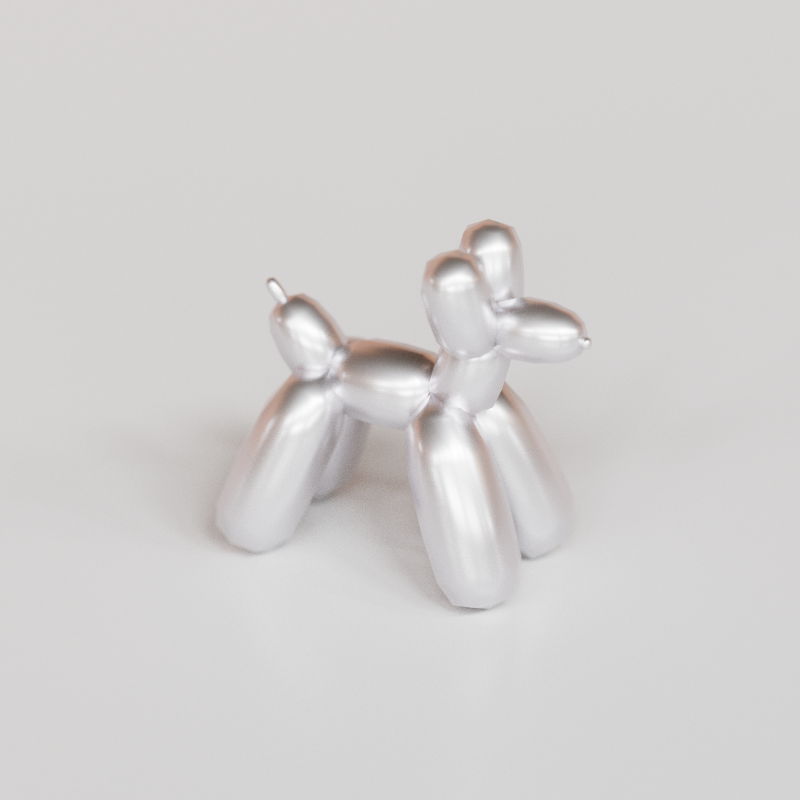}} &
	\raisebox{-0.5\height}{\includegraphics[width=0.19\textwidth]{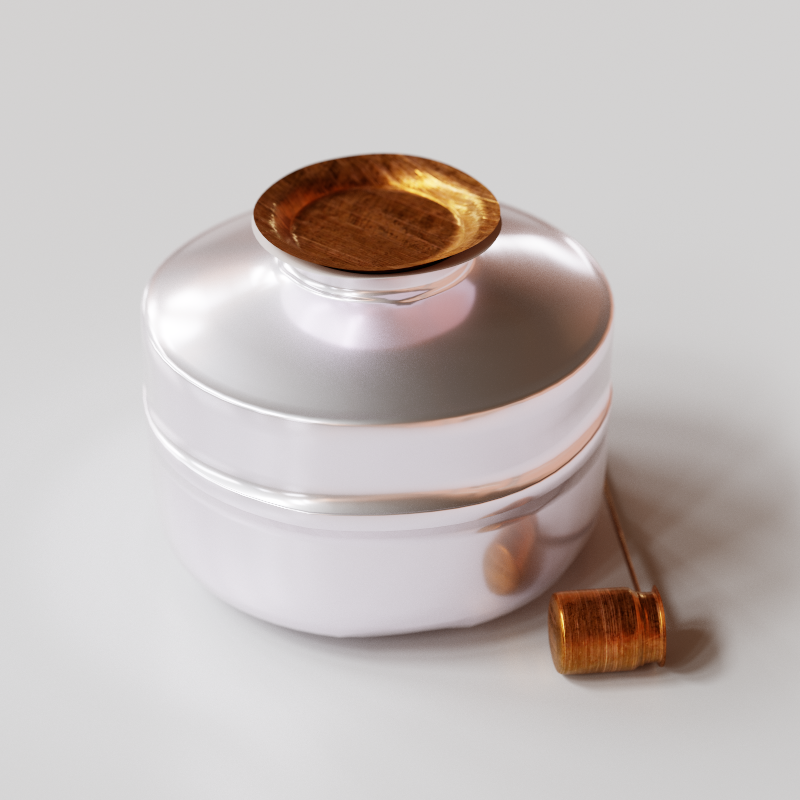}} &
	\raisebox{-0.5\height}{\includegraphics[width=0.19\textwidth]{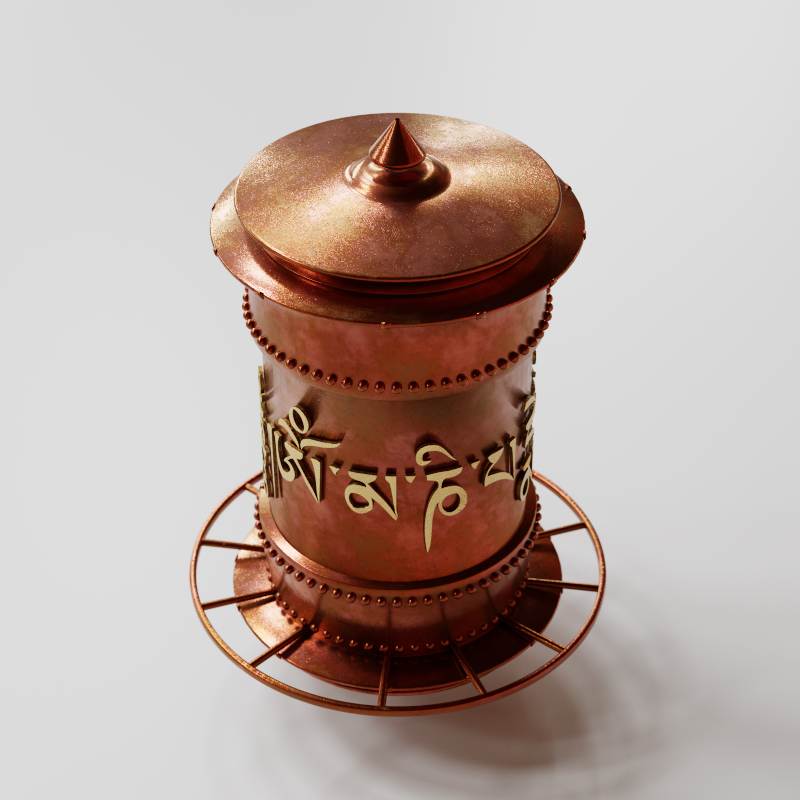}} &
    \raisebox{-0.5\height}{\includegraphics[width=0.19\textwidth]{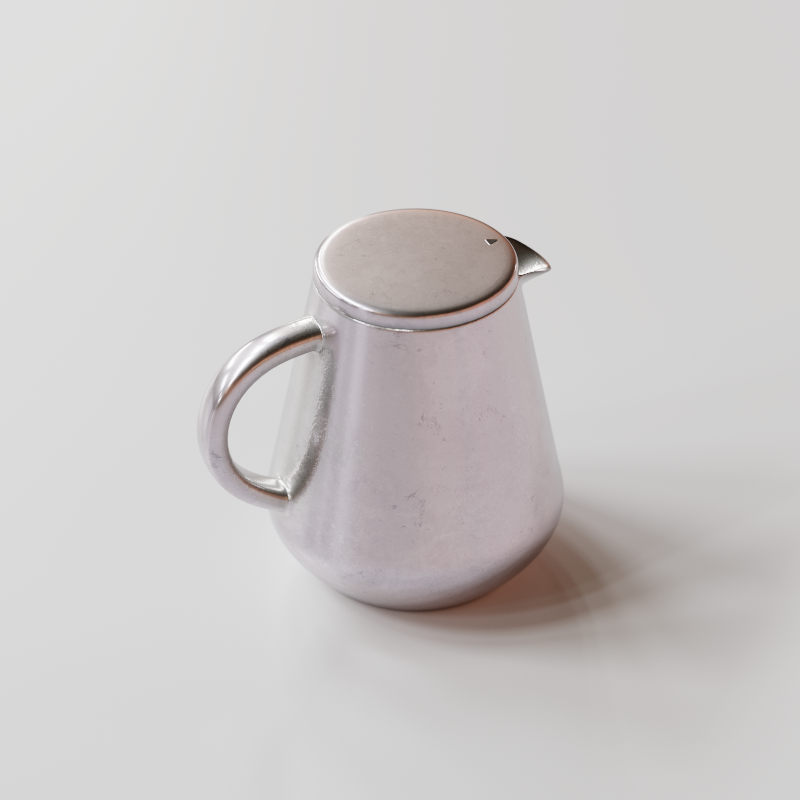}} \\

    \vspace{0.025in}
    
    \rotatebox[origin=c]{90}{Dust} &   
    \raisebox{-0.5\height}{\includegraphics[width=0.19\textwidth]{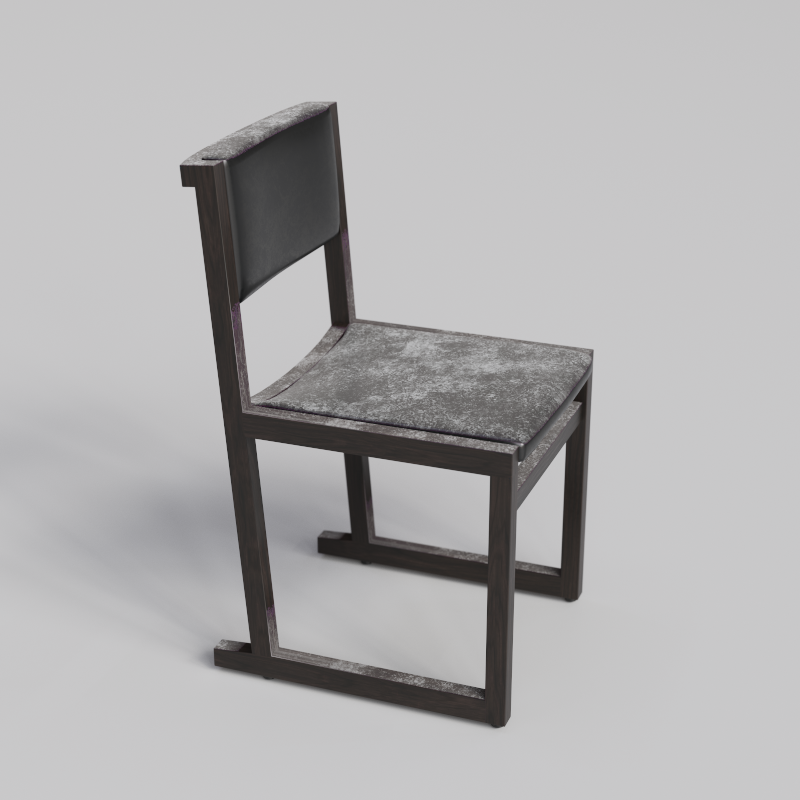}} &
	\raisebox{-0.5\height}{\includegraphics[width=0.19\textwidth]{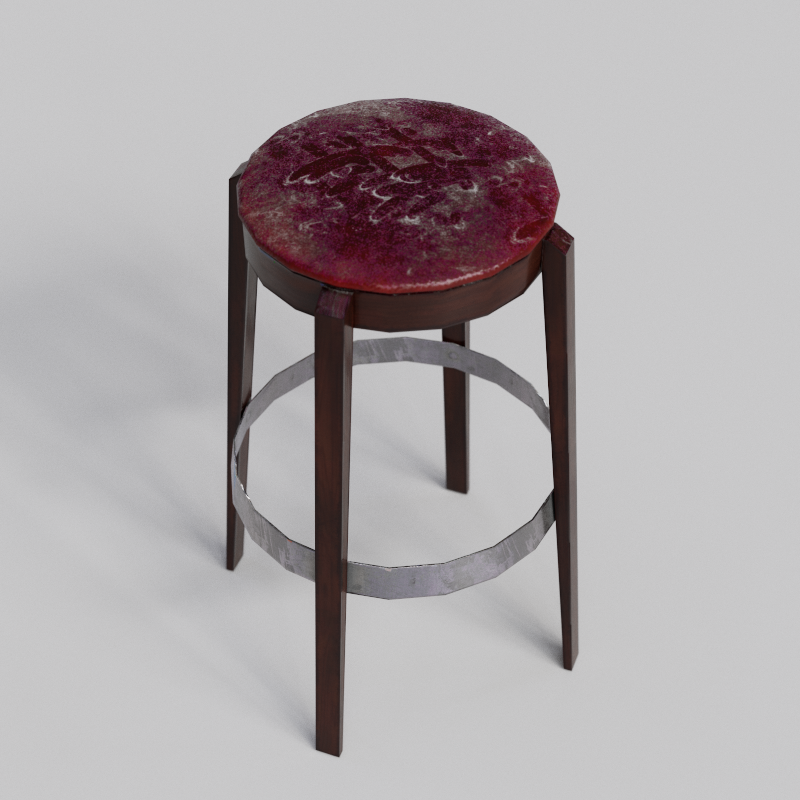}} &
	\raisebox{-0.5\height}{\includegraphics[width=0.19\textwidth]{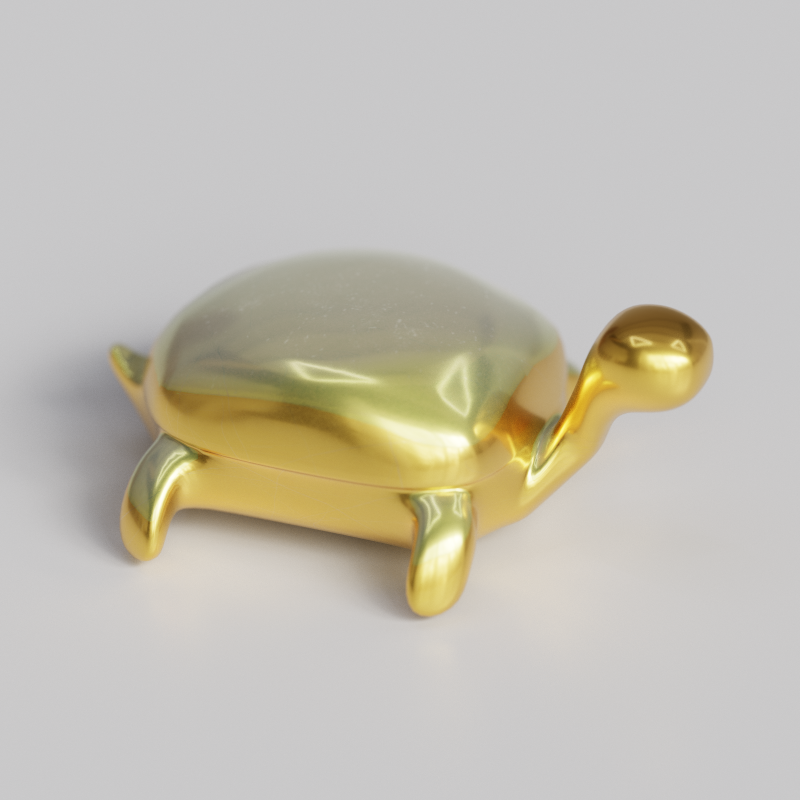}} &
	\raisebox{-0.5\height}{\includegraphics[width=0.19\textwidth]{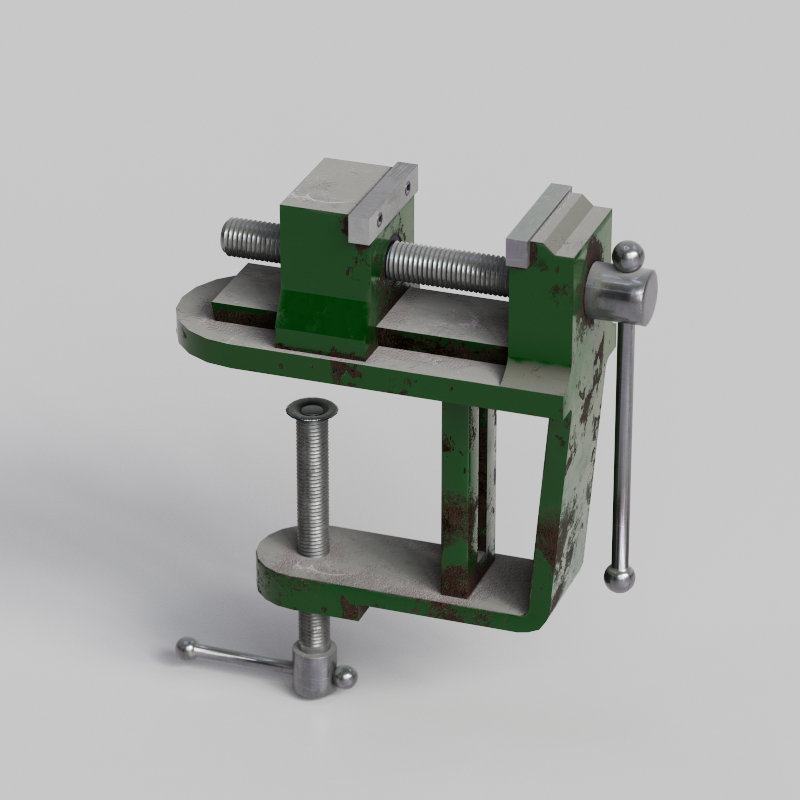}} &
    \raisebox{-0.5\height}{\includegraphics[width=0.19\textwidth]{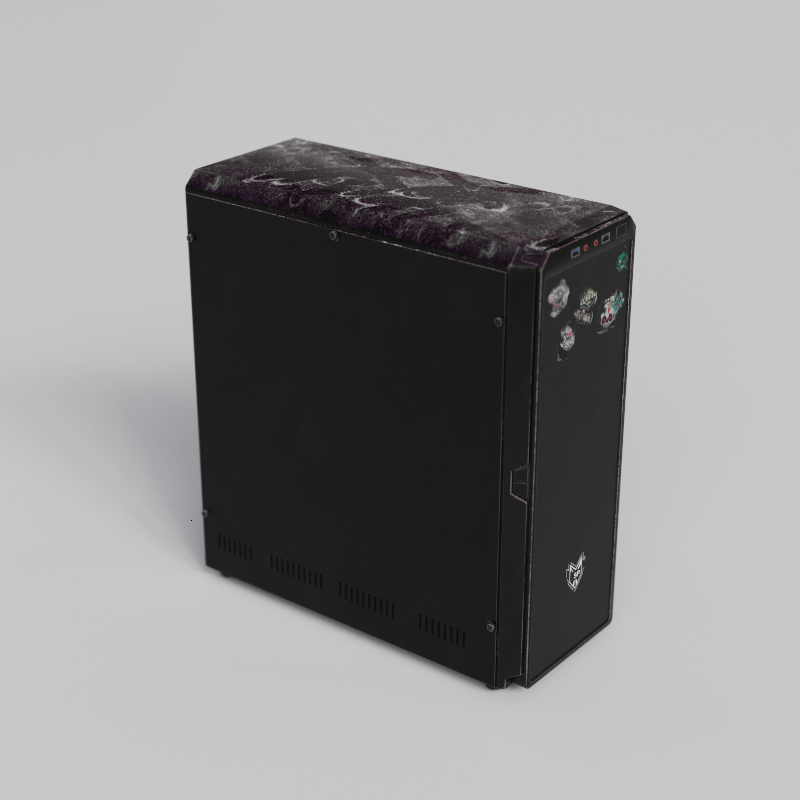}} \\
    
    \vspace{0.025in}
    
    \rotatebox[origin=c]{90}{Clearcoat} &
	\raisebox{-0.5\height}{\includegraphics[width=0.19\textwidth]{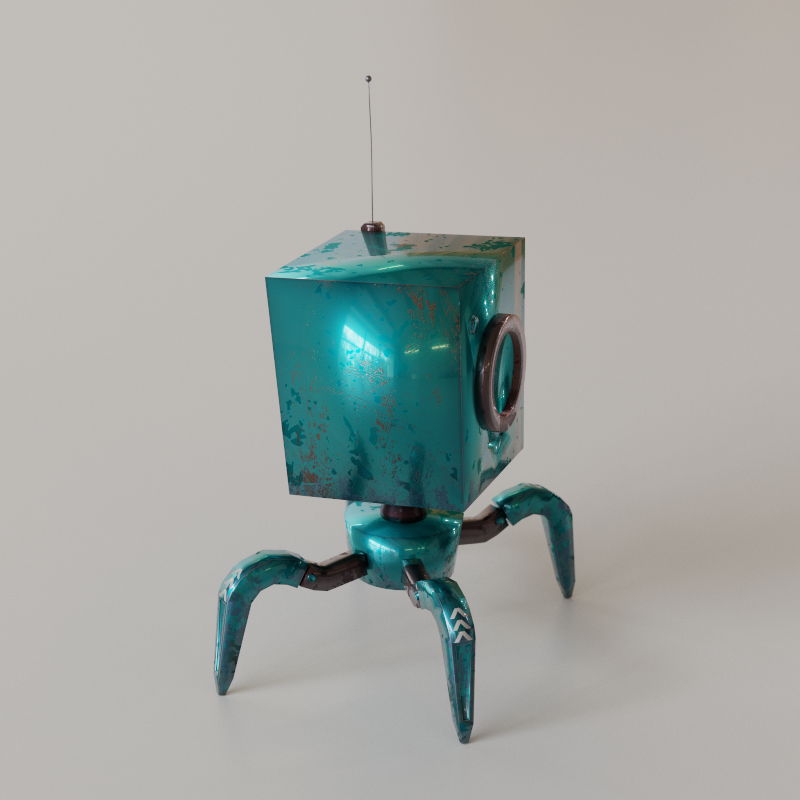}} &
	\raisebox{-0.5\height}{\includegraphics[width=0.19\textwidth]{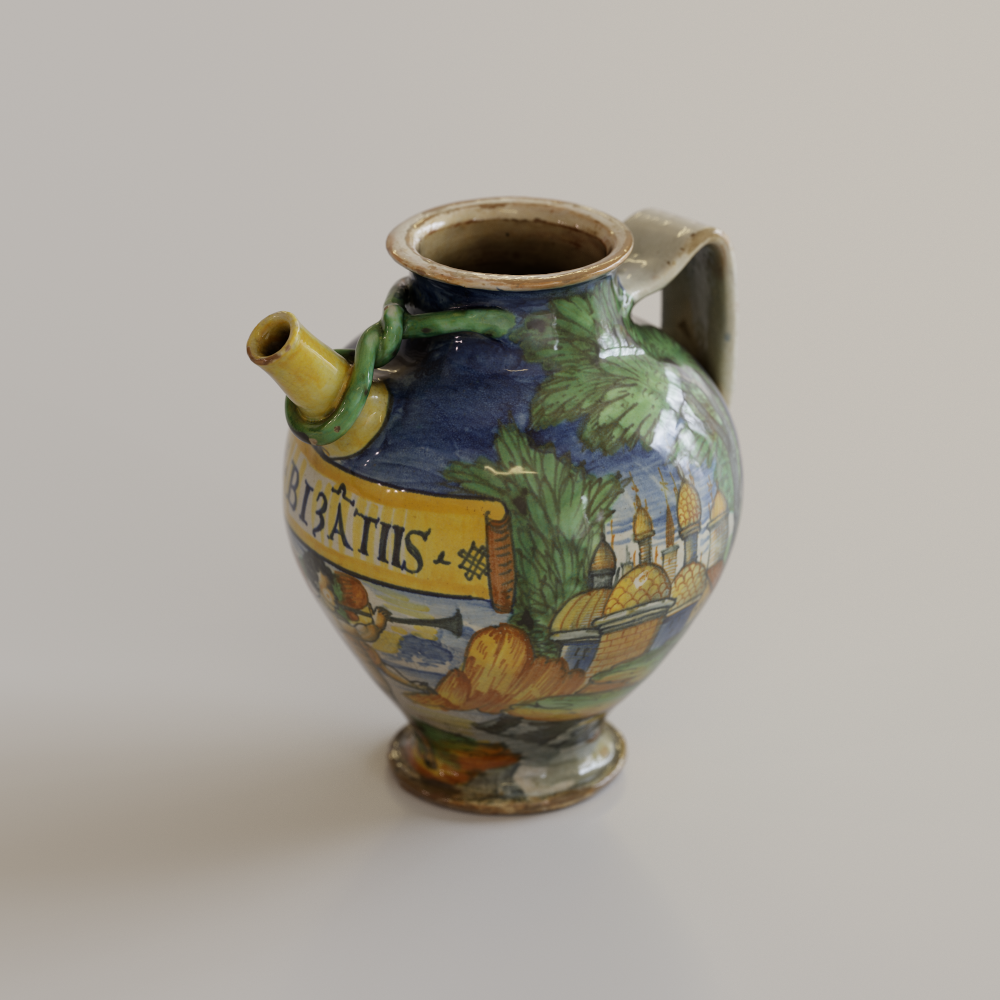}} &
	\raisebox{-0.5\height}{\includegraphics[width=0.19\textwidth]{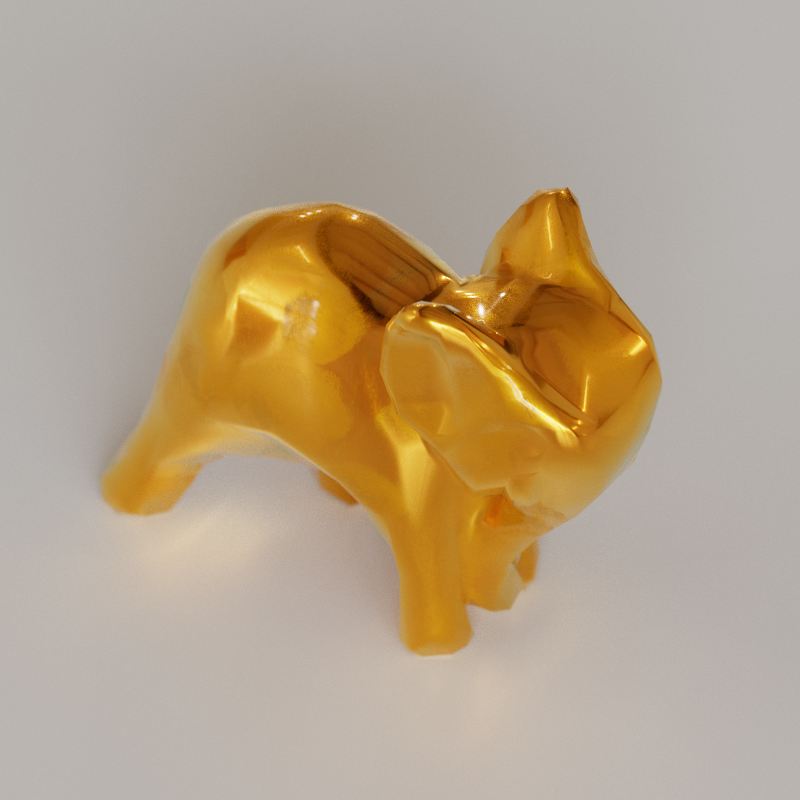}} &
	\raisebox{-0.5\height}{\includegraphics[width=0.19\textwidth]{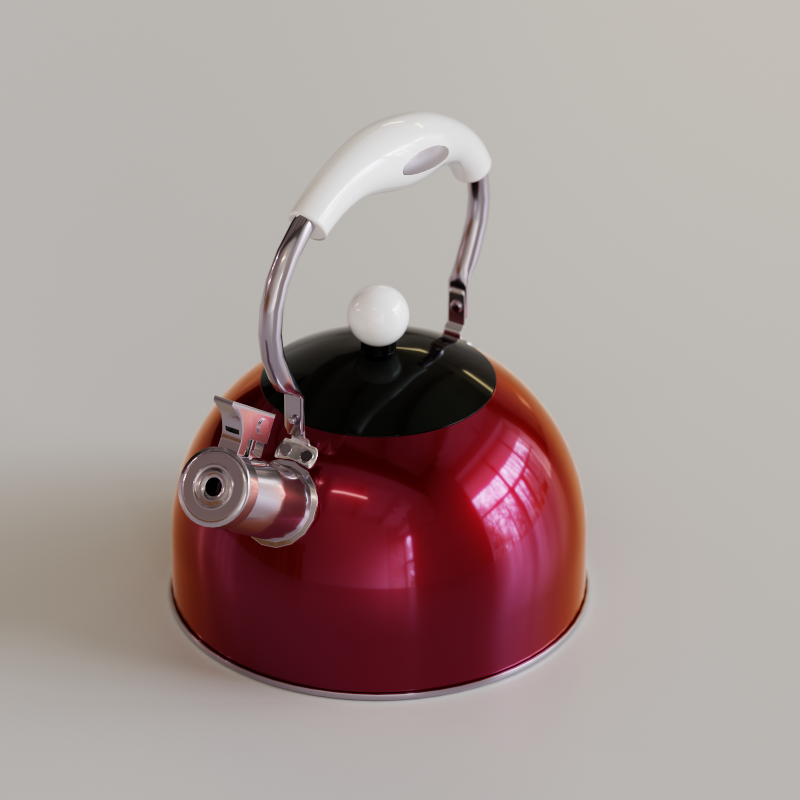}} &
    \raisebox{-0.5\height}{\includegraphics[width=0.19\textwidth]{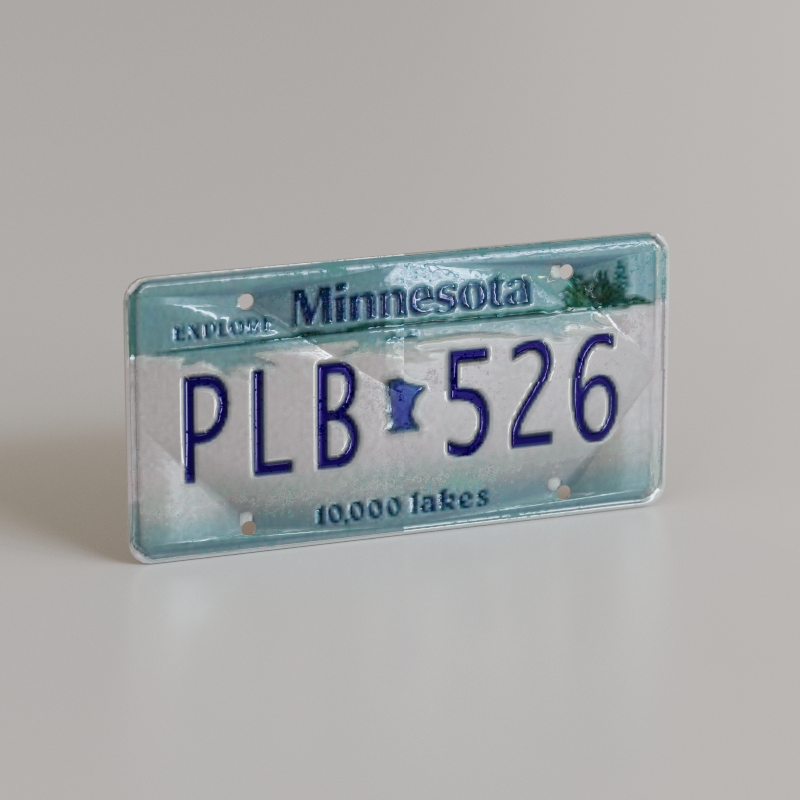}} \\

    \vspace{0.025in}

    \rotatebox[origin=c]{90}{Dust+Clearcoat} &
	\raisebox{-0.5\height}{\includegraphics[width=0.19\textwidth]{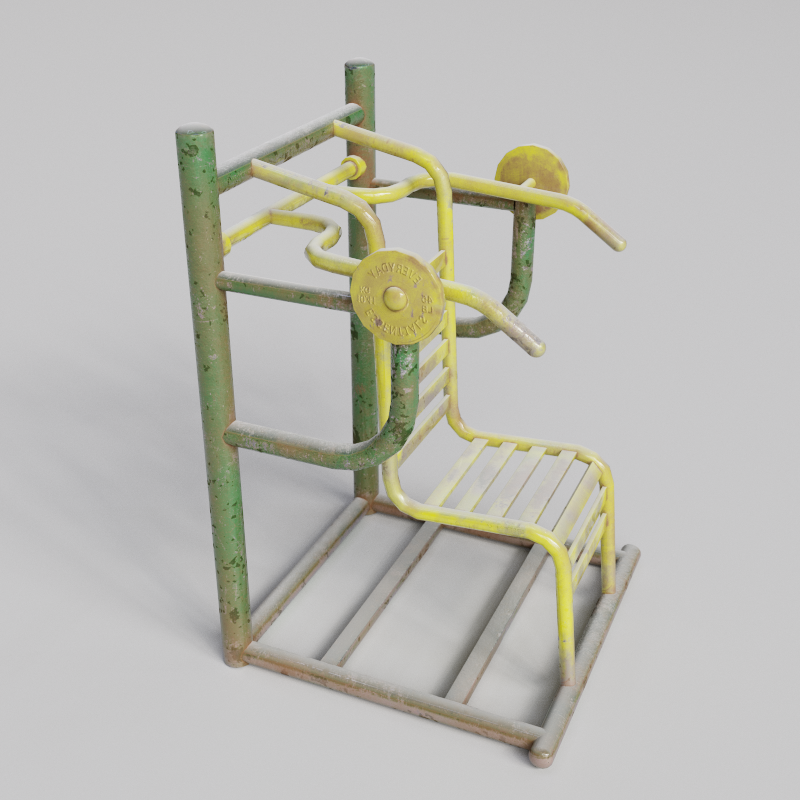}} &
	\raisebox{-0.5\height}{\includegraphics[width=0.19\textwidth]{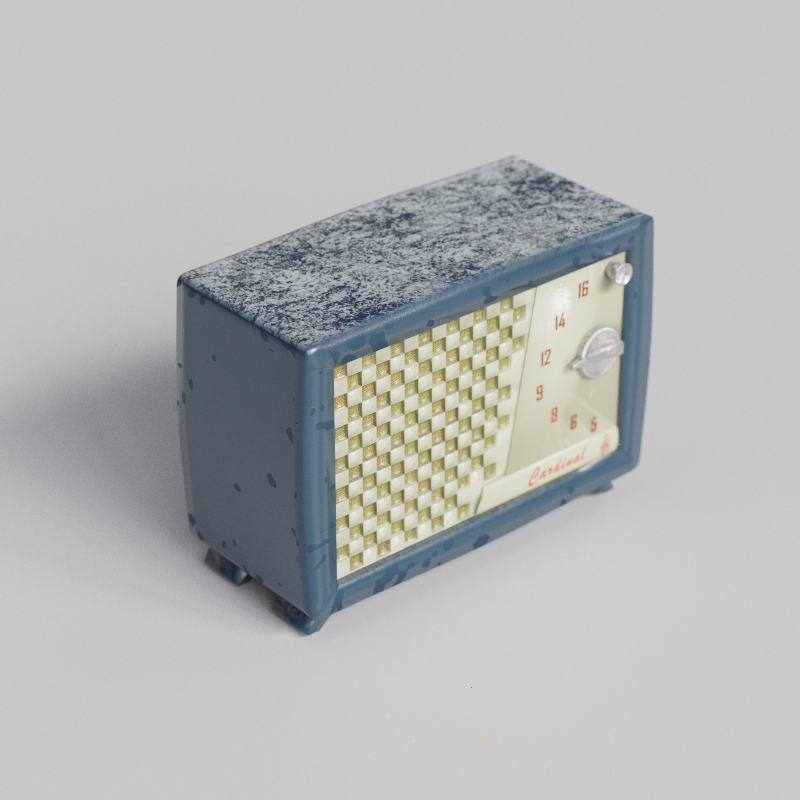}} &
	\raisebox{-0.5\height}{\includegraphics[width=0.19\textwidth]{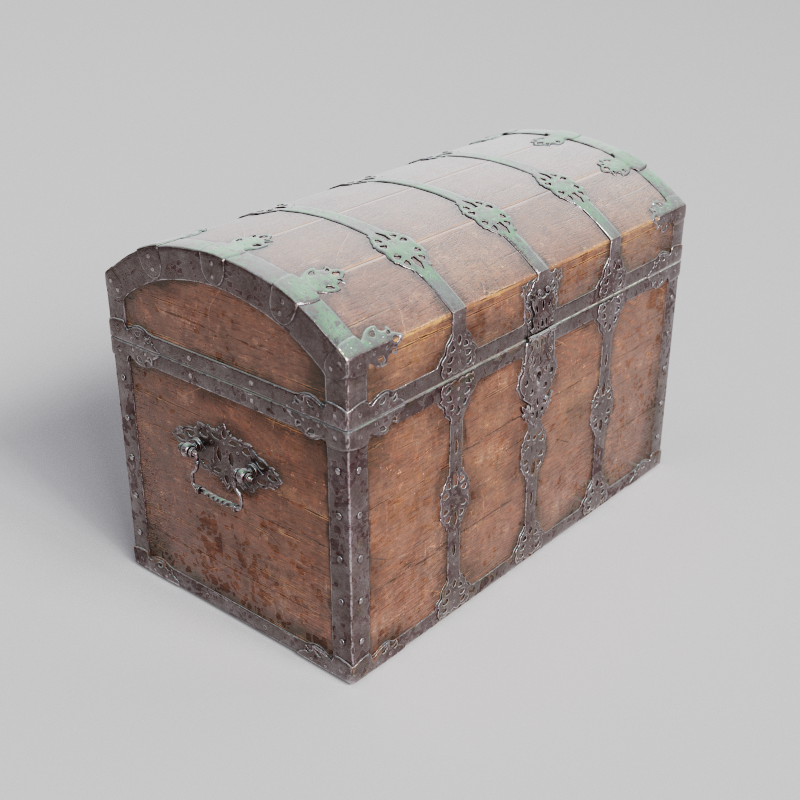}} &
	\raisebox{-0.5\height}{\includegraphics[width=0.19\textwidth]{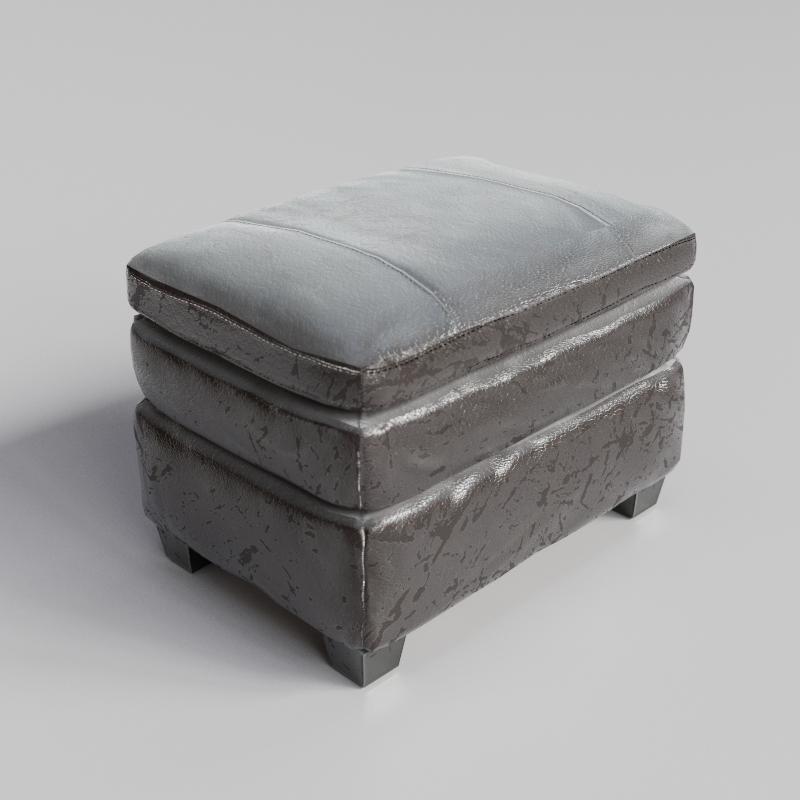}} &
    \raisebox{-0.5\height}{\includegraphics[width=0.19\textwidth]{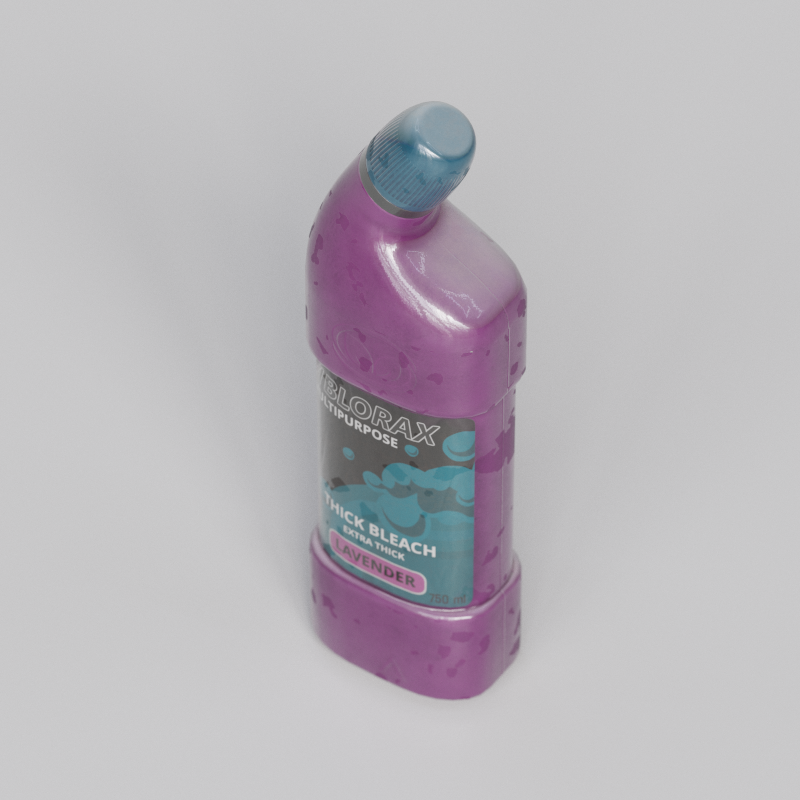}} \\

    \vspace{0.025in}

    \rotatebox[origin=c]{90}{Fuzz+Scatter} &
	\raisebox{-0.5\height}{\includegraphics[width=0.19\textwidth]{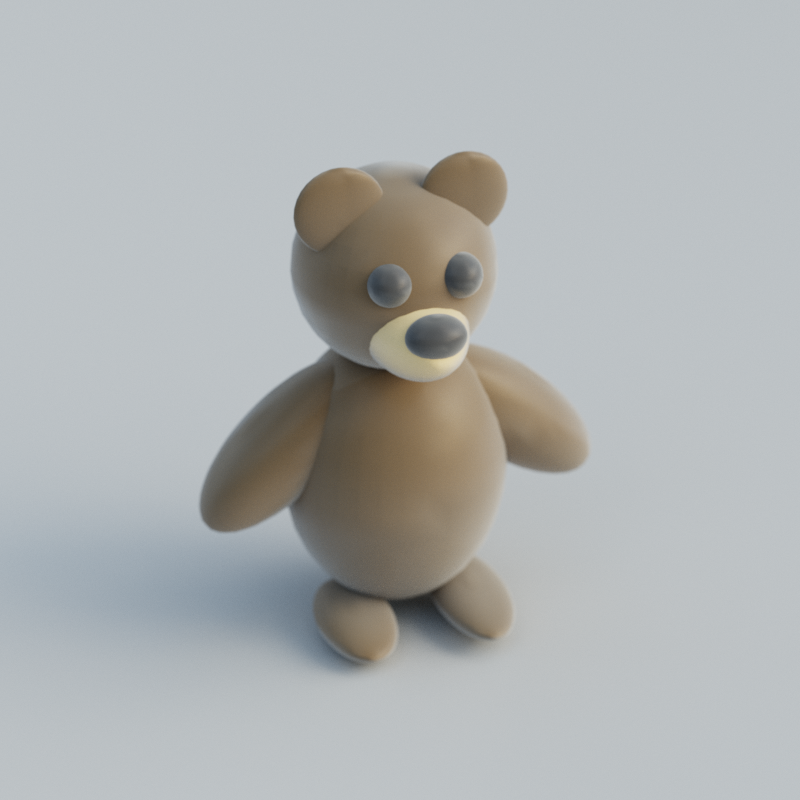}} &
	\raisebox{-0.5\height}{\includegraphics[width=0.19\textwidth]{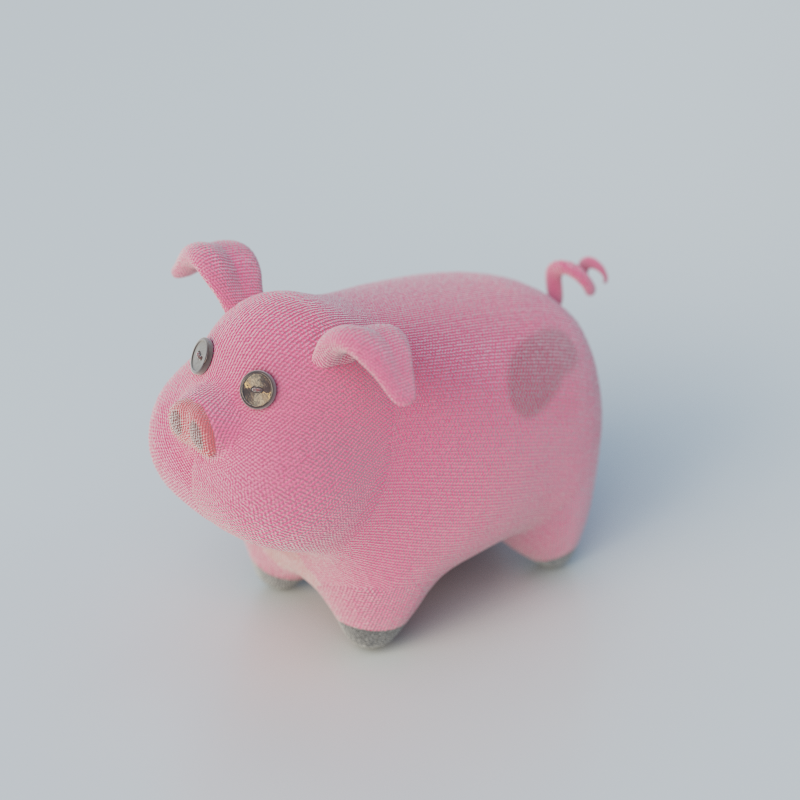}} &
	\raisebox{-0.5\height}{\includegraphics[width=0.19\textwidth]{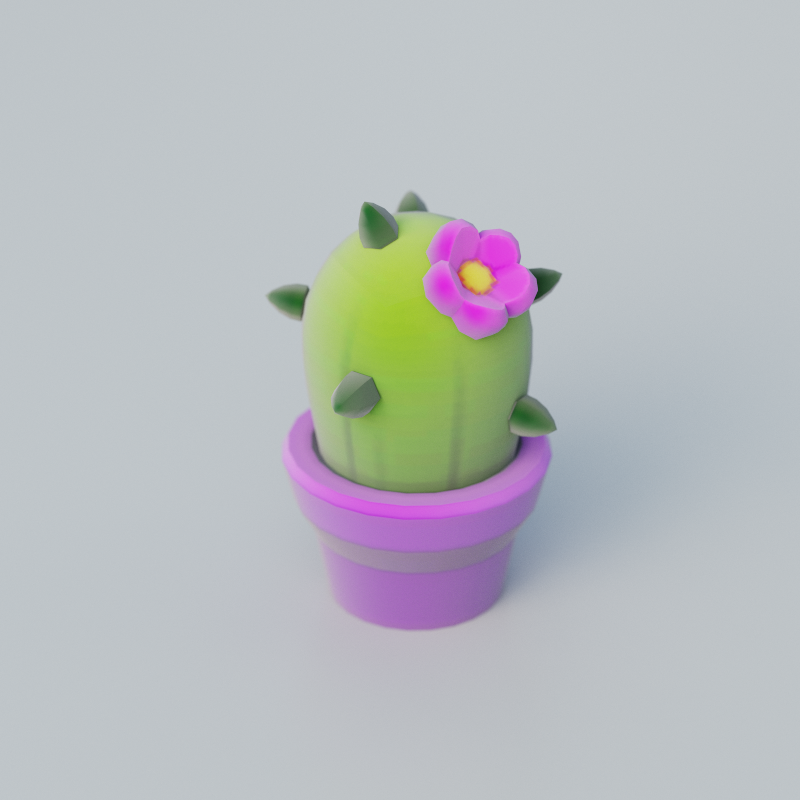}} &
	\raisebox{-0.5\height}{\includegraphics[width=0.19\textwidth]{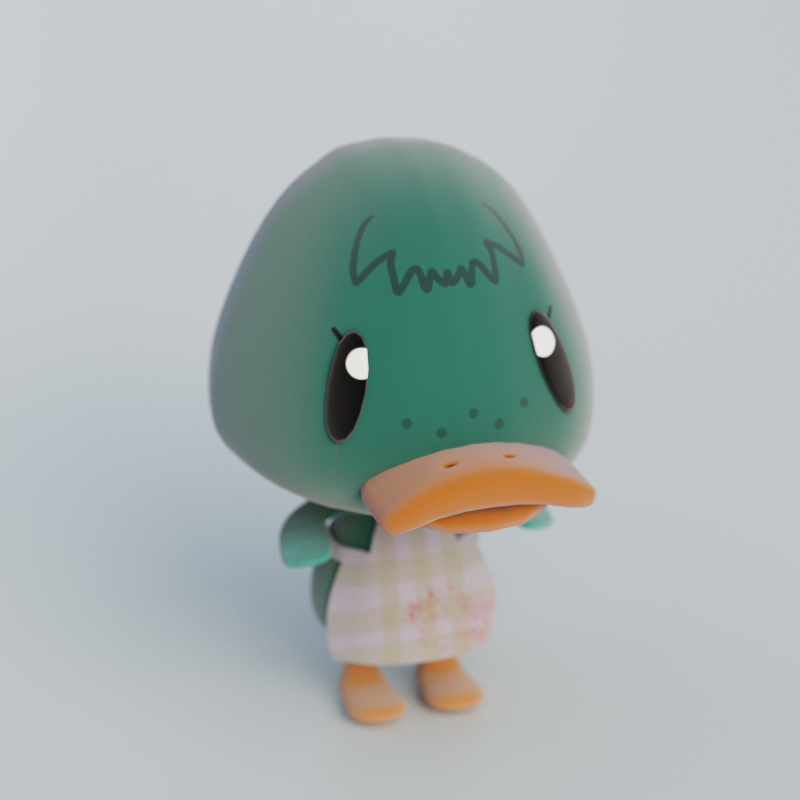}} &
    \raisebox{-0.5\height}{\includegraphics[width=0.19\textwidth]{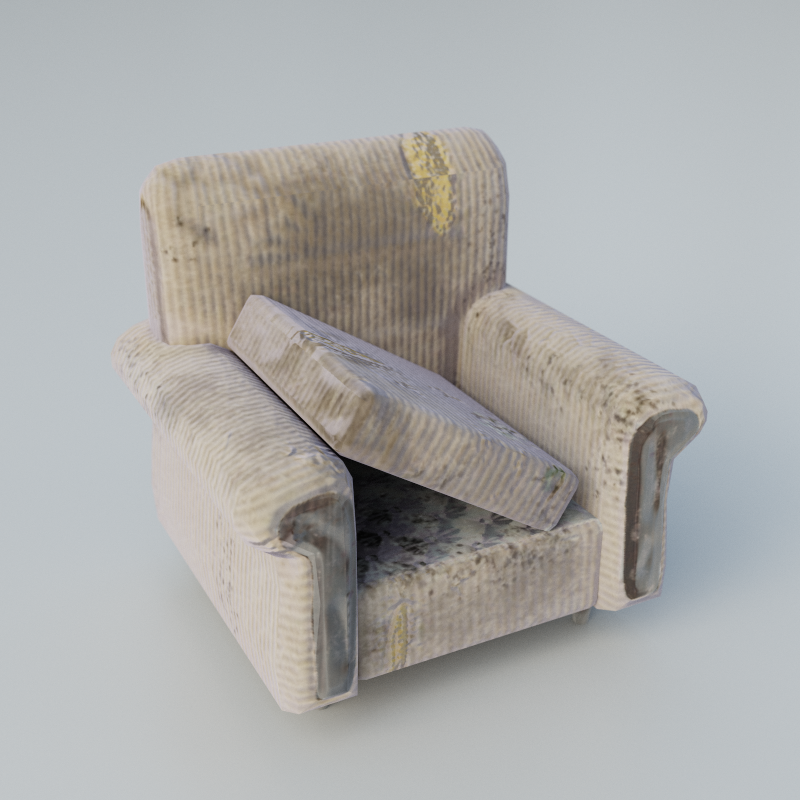}} \\

    \rotatebox[origin=c]{90}{Fuzz+Clearcoat+Scatter} &
	\raisebox{-0.5\height}{\includegraphics[width=0.19\textwidth]{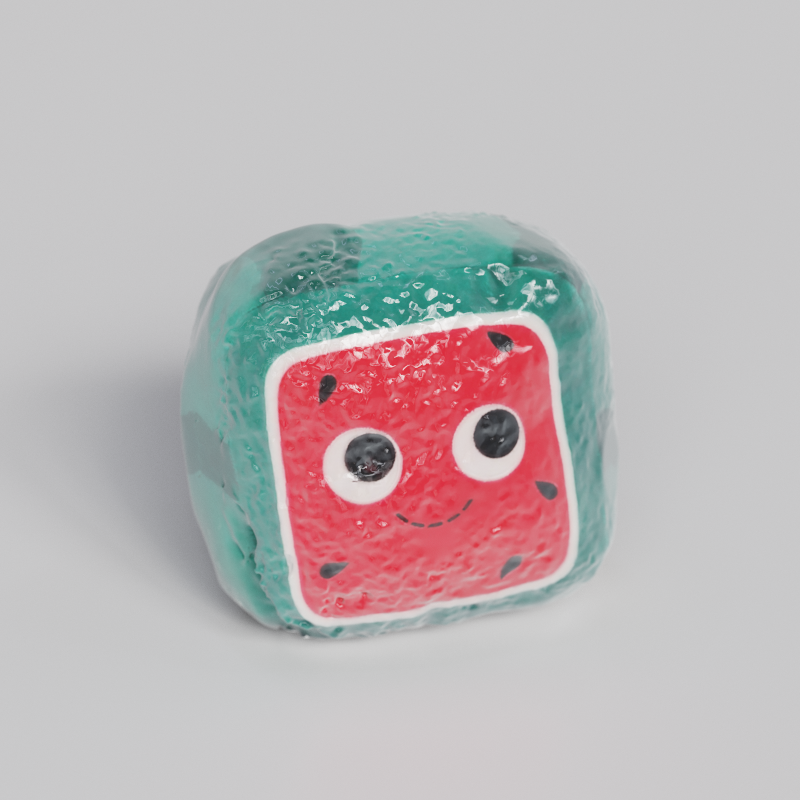}} &
	\raisebox{-0.5\height}{\includegraphics[width=0.19\textwidth]{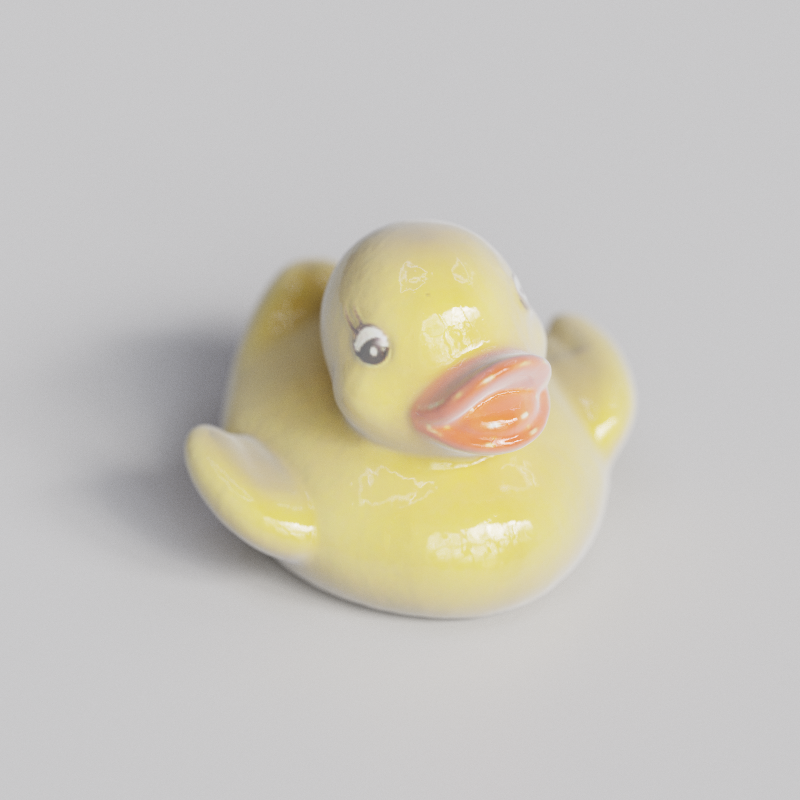}} &
	\raisebox{-0.5\height}{\includegraphics[width=0.19\textwidth]{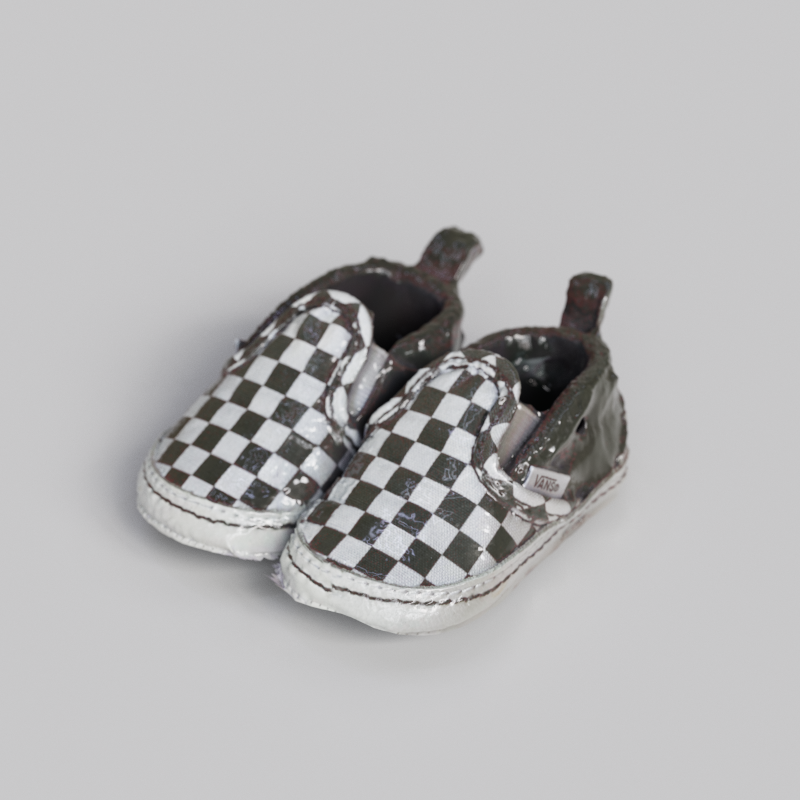}} &
	\raisebox{-0.5\height}{\includegraphics[width=0.19\textwidth]{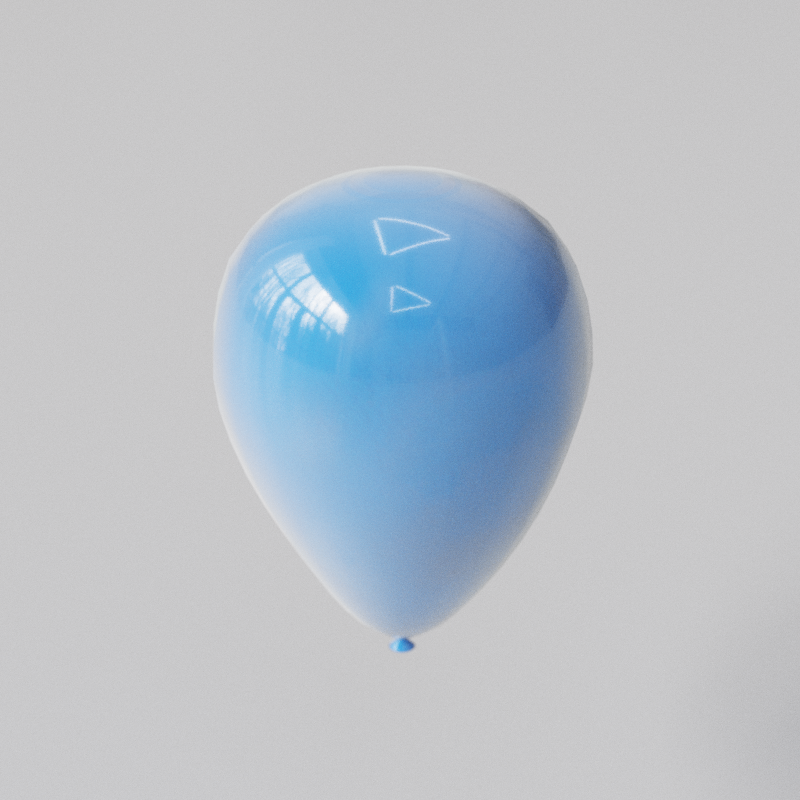}} &
    \raisebox{-0.5\height}{\includegraphics[width=0.19\textwidth]{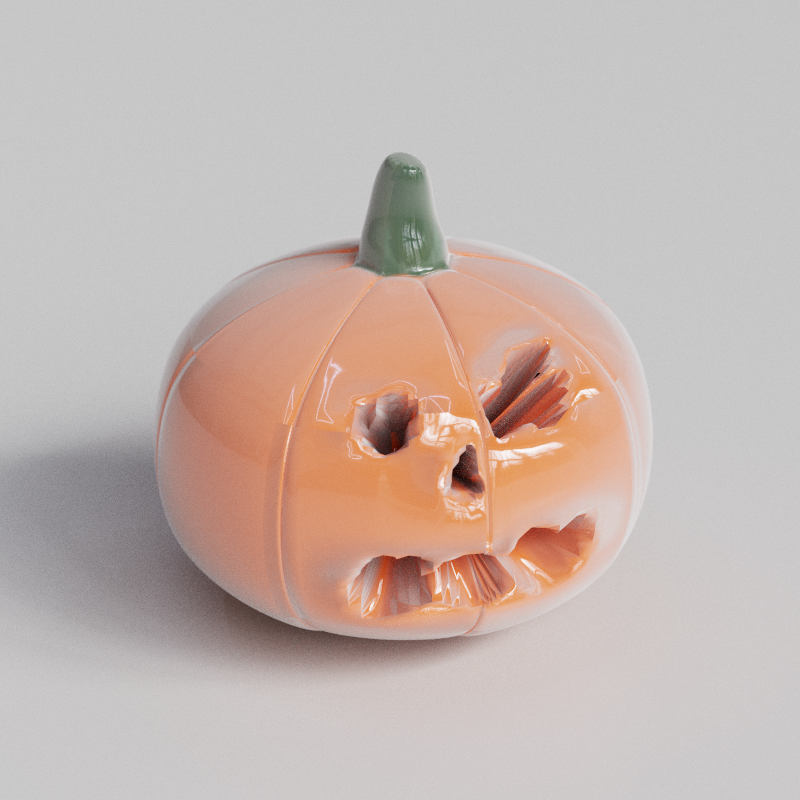}} 

\end{tabular}
\vspace*{-2mm}
\caption{
	3D Models from BlenderVault~\cite{litman2025materialfusion} and Objaverse~\cite{objaverse}
    with materials uplifted with our procedural data enhancement. The materials have been encoded into a compact neural material representation suitable for real-time rendering and large-scale data generation.
}
\label{fig:dataset}
\end{figure*}
}


\newcommand{\figMonkey}{
\begin{figure}
\centering
\begin{tikzpicture}
    \node at (0.0, 0.0) 
    {\includegraphics[width=0.32\linewidth]{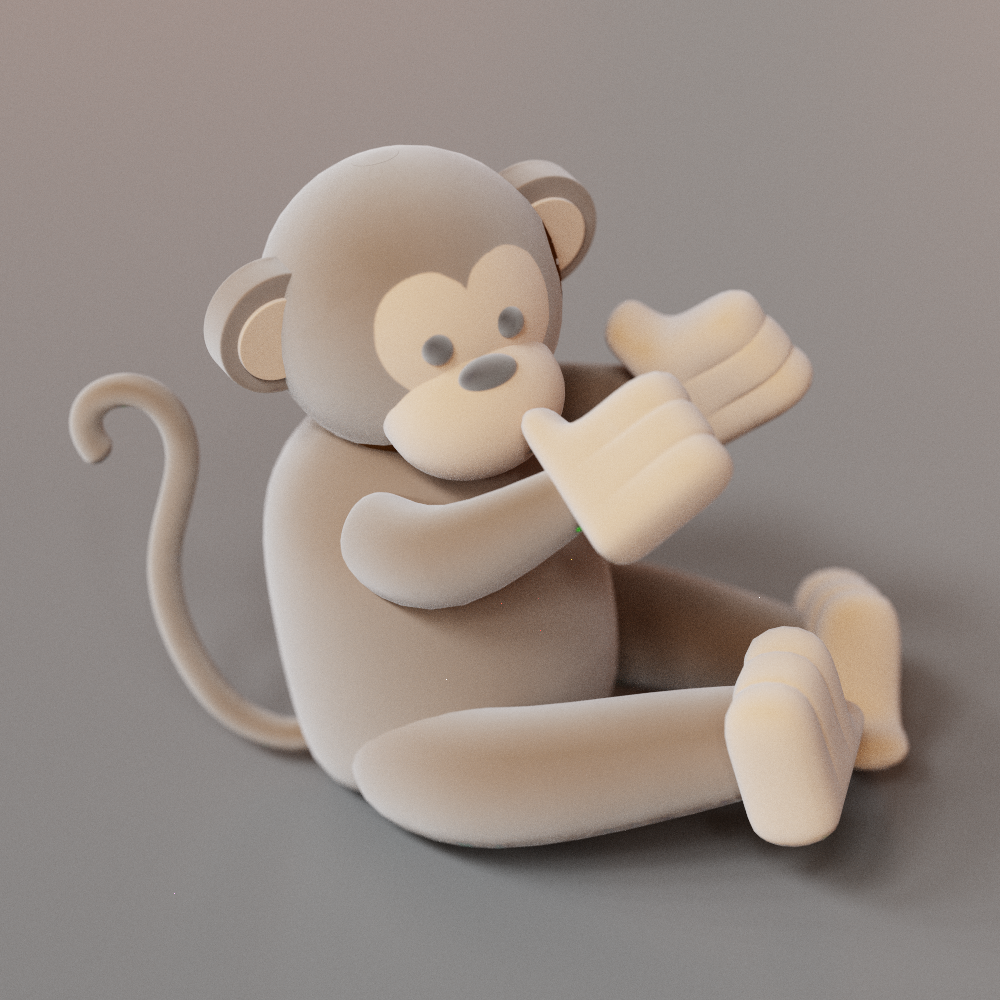}};
    \node at (-0.98, 1.15) {Fuzz};
    \node at (2.85, 0.0) 
    {\includegraphics[width=0.32\linewidth]{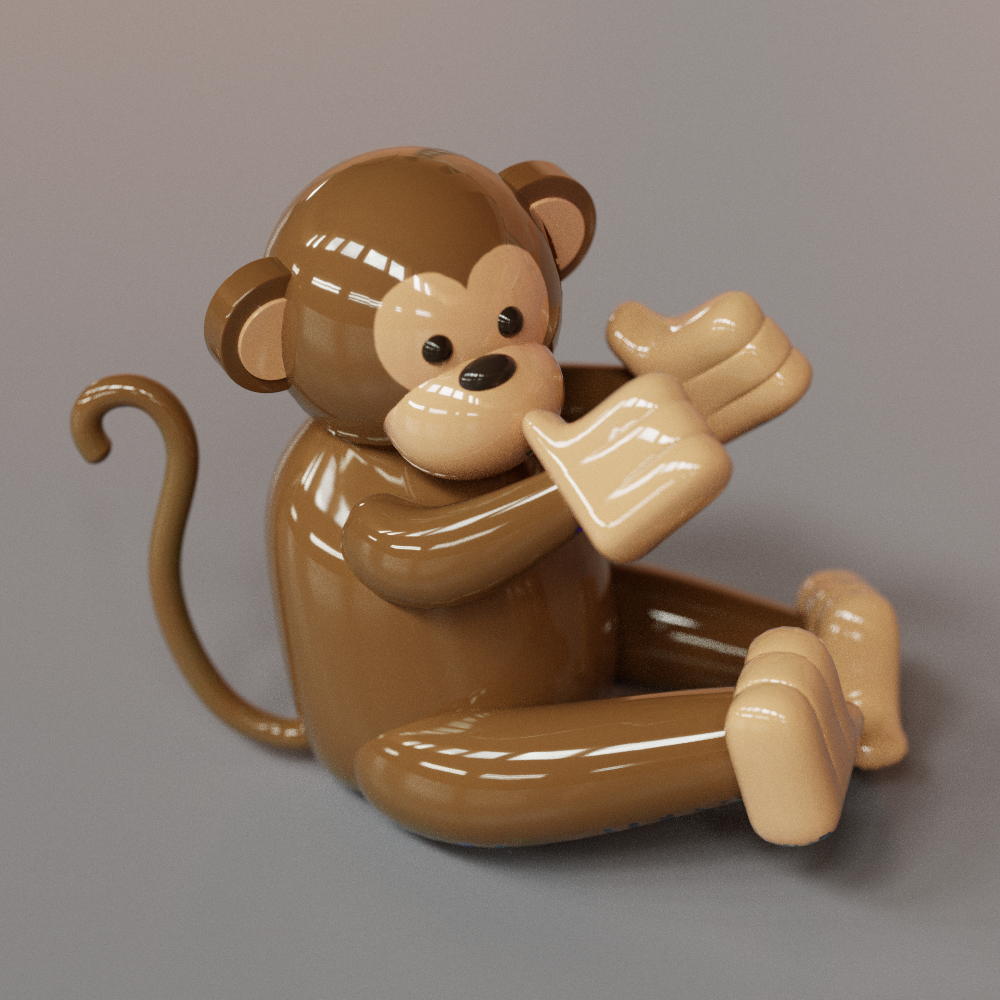}};
    \node at (2.15, 1.15) {Clearcoat};
    \node at (5.7, 0.0) 
    {\includegraphics[width=0.32\linewidth]{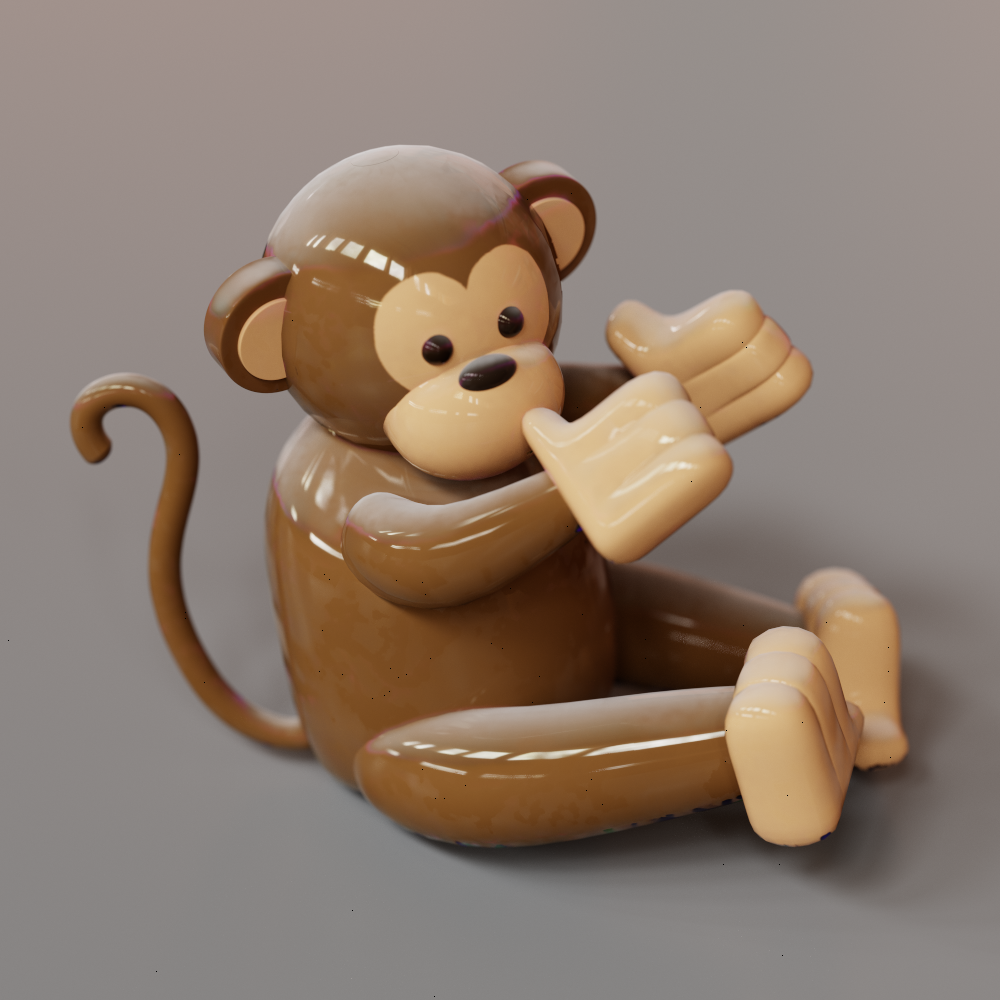}};
    \node at (5.5, 1.15) {Dust + Clearcoat};    
\end{tikzpicture}
\vspace*{-5mm}
\caption{Generated neural materials using three different prompts applied to the same 3D asset. Full text prompts are in the supplemental document.}
\label{fig:monkey}
\end{figure}
}


\newcommand{\figCarIssue}{
\begin{figure}
\centering
\begin{tikzpicture}
    \node at (0.0, 0.0) 
    {\includegraphics[width=0.48\linewidth]{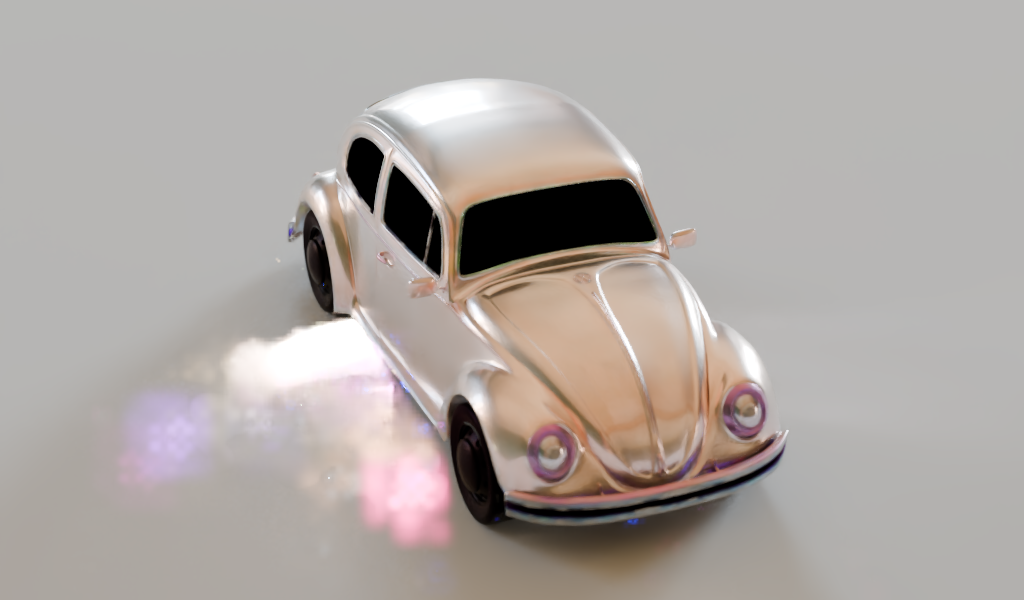}};
    \node at (-0.28, 1.05) {Generated neural material};
    \node at (4.2, 0.0) 
    {\includegraphics[width=0.48\linewidth]{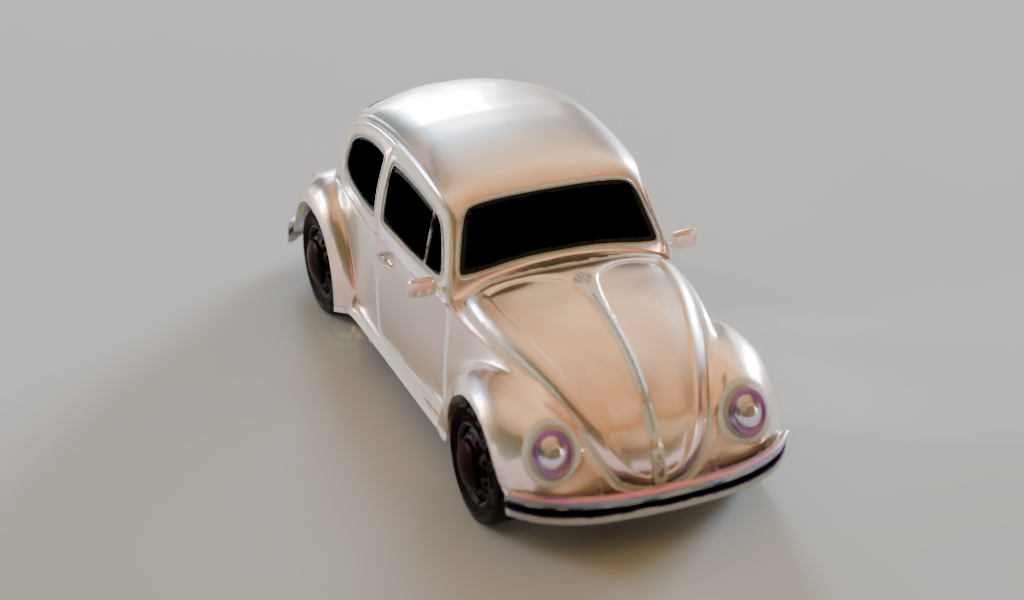}};
    \node at (3.09, 1.05) {w/ erosion fix};
\end{tikzpicture}
\vspace*{-5mm}
\caption{
We occasionally observe artifacts in generated neural materials, likely because the video model only approximates the manifold of valid latent textures. Assets with many UV islands can also introduce errors when projecting generated frames to texture space, especially in poorly covered silhouette regions. As shown above, we mitigate these artifacts by eroding uncovered regions inward by four pixels and using a diffuse material there.
}
\label{fig:car_issue}
\end{figure}
}


\newcommand{\figGlint}{
\begin{figure}
    \vspace*{-2mm}
    \centering
    \begin{tikzpicture}
    \node at (0.0, 0.0) 
    {\includegraphics[width=0.49\linewidth, trim={0 20mm 0 20mm}, clip]{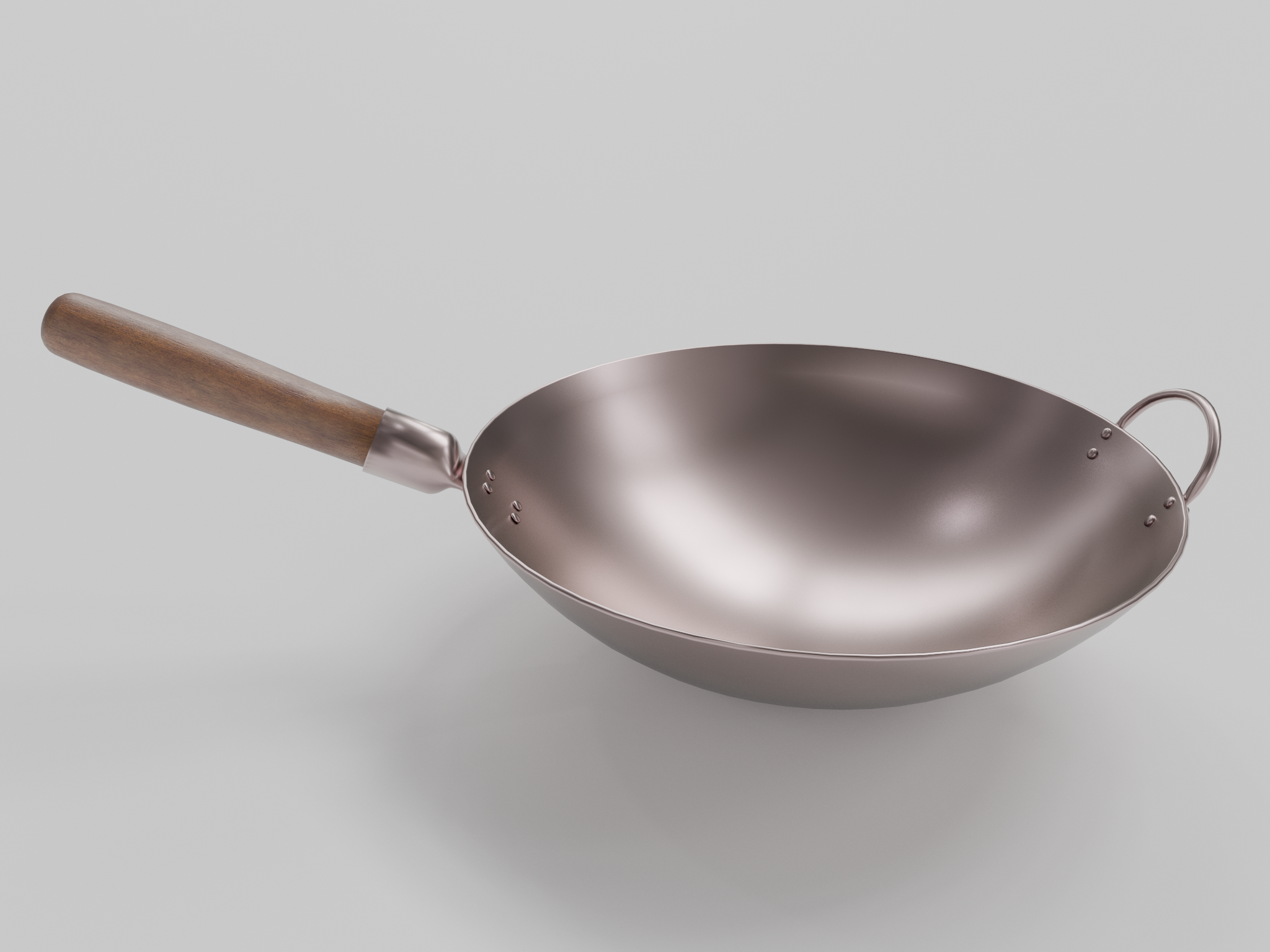}};
    \node at (4.27, 0.0) 
    {\includegraphics[width=0.49\linewidth, trim={0 20mm 0 20mm}, clip]{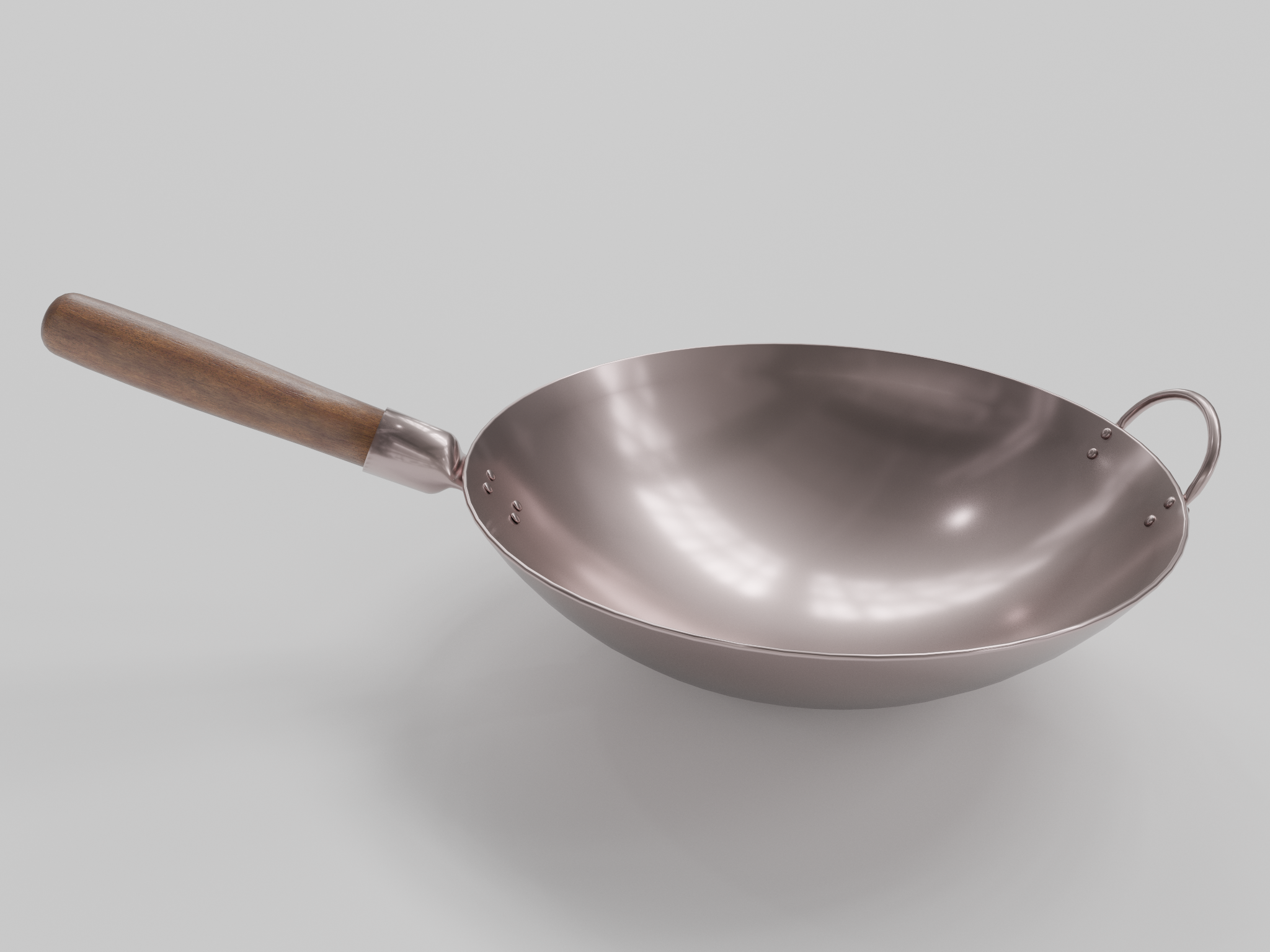}};
    \node at (0.0, -3.05) 
    {\includegraphics[width=0.49\linewidth, trim={0 20mm 0 20mm}, clip]{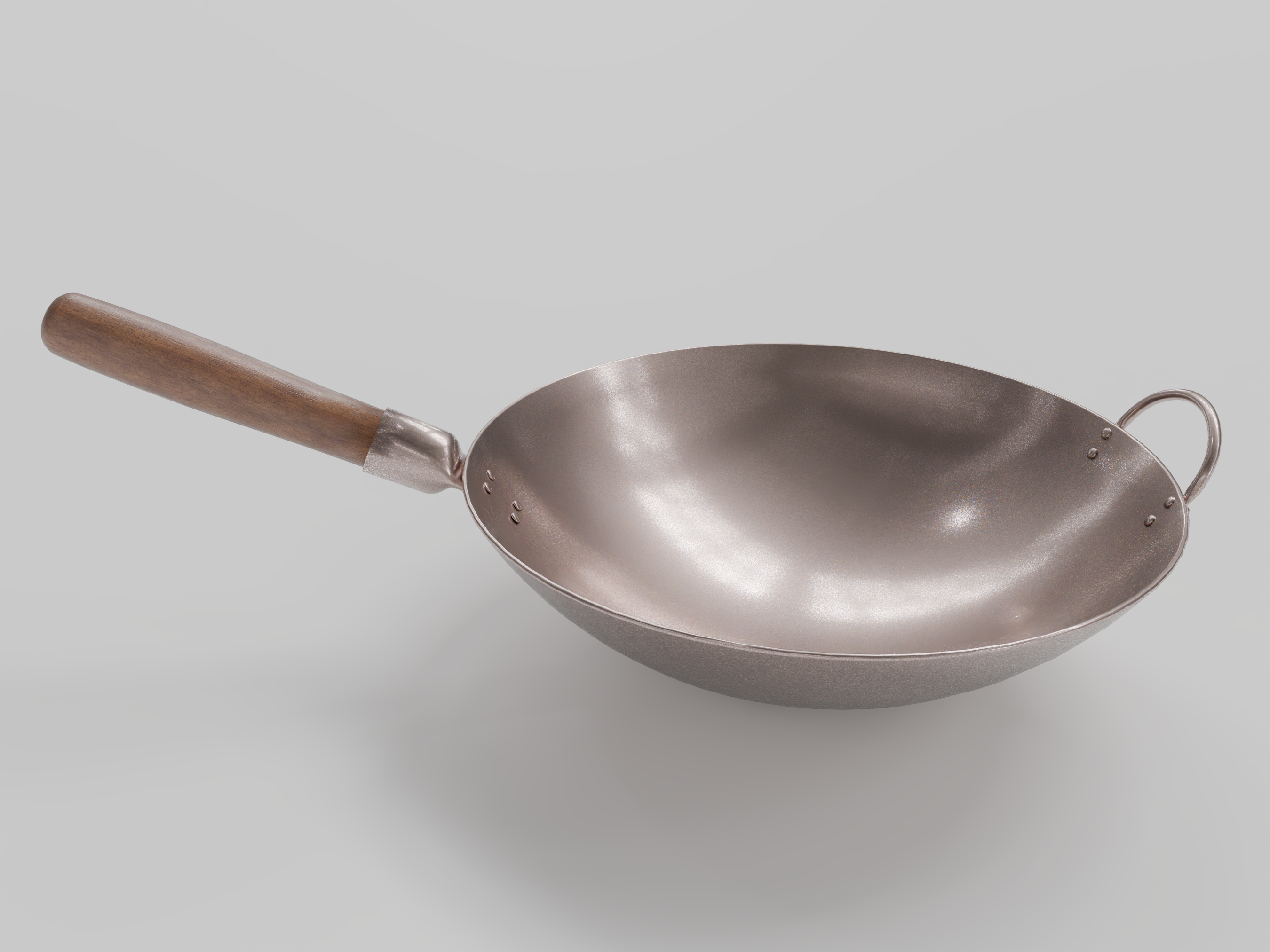}};
    \draw [draw=orange, line width=0.5mm] (-0.62, -3.21) rectangle ++(0.55, 0.42);
    \node at (4.27, -3.05)
    {\includegraphics[width=0.49\linewidth, trim={0 20mm 0 0mm}, clip]{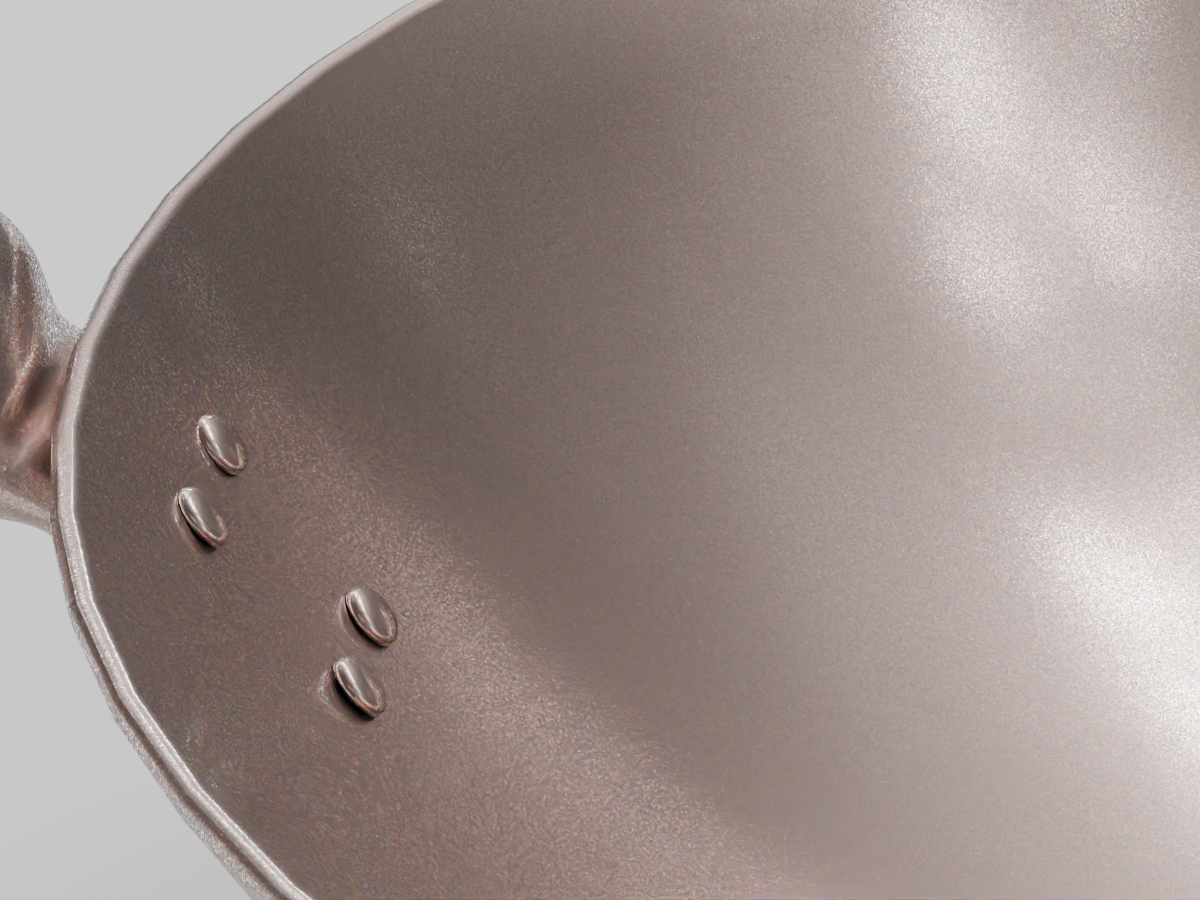}};
    \draw [draw=orange, line width=0.5mm] (2.2, -4.5) rectangle ++(4.15, 2.9);
    \node [black] at (-1.05, 1.1) {(a) Simple PBR};
    \node [black] at (2.8, 1.1) {(b) Haze};
    \node [black] at (-1.0, -1.85) {(c) Haze + Glint};
    \end{tikzpicture}
    \vspace*{-4mm}
    \caption{Procedural enhancement of a metallic material. (a) Simple PBR model (b) Our haze enhancement, producing both sharper highlights and an overall hazy appearance. (c) Haze and glint enhancement (zoom in to compare with ``Haze''), with closeup of colored glints on the right.}
    \label{fig:wave}
\end{figure}
}

\maketitle

\section{Introduction}

Generative models recently emerged as a new way of authoring materials. However, the quality of these models is inherently bound to the quality and coverage of input training data. While large datasets exist for simplified appearance models using the basic combination of a diffuse and a GGX specular BSDF~\cite{10.5555/2383847.2383874}, these models fail to encompass the visual richness of real-world surfaces.

\begin{figure}
    \centering
    \includegraphics[width=0.24\linewidth]{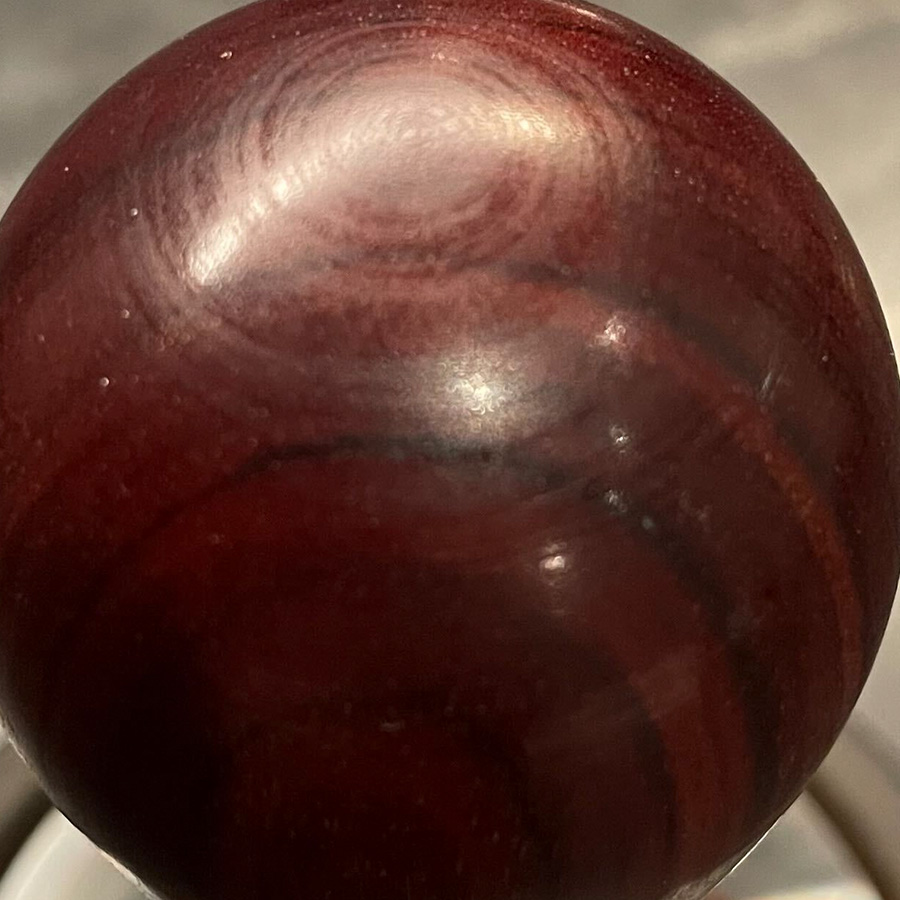}
    \includegraphics[width=0.24\linewidth]{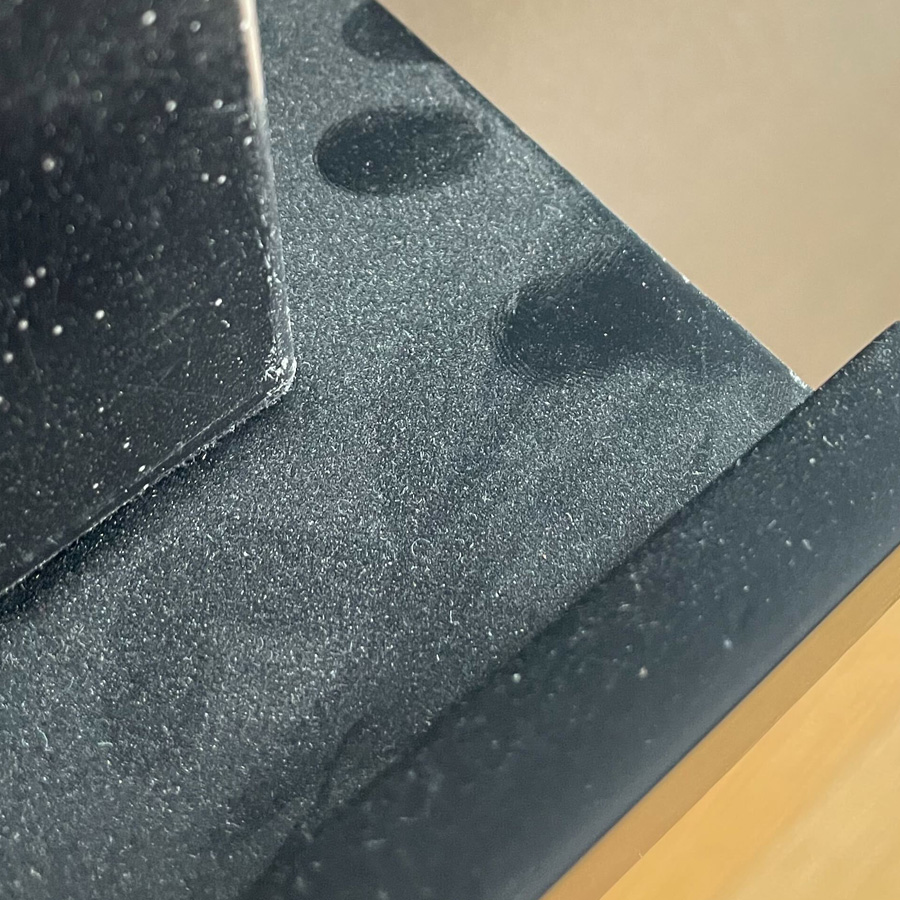}
    \includegraphics[width=0.24\linewidth]{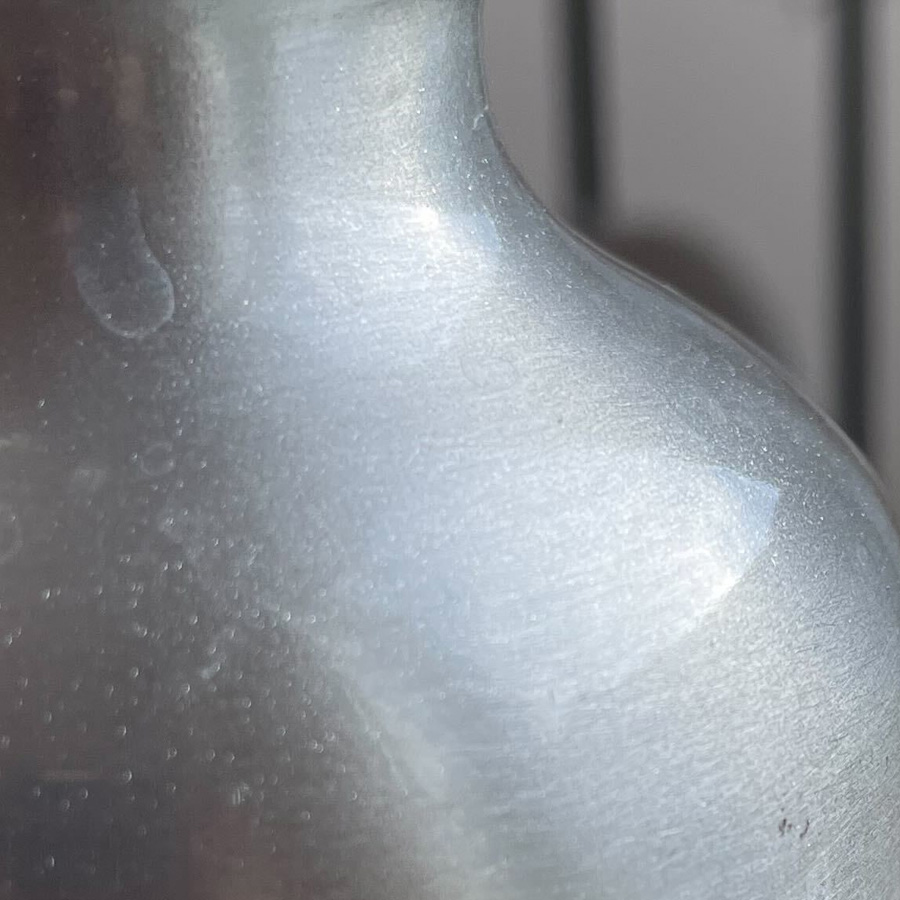}
    \includegraphics[width=0.24\linewidth]{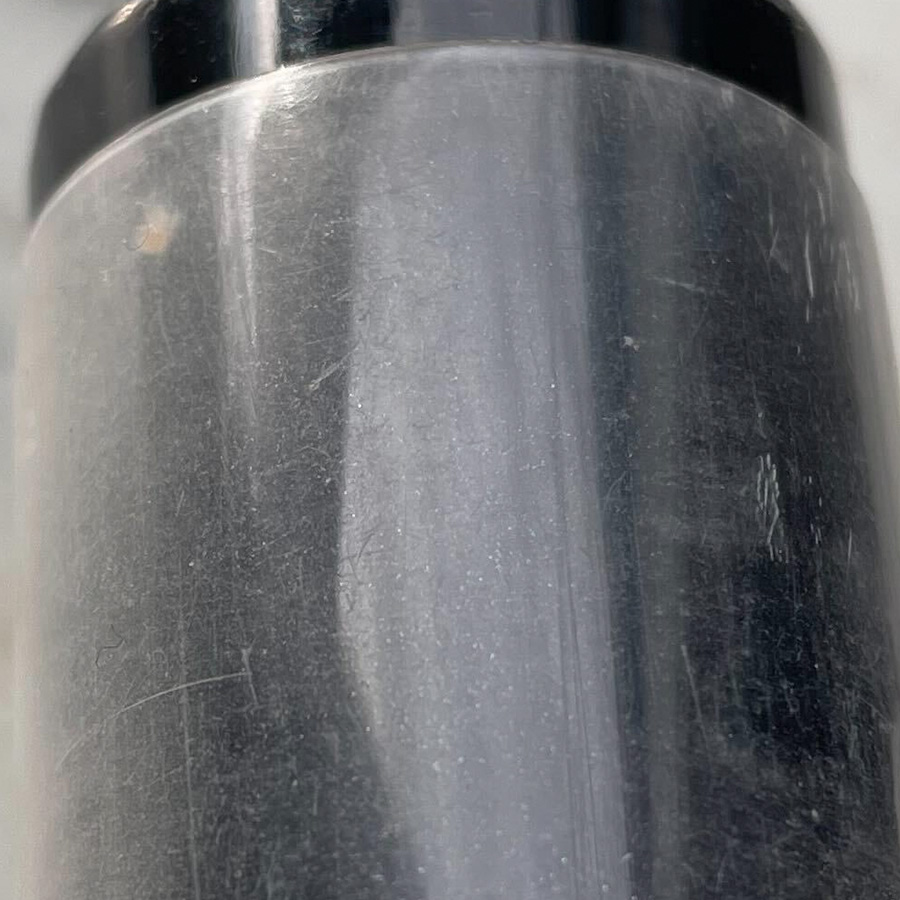}
    \caption{Simple PBR material models cannot reproduce complex surface effects of real materials such as (from left to right) double highlights (haze), dust, clearcoat, or thin translucency.}
    \label{fig:photos_real}
\end{figure}

As can be seen in Fig.~\ref{fig:photos_real}, real-world materials show effects like clearcoat, dust and other imperfections, which are necessary to achieve realistic appearances but require more expressive models. Such models include OpenPBR~\cite{10.1145/3744199.3744632}, the Disney Principled BSDF~\cite{Burley2012}, or even custom BSDFs and layering solutions like MaterialX~\cite{MaterialXSiggraph}, and are commonly used in VFX. However, large-scale datasets for these high-fidelity models do not exist. Since manually authoring complex materials is prohibitively expensive, the field lacks the data necessary to train the next generation of expressive material synthesizers. 

Unlike prior work that focuses on enhancing texture resolution~\cite{gauthier2024matup} or detail while retaining simplified shading models, leaving the BSDF representation untouched~\cite{Hadadan2025GenerativeDetail}, our goal is to expand the appearance space of the BSDF itself. To achieve this, we propose a procedural dataset augmentation pipeline that uses simple, widely available low-dimensional material data as input and uplifts it into a more expressive layered BSDF representation, guided by physical principles and domain priors. We further encode our uplifted data into a compact, 6D universal neural representation, with a regularized latent space suitable for downstream generative applications.

\begin{figure}
    \centering
    \includegraphics[width=0.99\linewidth]{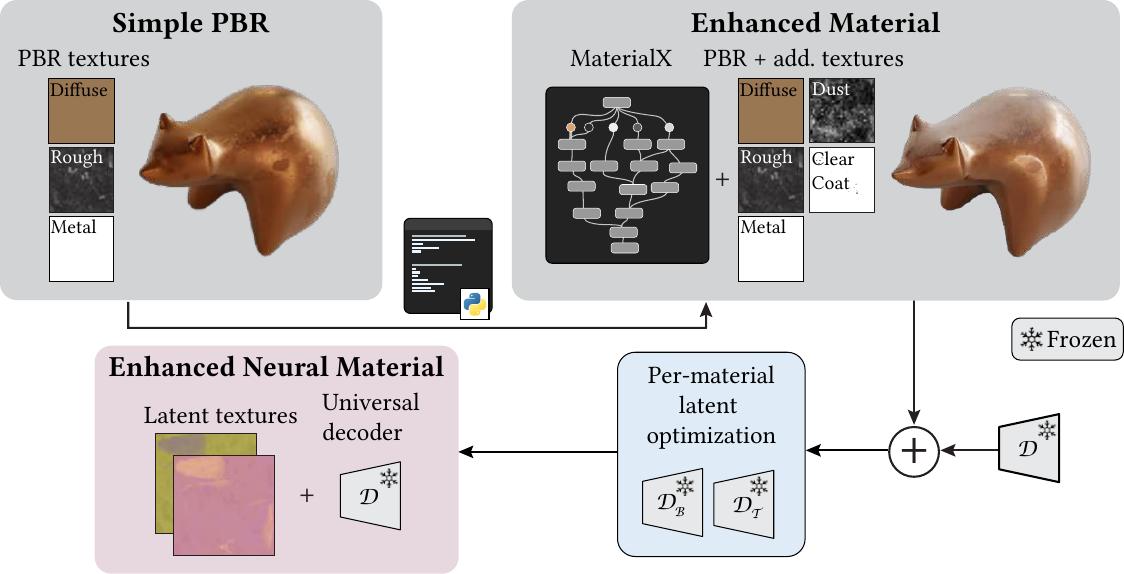}
    \caption{Our uplifting pipeline. Procedural rules create an enhanced material as MaterialX graph while reusing existing textures. We use a universal decoder $\mathcal{D}$ with frozen weights and optimize only the latent textures.}
    \label{fig:block}
\end{figure}

The contribution of this work is a pipeline (Fig. \ref{fig:block}) for constructing a large dataset of expressive neural materials from existing PBR datasets: we describe our per-material procedural enhancement process, explaining design decisions (Section \ref{sec:enhance}), and discuss our neural compression of the enhanced materials (Section \ref{sec:neural}). As an example demonstration, we use our resulting dataset to train a generative model that can produce materials with improved visual richness compared to simple PBR materials.

\section{Related Work}
Our work builds on prior advances in procedural material modeling, neural materials, and generative models, as reviewed in this section.

\subsubsection*{\textbf{Procedural Material Enhancement}}
Computer graphics has long relied on specialized models to capture distinct surface appearances. Dedicated models exist for e.g. fuzz~\cite{10.1145/3532836.3536240}, dust~\cite{10.1145/3658224}, haze~\cite{https://doi.org/10.1111/cgf.13475}, and clearcoat~\cite{10.1145/1321261.1321292}, but these effects are often underrepresented in training data for generative material models. Structural enhancements on analytical materials are typically restricted to microscale effects like the addition of flakes~\cite{10.1145/2897839.2927391, 10.1007/s11390-024-4123-3, Chermain2020Procedural}, scratches~\cite{10.1145/3130800.3130840, 10.1145/2897824.2925945} and other imperfections~\cite{10.1145/2815618, 10.1145/3355089.3356525}. To expand the appearance space, materials can be synthesized from measured data~\cite{M3ashy2026}, but such data lacks spatial variation. We consider texture resolution improvements or techniques that purely operate on PBR textures like~\cite{chen2025pbrsr} orthogonal to our work; these textures are inputs to our uplifting. 

Our uplifted BRDF is a deterministic, analytic model that avoids precomputation \cite{10.1145/2601097.2601139, 10.1145/3197517.3201321}, making training and evaluation efficient relative to position-free layering \cite{10.1145/3272127.3275053}. Neural layering is another option \cite{10.1145/3618365, 10.1145/3528233.3530732}, but our analytic form is simpler and easier to extend to new materials. Volumetric methods \cite{10.1145/3546940} are less useful, as input data is lacking. Unlike fixed-structure models such as OpenPBR, our model is easy to extend through adding BSDFs.

\subsubsection*{\textbf{Neural Materials}}
Neural materials~\cite{kuznetsov2021neumip,zeltner2024nm} replace analytical BSDFs and textures with small neural networks and corresponding latent textures. They provide a powerful way of representing self-shadowing, fibers, and complex micro-structures. Recently, there is also a \emph{generative} neural 
material approach~\cite{raghavanmullia2025genneumat} for 2D material patches, which combines a universal MLP defining a basis for neural materials and a conditional diffusion model for generating neural material latents from text or images. In this work, we extend the generative neural material approach to texture-mapped 3D objects, creating object-specific materials with enhanced expressivity.

\subsubsection*{\textbf{Diffusion Models}}
Image diffusion models add noise to an image through a sequence of diffusion steps. They are trained to reverse this process, enabling sample generation by iterative denoising starting from Gaussian noise. Many generative models have been developed based on similar principles~\cite{sohl2015deep, ho2020denoising, dhariwal2021diffusion}. Video diffusion models~\cite{blattmann2023videoldm, blattmann2023svd, hong2023cogvideo, yang2024cogvideox, cosmos_short, wan2025} extend image-based diffusion to the temporal domain, enabling video generation from text or images. 
In our application of generative neural materials, we build upon the Cosmos~\cite{cosmos_short} video diffusion model.

\subsubsection*{\textbf{Material Extraction using Diffusion}} 
Various approaches combine image diffusion models with inpainting, or texture refinement~\cite{richardson2023texture,chen2023text2tex,zeng2024paint3d,yeh2024texturedreamer}. By finetuning image diffusion models, conditioning on geometry and lighting~\cite{Zhang2024dreammat,deng2024flashtex}, light and material disentanglement are improved.

Another line of research leverages databases of PBR materials, and learns to project the input (image or text) onto the known representation~\cite{zhang2024mapa,ceylan2024matatlas,fang2024makeitreal}. These methods are limited by the expressiveness of their material databases, but benefit from improved regularization.

By finetuning multi-view image diffusion models~\cite{shi2023MVDream} conditioned on normal maps, per-object materials can be generated from text or image inputs.
These approaches~\cite{zhang2024clay,boss2025sf3d} generate a set of views of PBR texture maps, 
which are projected into texture space. Recent methods~\cite{feng2025romantex,he2025materialmvp,shao2025mvpainter,yang2025pandora3d,engelhardt2025svim3d,seed3d,chen2024primx} 
extend this approach with additional conditioning (e.g. depth and/or world space positions).

A common limitation is lack of view consistency, which may result in blurring in the extracted textures. By finetuning \emph{video models} for object rotation, view consistency is improved~\cite{voleti2024sv3d,yang2024hi3d,munkberg2025videomat,hasselgren2026videomatgen}. The view consistency problem can be avoided by having the diffusion model operate directly in 3D space~\cite{xiang2024structured,yu2024texgen,xiang2025trellis2}, but these approaches require large-scale supervision with high quality 3D models.
\section{Procedural Enhancement of Simple PBR Materials}
\label{sec:enhance}

\begin{figure}
    \centering
    \includegraphics[width=0.99\linewidth]{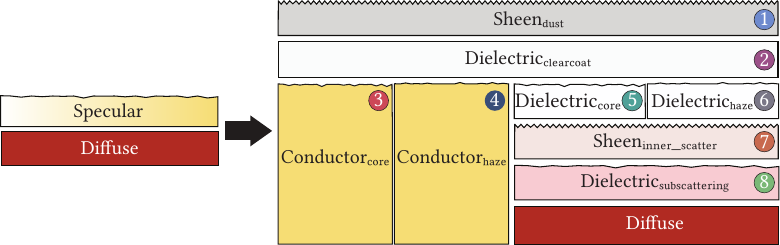}
    \caption{A schematic illustration of our enhanced BRDF. The specular component includes eight lobes, while the diffuse component is unmodified. Lobes 3, 4, 5 and 6 are the basic core and haze reflection lobes, whereas lobe 1, 2, 7 and 8 are what we call ``decorator lobes.''}
    \label{fig:illustration_bsdfs}
\end{figure}

Our goal is to build a dataset of physically plausible materials that spans a wide range of appearances. We represent these materials using a multi-lobe BRDF capable of expressing diverse visual phenomena (Section~\ref{sec:new_model}), and construct them by procedurally enhancing simple PBR materials from existing datasets (Section~\ref{sec:procen}).

\subsection{Generalized Non-Diffuse Reflection Model}
\label{sec:new_model}

A simple PBR material consists of a Lambertian diffuse lobe and a GGX specular lobe; here, we consider PBR materials parameterized by base color, roughness, and metalness maps~\cite{karis2013real,gltf2}. In our proposed BRDF model, shown in Fig.~\ref{fig:illustration_bsdfs}, we replace the single GGX specular lobe with a multi-lobe non-diffuse component while preserving the Lambertian diffuse term. Maintaining separate diffuse and non-diffuse components in our design enables flexible editing of each part (see Section~\ref{sec:wave}), and makes rendering tasks such as denoising (requiring albedo) remain straightforward.

Our non-diffuse component consists of eight lobes combined via coating and mixing, organized into conductive and dielectric branches. The conductive branch models metallic reflection with two lobes of different roughness, denoted \emph{core} and \emph{haze}; the dielectric branch similarly contains two lobes layered above the original diffuse component, with optional scattering lobes in between. The two branches are combined through mixing, and additional decorator lobes may be layered on top. The formulation and parameters used for each lobe are guided by domain knowledge from material science and VFX production materials~\cite{Kulla2017, Matusik:2003, 10.5555/2383654.2383671, 10.1145/3543664.3543675,iors}.

\newcommand{\fcircle}[2]{%
\tikz[baseline=(char.base)]{
  \node[shape=circle, draw=#1, fill=#1, text=white, inner sep=1pt] (char) {#2};
}}
\definecolor{deepteal}{HTML}{6B8ECA}
\definecolor{royalpurple}{HTML}{9E6287}
\definecolor{oxfordblue}{HTML}{cc5e5e}
\definecolor{mutedplum}{HTML}{3A5477}
\definecolor{steelblue}{HTML}{4A9E9A}
\definecolor{burntsienna}{HTML}{807D88}
\definecolor{amethyst}{HTML}{C8795C}
\definecolor{darkslate}{HTML}{82B384}

\subsubsection*{\textbf{Core and Haze Lobes}}
\fcircle{oxfordblue}{3}\fcircle{mutedplum}{4}\fcircle{steelblue}{5}\fcircle{burntsienna}{6}
In the specular component of a simple PBR model, base color and metalness are typically used together to define an effective Fresnel reflectance (e.g.~\cite{karis2013real}); moreover, a single surface roughness is used. In contrast, every material we construct includes two conductive GGX lobes and two dielectric ones. We model metallic and non-metallic reflection separately using conductive and dielectric GGX lobes to preserve their respective Fresnel behavior. In addition, both conductive and dielectric branches adopt two correlated roughness scales (see Section~\ref{sec:procen}): a lower-roughness \emph{core} lobe that produces a sharp highlight, and a higher-roughness \emph{haze} lobe that captures blurrier reflection. Together, these lobes produce characteristic double-highlight appearances that often make materials more realistic, as shown in Fig.~\ref{fig:teaser} (Haze) and Fig.~\ref{fig:photos_real} (leftmost).

\subsubsection*{\textbf{Decorator Lobes}}
To expand the range of achievable appearances, we also include four optional decorator lobes, each modeling a distinct type of surface or volume scattering, in thin media above or below the main specular reflection.

\smallskip

\textit{Dust/Fuzz} \fcircle{deepteal}{1} Our top-layer decorator is a sheen lobe~\cite{Kulla2017} that captures strongly view-dependent effects coming from dust or fuzz (short hair). The strength and angular dependence of this lobe are controlled by a weight and a thickness parameter. We use the term ``dust'' if the weight of this lobe is modulated by the object's surface normal, and use ``fuzz'' if the weight is uniform.

\smallskip

\textit{Clearcoat} \fcircle{royalpurple}{2} Clearcoat is modeled as a low-roughness GGX lobe under the dust/fuzz lobe. Coated upon both the conductive and dielectric GGX lobes, it models a thin, glossy coating like glazing. The lobe is controlled by a binary weight (1 if present) and an IOR.

\smallskip

\textit{Inner Scattering} \fcircle{amethyst}{7} We use another sheen lobe beneath the dielectric GGX lobes to model view-dependent scattering within thin layers. It primarily affects grazing angle appearances by introducing soft, retro-reflective scattering. The lobe is controlled by a weight, a thickness parameter, and a scatter color.

\smallskip

\textit{Subcutaneous Scattering} \fcircle{darkslate}{8} This is a rough, tinted GGX lobe under the inner scatter lobe, modeling subtle, low-order scattering from the otherwise opaque-looking bulk material and producing a soft, gently colored highlight. It is controlled by a binary weight, a high roughness, and a scatter color loosely coupled to the base color. 

\smallskip

Details on all lobes, their parameter ranges and constraints, and isolated effects of each lobe are in our supplemental material (Sec1).

\begin{figure}
    \centering
    \includegraphics[width=0.99\linewidth]{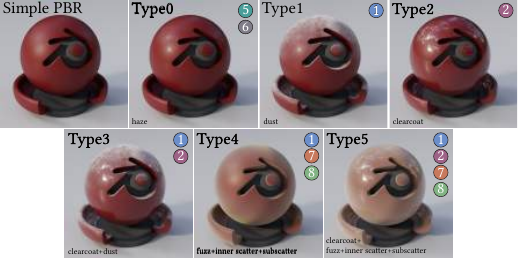}
    \caption{Enhancement effects used in our dataset as compared to the simple PBR model. Our haze enhancement (Type~0) is applied to all materials, while Types~1–-5 correspond to different combinations of decorator lobes.}
    \label{fig:types}
\end{figure}
\subsection{Procedural Enhancement of Simple PBR Materials}
\label{sec:procen}

We now describe how a specific simple PBR material instance is procedurally enhanced into a multi-lobe material from Section \ref{sec:new_model}.

In order to more accurately represent specular reflection from both conductors and dielectrics and improve upon the limited appearance of single-roughness highlights, our baseline enhancement---on all materials---replaces the single GGX lobe with four specular lobes, mapping the original base color, roughness, and metalness textures to the parameters of our core and haze lobes.

First, we use metalness to mix the conductive and dielectric branches without further modification, preserving its intended meaning as an indicator of whether a material is metallic or dielectric. While this may allow ``half-metal'' materials (e.g. metalness $0.5$), in practice, most maps in PBR datasets are near-binary. The complex-valued IOR of the conductive GGX lobes are derived from the simple PBR base color, following~\cite{Gulbrandsen2014Fresnel}, while the IOR of the dielectric GGX lobes is sampled separately from a bounded range.

Moreover, we use the single roughness texture from the simple PBR material to compute two correlated roughness scales for our core and haze lobes. Let $r \in (0,1)$ be the input roughness, typically converted to GGX roughness $\alpha$ via $\alpha = r^{2}$. We instead construct
\begin{equation}
\alpha_{\text{core}} = r^{p}, \qquad \alpha_{\text{haze}} = s \cdot r^{q},
\label{equ:haze}
\end{equation}
where $p > 2$, $q < 2$, and $s > 1$ are per-material scalars sampled within their ranges (see supplemental Sec2). In our design, the same $(\alpha_{\text{core}}, \alpha_{\text{haze}})$ pair is shared by both dielectric and conductive GGX lobes to preserve the artistic intention of the input roughness data.

This process defines our \emph{haze-only} enhancement, denoted as \textit{Type~0}. We include five other appearance types by enabling additional decorator lobes (Fig.~\ref{fig:types}), with the last two types (4--5) applied only to dielectric materials. These types span a range of additional appearances but do not exhaust all possible combinations. Parameters of decorator lobes (e.g. dust thickness, clearcoat IOR) are sampled from physically meaningful ranges to ensure realistic results, with some (e.g. subcutaneous scatter color) loosely tied to the input PBR parameters to maintain consistency with the bulk material.

\smallskip

\textit{Type~1: Dust} \fcircle{deepteal}{1} We enable the dust/fuzz lobe by setting its weight nonzero. To simulate dust accumulation, this lobe weight is driven by the world-space normal vectors of the object, favoring upward-facing surfaces, and further shaped by spatially varying masks that introduce features such as fingerprints on dust applied through triplanar mapping. Examples of dust variation masks are in Fig. \ref{fig:masks}.

\smallskip

\textit{Type~2: Clearcoat} \fcircle{royalpurple}{2} We enable the clearcoat lobe by setting its weight to 1 across the material, and optionally modulate the weight texture with binary masks, to simulate small damages on the coating.

\smallskip

\begin{figure}
    \centering
    \includegraphics[width=0.24\linewidth]{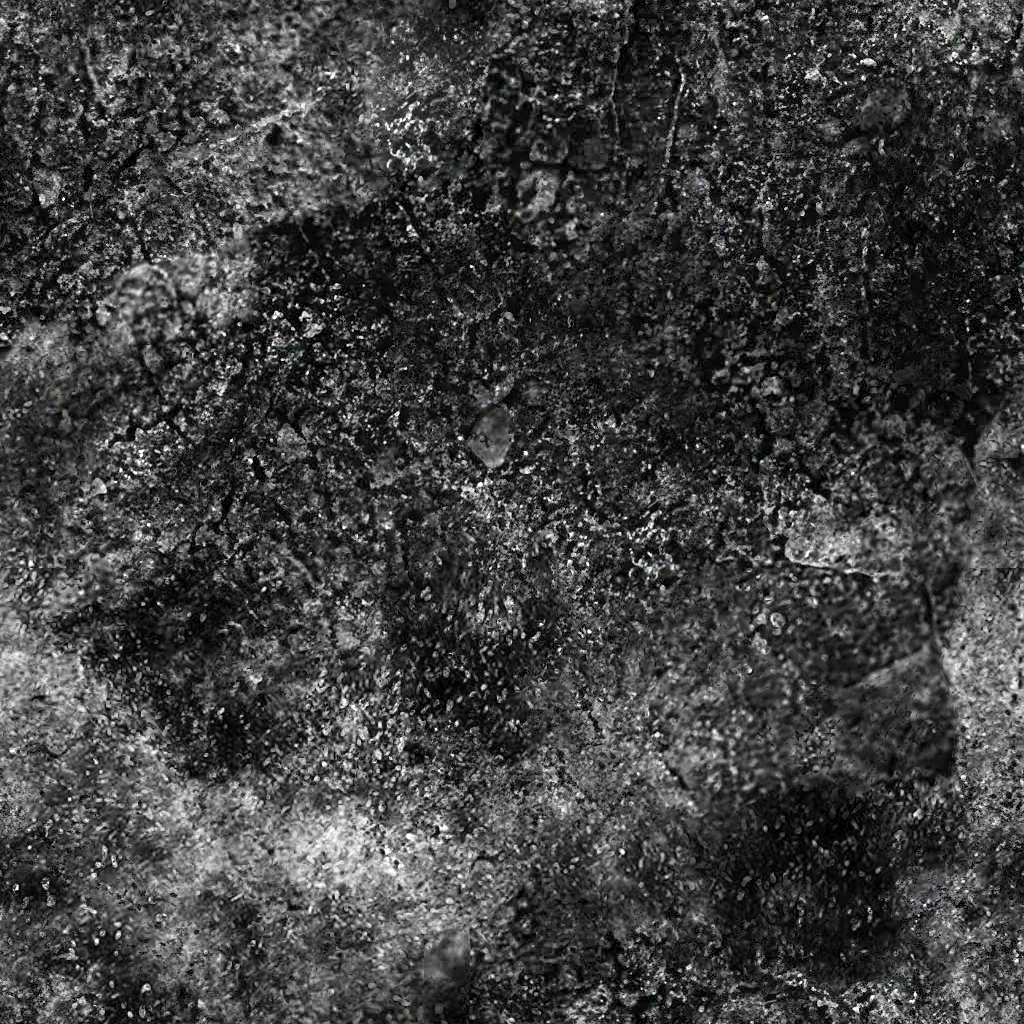}
    \includegraphics[width=0.24\linewidth]{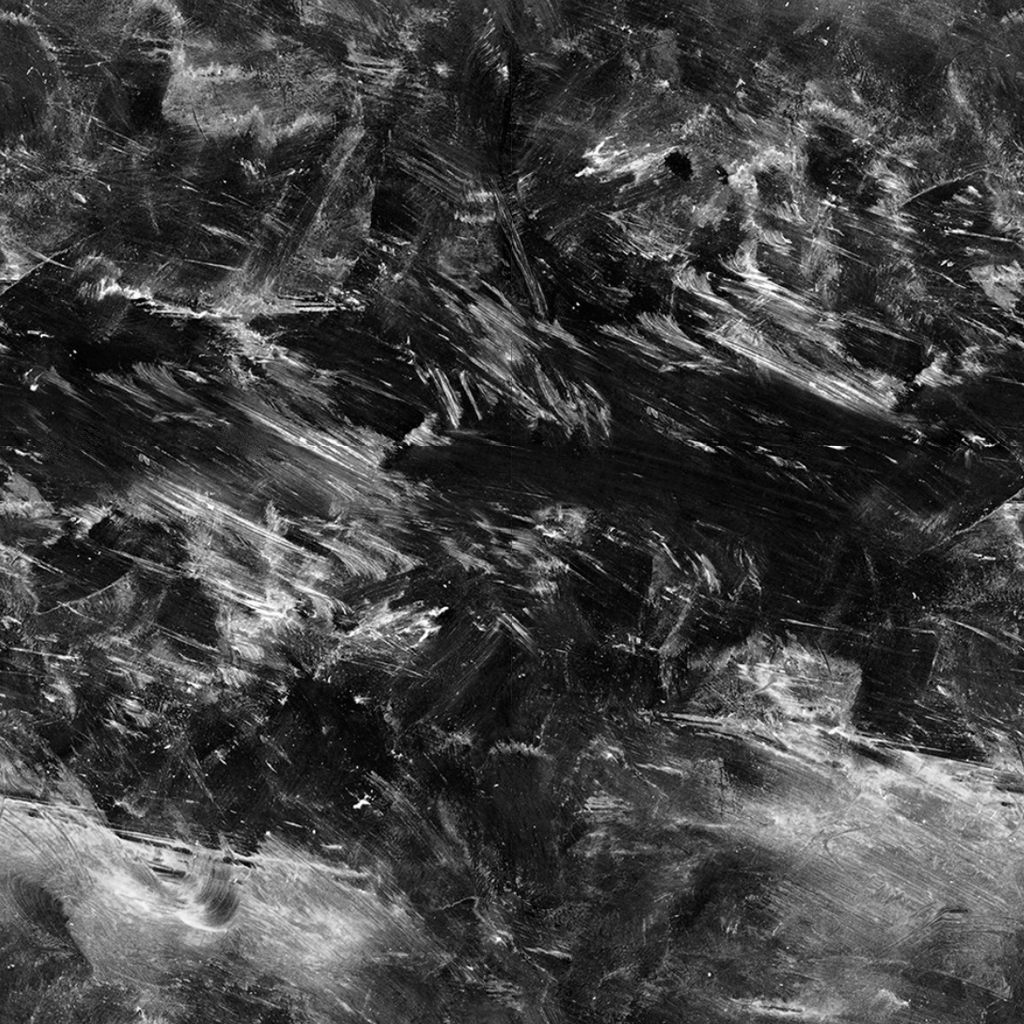}
    \includegraphics[width=0.24\linewidth]{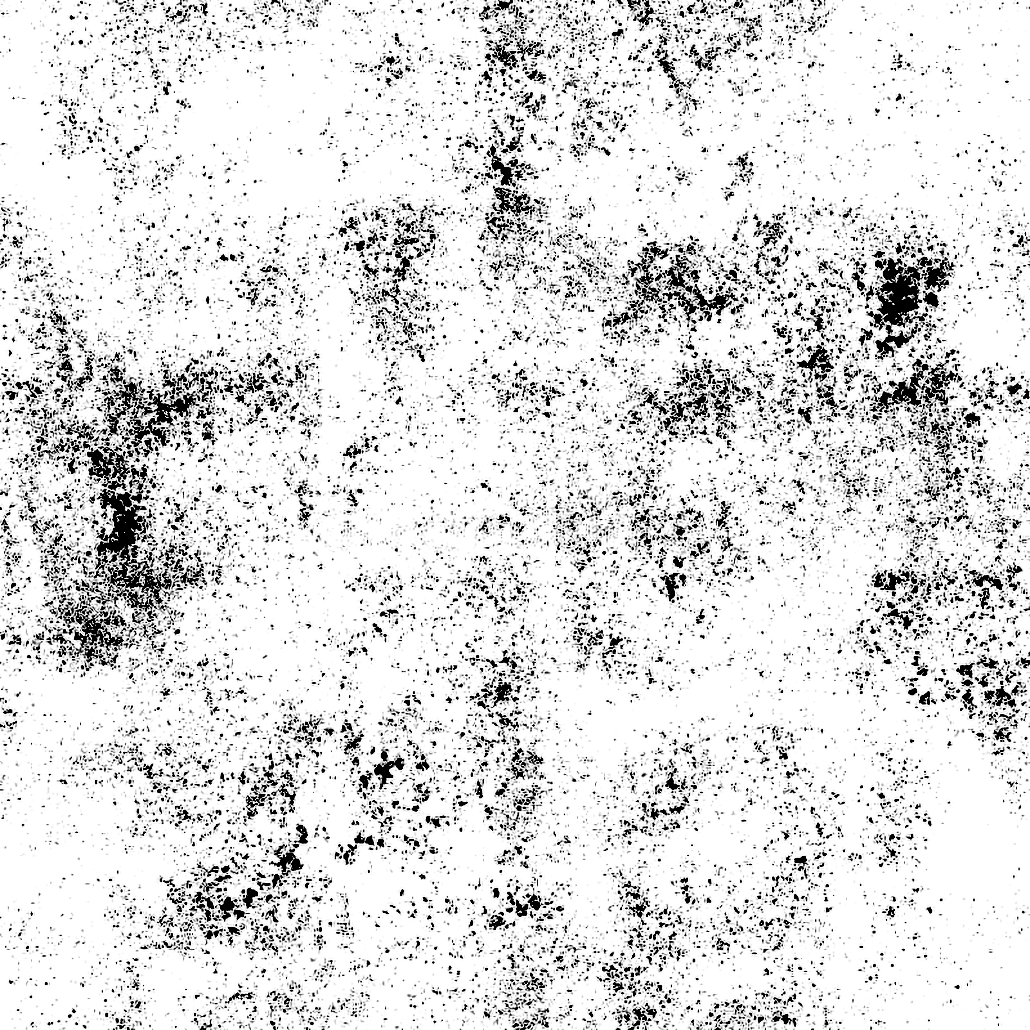}
    \includegraphics[width=0.24\linewidth]{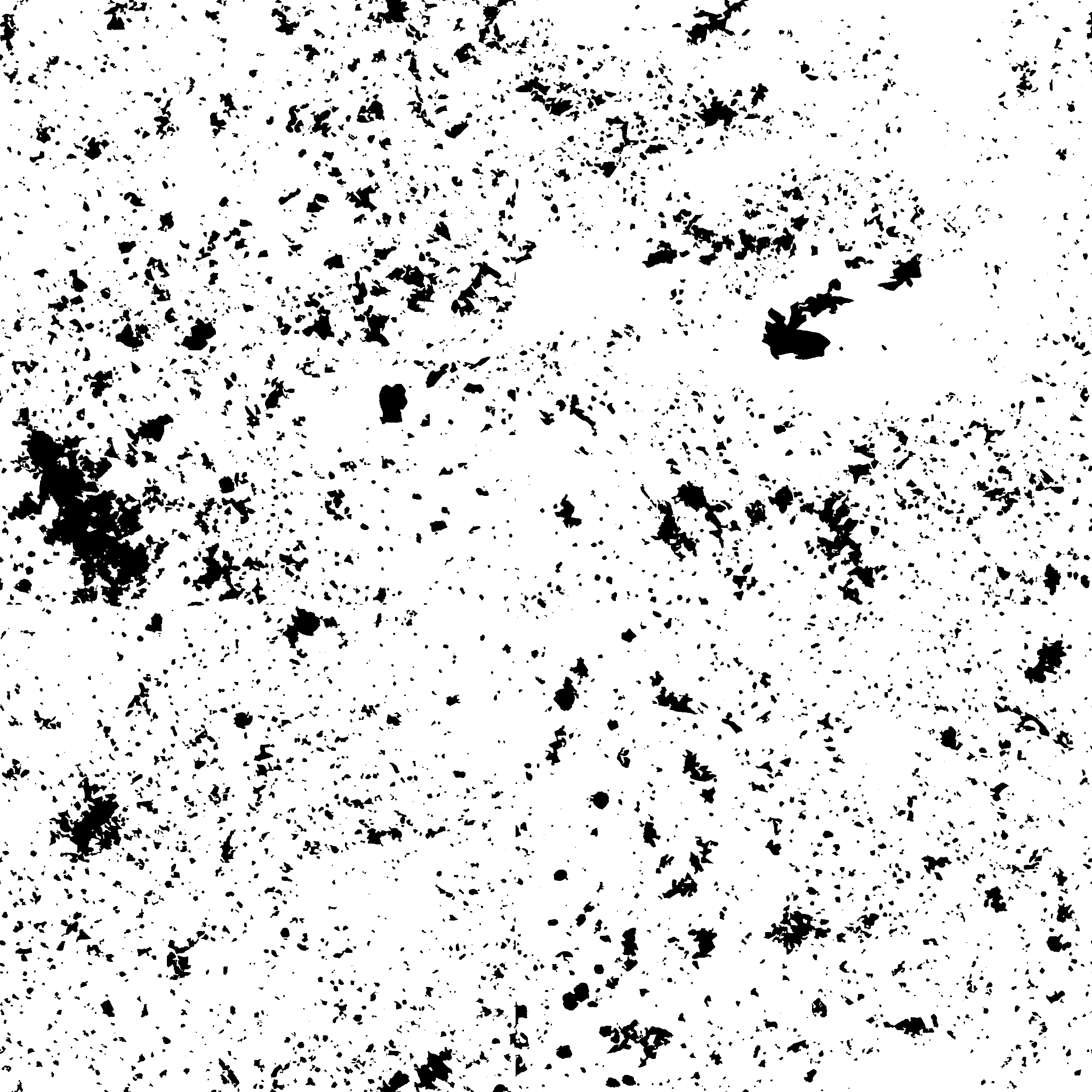}
    \caption{Example masks that modulate spatial variations in dust (left images) and clearcoat (right images), simulating fingerprints or ripple-like patterns on dust layers and dents or scratches on shiny coatings.}
    \label{fig:masks}
\end{figure}

\textit{Type~3: Dust + Clearcoat} \fcircle{deepteal}{1}\fcircle{royalpurple}{2} Combining Type~1 and Type~2, producing surfaces that have a glossy coating and a dust layer.

\smallskip

\textit{Type~4: Fuzz + Inner- and Sub-Scatter} \fcircle{deepteal}{1}\fcircle{amethyst}{7}\fcircle{darkslate}{8} We apply the dust/fuzz lobe with uniform weight over dielectric objects to produce a fuzzy look (see Fig.~\ref{fig:teaser}, peach fuzz). To further soften the appearance, we also enable the inner and subcutaneous scatter lobes in the dielectric branch. The scatter color, shared by both lobes, is derived from the base color via gentle desaturation.

\smallskip

\textit{Type~5: Fuzz + Clearcoat + Inner- and Sub-Scatter} \fcircle{deepteal}{1}\fcircle{royalpurple}{2}\fcircle{amethyst}{7}\fcircle{darkslate}{8} All lobes are enabled, with the dust/fuzz lobe applied uniformly. The inner- and sub-scattering lobes soften the appearance at grazing angles. If the fuzz weight is high, the fuzz reduces the sharpness of the clearcoat, yielding a milky or gel-covered appearance.

\smallskip

Details of each enhancement type are in the supplemental material (Sec2), along with comparisons to simple PBR appearances. Currently, we allow any enhancement type on any input material, which might produce uncommon appearances. A natural extension is to guide our type selection using object/material recognition.
\section{Neural Representation of Enhanced BRDFs}
\label{sec:neural}

Our enhanced BRDF contains a Lambertian diffuse lobe and a multi-lobe non-diffuse component. The non-diffuse component requires 22 parameters (supp-Sec1) and is impractical to use directly for generative material applications. Instead, we represent the non-diffuse component as a neural material with a compact 6D latent space---compressing the 22 parameter textures into two RGB latent textures---and combine it with the diffuse lobe to form the full BRDF.

\subsection{Neural Material Formulation}
\label{sec:universal}

As in prior work on neural materials \cite{zeltner2024nm}, we train a single encoder–decoder MLP pair to represent the space of non-diffuse materials expressible by our multi-lobe model. After training, the network weights are frozen \cite{raghavanmullia2025genneumat}, defining a latent space shared across material instances. Individual materials are then represented by distinct latent textures in this 6D space.

Our universal decoder MLP consists of four hidden layers with 64 neurons each, and uses an exponential activation function. Given a 6D latent vector, which represents a non-diffuse material, along with a pair of queried directions $(\omega_i, \omega_o)$, the decoder outputs a BRDF value $f_{\text{neu}}(\omega_i, \omega_o)$, as well as a transmission albedo $T_{\text{neu}}(\omega_i)$ and a luminance reflectance value $R_{\text{neu}}(\omega_i)$.

In particular, $T_{\text{neu}}(\omega_i)$ represents the fraction of incident energy transmitted through the non-diffuse material $f_{\text{neu}}$; it is used to enforce energy conservation when layering $f_{\text{neu}}$ above a Lambertian diffuse lobe to build the full, enhanced BRDF. Specifically, using $c$ to denote the PBR base color, our combined BRDF is
\begin{equation}
f(\omega_i,\omega_o)
=
f_{\text{neu}}(\omega_i,\omega_o)
+
T_{\text{neu}}(\omega_i)\,\frac{c}{\pi},
\label{eq:energy}
\end{equation}

Moreover, $R_{\text{neu}}(\omega_i)$ is a scalar measure of reflected energy in $f_{\text{neu}}$, used for importance sampling \textit{between} this non-diffuse component and the diffuse lobe during rendering. In addition to the evaluation decoder, we also train a sampling decoder following \cite{zeltner2024nm}, used for importance sampling \textit{within} the non-diffuse BRDF.
\subsection{Universal MLP Training}
We next describe our data used for training the universal neural material, as well as practical challenges encountered in the training.

\subsubsection*{\textbf{Training Data}}
Training the universal neural material does not require a large curated material database, as our  goal is to learn the directional behavior of BRDFs represented by our multi-lobe model rather than how material parameters may vary across a textured surface. Thus, we generate training materials procedurally, matching the statistical distributions of the materials we hope to represent.

\begin{figure}
    \centering
    \includegraphics[width=0.99\linewidth]{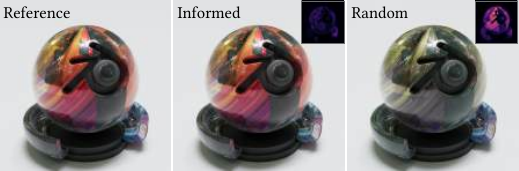}
    \caption{Comparison of informed vs. random training data and their \FLIP error images \cite{Andersson2020}. We reconstruct the same reference material with both approaches. The network built with informed data captures the reference material significantly better than the one with random data.}
    \label{fig:trainingdata}
\end{figure}

Our training data comprises five material classes. Four are \emph{themed} classes targeting specific appearance effects: \emph{haze}, \emph{dust}, \emph{clearcoat}, and \emph{inner/subcutaneous scattering}. These classes enable corresponding decorator lobes alongside optional additional lobes, allowing the network to learn both isolated and combined effects. Each themed class contains an $8192 \times 8192$ grid of material data points, generated from a total of 4096 distinct base materials per theme. For each base material, all parameters (except specular color) are held constant while roughness is densely sampled, yielding 16384 ($128 \times 128$) roughness variations per base material. To ensure physical plausibility and prevent the network from wasting capacity on unlikely combinations, base material parameters are sparsely sampled from a predefined data distribution. In Fig. \ref{fig:trainingdata}, we compare training with our physically informed, biased data distribution and with purely random material data, where both setups use the same training time, network size, and number of material samples.

In addition, we include a fifth \emph{infill} class to populate the latent space more uniformly (see ``latent space regularization''). In this class, each cell in the $8192 \times 8192$ parameter grid corresponds to a material whose parameters are sampled independently and uniformly in their ranges. This gives rise to another 67M randomly parameterized materials. Visualization of all material classes, physically informed training data preparation, and training details are in the supplemental document.

\subsubsection*{\textbf{Latent Space Regularization}}
We aim to obtain a neural material with a latent space suitable for downstream generative material creation. However, our appearance space is substantially larger than that of prior work (e.g. \cite{raghavanmullia2025genneumat}) due to our material uplifting. For our generative process we need a low-dimensional latent space, but compressing 22 parameters into six channels leads to an entangled mapping from original parameters to latent channels, making it challenging to build a well-behaved latent space.

Early on, as we finetuned diffusion models (Section \ref{sec:video_model}) on small datasets, diffusion-generated latent codes sometimes decoded to invalid materials (producing Infs or other artifacts), spoiling up to $50\%$ of our outputs. This suggested that the neural material latent space had useful materials packed closely against invalid regions, making small errors in generation result in unusable outputs. To resolve this, we made two training adjustments to help smooth the mapping between materials and latent codes. 

First, adding the already mentioned \textit{infill} class containing random input materials helped populate the latent space more uniformly. Since this class is only $20\%$ of our training data, the reconstruction quality was not significantly affected by including random, possibly implausible materials. Second, we injected small uniform noise in the range of ±0.005 into latent values when evaluating the reconstruction loss. This requires the decoder to accurately reconstruct BRDFs from slightly perturbed latent codes and made our neural model less sensitive to perturbations in latents and more robust for generative use. The bound of 0.005 was chosen experimentally based on our training data, as the largest value we could adopt without leading to a large increase in training loss. While we did preliminary experiments with a learned jittering noise range instead of a fixed one, we did not observe improvements over our simple jittering strategy. Overall, we found that combining random infilling and latent jittering largely eliminated artifacts in material generation.
\subsection{Per-Material Optimization}
\label{sec:individual}

With a pretrained universal decoder, we can then convert analytical materials enhanced from simple PBR inputs, initially described by many parameter maps, into neural materials specified with 6D latent codes---only two RGB textures. For each individual material, we optimize its latent textures against the frozen universal decoder \footnote{Latent codes directly computed with the analytical parameters and the universal encoder are not sufficiently accurate for materials not present in the training data.}. 

\figDataset

\figSystem

Given analytical parameter textures that describe an enhanced non-diffuse BRDF, we need to minimize the errors between the decoded outputs from the latents and their analytically computed references. Specifically, we use an $\ell_1$ loss that combines three weighted error terms to encourage accurate reconstruction of the BRDF, transmission albedo, and luminance reflectance, across all texels and over all incident–outgoing direction pairs:
\begin{equation}
\mathcal{L}
= \lambda_f \, \lVert f_{\text{neu}} - f_{\text{ref}} \rVert_1
+ \lambda_t \, \lVert T_{\text{neu}} - T_{\text{ref}} \rVert_1
+ \lambda_r \, \lVert R_{\text{neu}} - R_{\text{ref}} \rVert_1 ,
\label{eq:combined_loss}
\end{equation}

\begin{table}
\centering
\caption{Best, medium, and worst \FLIP error for each type.}
\begin{tabular}{lccc}
\hline
Type & Best & Medium & Worst \\
\hline
Type 0 & 0.0487 & 0.0920 & 0.1871 \\
Type 1 & 0.0312 & 0.0727 & 0.1670 \\
Type 2 & 0.0649 & 0.0858 & 0.1192 \\
Type 3 & 0.0554 & 0.0877 & 0.1664 \\
Type 4 & 0.0276 & 0.0643 & 0.1705 \\
Type 5 & 0.0454 & 0.0807 & 0.1754 \\
\hline
\end{tabular}
\label{tab:type_values}
\end{table}

In practice, we found that the BRDF term must dominate the loss, as errors in the BRDF directly lead to appearance discrepancies. When the BRDF is well fitted, the remaining quantities also tend to be reconstructed accurately, as they are strongly correlated with the BRDF. In particular, luminance reflectance is derived from integrating the BRDF and therefore requires only a small weight, while transmission albedo is less coupled to the BRDF and benefits from a slightly higher weight. These observations led to our choice of $\lambda_f = 0.95$, $\lambda_t = 0.04$, and $\lambda_r = 0.01$. To evaluate the fitting quality, we randomly selected 50 examples from each enhancement type, rendered reference and reconstructed materials on preview balls, and generated \FLIP error images. We computed the mean \FLIP value in each error image and report the best, medium, and worst errors in Table \ref{tab:type_values} (more visualizations in supplemental material).

The latent textures optimized according to Eq. \ref{eq:combined_loss} fully specify a non-diffuse BRDF as described in Section \ref{sec:universal}, and importance sampling of $f_{\text{neu}}$ is handled by the sampling decoder\footnote{During per-material optimization, we did not include sampling-related terms in the loss. The latent textures still produce effective sampling parameters when decoded, thanks to the correlation between BRDF shapes and importance sampling schemes.}, as in \cite{zeltner2024nm}. Notably, because $f_{\text{neu}}$ and $T_{\text{neu}}$ in Eq. \ref{eq:energy} are neural approximations, 
we might still violate energy conservation due to fitting errors (e.g. overestimated $T_{\text{neu}}$). Thus, we filter combined materials with a (weakened) white furnace test: we render objects under a unit-valued environment map and discard materials that exhibit energy gain (pixel values $>1$), tolerating small violations due to Monte Carlo noise in rendering. In practice, we find that around $90\%$ of our materials pass this white furnace test, and these valid materials form our dataset of expressive neural materials.
\section{Experimental Downstream Application}
\label{sec:application}

We now demonstrate an example application of the dataset built with our proposed pipeline, using the data to finetune a video diffusion model to generate neural material latents on 3D objects.

\subsection{A Dataset of 3D Objects with Neural Materials}
\label{sec:dataset}

With our enhancement and neural fitting processes discussed earlier, we constructed a dataset of neural materials on 10,640 3D models from Objaverse~\cite{objaverse} and BlenderVault~\cite{litman2025materialfusion}. Our latent optimizations, combined, took 22 hours on a multi-user cluster with a total of 80 L40 and L40S GPUs. For each asset, we create a Lambertian lobe using the base color texture, scale it with $T_{\text{neu}}$, and replace the GGX specular lobe with an enhanced non-diffuse component, represented by two RGB latent textures. Each asset adopts one of the enhancement types in Section \ref{sec:procen}, with approximate proportions of $10\%, 20\%, 20\%, 20\%, 10\%$ and $20\%$ for Types~0--5. Fig.~\ref{fig:dataset} shows selected objects with uplifted materials.
\subsection{Generative Neural Materials}
\label{sec:video_model}

Inspired by Munkberg~et~al.~\shortcite{munkberg2025videomat}, we synthesize neural materials on 3D shapes by leveraging a recent video diffusion model architecture~\cite{hasselgren2026videomatgen} to produce the neural material textures.

\figGenMat

As shown in Fig.~\ref{fig:system}, our pipeline takes a 3D model with a valid texture parameterization and a text prompt describing the material as input. We assume the 3D model has a base color texture; if missing, it can be generated from the prompt using any previous method~\cite{munkberg2025videomat,xiang2025trellis2}.
The diffusion model generates multiple views of material intrinsics: G-buffers of neural material textures, conditioned on corresponding input geometry (views of surface normals and world space positions). 
Finally, we project the intrinsic views into texture space using texture splatting, with bilinear filtering, view averaging, and inpainting for uncovered regions. This allows us to bake down a single set of neural textures that, when combined with the existing base color map, defines a full, enhanced material.

\subsubsection*{\textbf{Dataset}}
For each object in our dataset, we render a video with 16 frames at $512 \times 512$ resolution, using a path tracer with three bounces and Blender AgX tonemapping. We use black backgrounds and smooth camera orbits, and adopt the ``Boiler Room'' light probe from Poly Haven~\cite{polyhaven} for all objects. We only use the shaded videos to generate captions using Qwen2.5-VL-7B~\cite{qwen25}, and the fixed light probe avoids prompt noise due to lighting variation; these prompts are then augmented with keywords describing enhancement types. We also render intrinsic maps (normals, world space positions, base colors and neural material textures\footnote{We mean neural latents, but try to avoid confusion with the diffusion latent space.}). We use these data to finetune a recent Diffusion Transformer (DiT) video model, Cosmos~\cite{cosmos_short}, for neural material generation.

\subsubsection*{\textbf{Video Diffusion Model Architecture}}
Here we use \texttt{Cosmos-1.0-} \texttt{Diffusion-7BVideo2World}\footnote{https://github.com/NVIDIA/Cosmos}, which supports text and image guided video generation at a resolution of 1280$\times$704 pixels and 121 frames.
The base model leverages the pretrained \texttt{Cosmos-1.0-Tokenizer-} \texttt{CV8x8x8} to encode and decode RGB videos to and from latent space.

\subsubsection*{\textbf{Finetuning}}
The model comprises the pretrained VAE encoder-decoder pair (the Cosmos Tokenizer mentioned above), $(\vaeEncoder, \vaeDecoder)$, and a transformer-based denoising function, $\diffusionModelFn$. Given an input video $\videoInput$ consisting of normals, world space positions, and base colors for $N$ views of a 3D model, our goal is to train $\diffusionModelFn$ to denoise views of neural material textures conditioned on $\videoInput$. 
We finetune the embedding layer (extended from the base model to support our input conditions) and all DiT layers on 64 A100 GPUs for 20K iterations, using the AdamW optimizer and a learning rate of $5e{-5}$.

We use $\vaeEncoder$ to encode the input conditions, $\videoInput$, into a latent tensor, $\textbf{z}^{\videoInput}$. The target latent variable, $\textbf{z}_0^{\matmap}$, for our dataset is constructed by encoding the neural material texture, $\textbf{y}$, which has six channels, using $\vaeEncoder$.
This encoder expects RGB input, so we encode three channels at a time and concatenate the encoded results in the frame dimension:
\begin{equation}
\textbf{z}_0^{\matmap} = [\vaeEncoder(\textbf{y}_{0,1,2}),\vaeEncoder(\textbf{y}_{3,4,5})].
\end{equation}

Noise, $\diffusionNoise$, is introduced to $\textbf{z}_0^{\matmap}$, representing the neural material texture, to produce $\textbf{z}_\tau^{\matmap}$. 
The parameters, $\theta$, of the diffusion model $\diffusionModelFn$ are optimized by minimizing the objective function: 
\begin{eqnarray}
\hat{\textbf{z}}^{\matmap}(\theta) &=&  \diffusionModelFn ([\textbf{z}_\tau^{\matmap}, \textbf{z}^{\videoInput}]; \typeEmb, \tau) \label{eq:denoiser} \nonumber \\
\mathcal{L}(\theta) &=& \mathbb{E}_{\textbf{z}_0^{\matmap}\sim\dataDistribution,\diffusionNoise\sim \mathcal{N} (0,\sigma^2 I)} \left\| \hat{\textbf{z}}^{\matmap}(\theta) - \textbf{z}_0^{\matmap} \right\|_2^2
\label{eq:objective}
\end{eqnarray}
where $[\,,]$ denotes concatenation in the channel dimension, $\typeEmb$ is the encoded text prompt (encoded by T5-XXL~\cite{raffel2023t5}) and $\tau$ is the timestep. We increase the input feature count of the input embedding layer of $\diffusionModelFn$ to account for our additional input conditions, $\textbf{z}^{\videoInput}$. We use the denoising score matching loss with uncertainty-based weighting from Cosmos~\cite{cosmos_short}, applied to the predicted latent $\hat{\textbf{z}}^{\matmap}(\theta)$ and the corresponding  target latent $\textbf{z}_0^{\matmap}$.

\subsubsection*{\textbf{Evaluation}}
We evaluate our generative pipeline on a test set of 16 3D assets with hand-crafted neural materials. We first render images of test objects with their reference neural materials, and annotate these images with captions from Qwen2.5-VL-7B and enhancement type keywords. We then generate neural materials using these text prompts, repeating with five random seeds and subjectively picking the best results, and render 16 views of each asset with the generated material. On our $16\times16=256$ images, we report the CLIP-based Fr\'echet Inception Distance (CLIP-FID)~\cite{Kynkaanniemi2022}, CLIP Maximum-Mean Discrepancy (CMMD)~\cite{jayasumana2024cmmd}, Learned Perceptual Image Patch Similarity (LPIPS)~\cite{zhang2018perceptual}, and the Peak Signal-to-Noise Ratio (PSNR) in the last row of Table~\ref{tab:matgen} ($\textrm{Our}_{\textrm{ref}}$). Some examples of generated neural materials are shown in Fig.~\ref{fig:genmat}, and Fig.~\ref{fig:car_issue} shows a failure case. Moreover, by editing the enhancement keyword in the prompt, we can create different neural material variants on an object, as in Fig.~\ref{fig:monkey}.

\figCarIssue
\figMonkey

We compare against TRELLIS.2~\cite{xiang2025trellis2}, a recent image-guided PBR material generation method. We use their PBR texture generation pipeline with known geometry, and for each test set example, input a shaded image rendered with our reference neural material, and use TRELLIS.2 to generate PBR material maps. TRELLIS.2 produces base color, roughness, and metalness textures, and our quantitative evaluation on these PBR materials are reported in Table~\ref{tab:matgen}, the TRELLIS.2 row. Since a major difference is that we assume a known base color and generate the non-diffuse components, when comparing with TRELLIS.2, we also compose full materials from our generation in another version, by combining TRELLIS.2 generated base colors and our generated neural latents. The evaluation of this material variant is in Table~\ref{tab:matgen}, the $\textrm{Our}_{\textrm{trellis}}$ row; note that the $\textrm{Our}_{\textrm{ref}}$ variant mentioned earlier instead combines our generated neural latents with known base colors. See the supplemental material for additional details and visual comparisons.

\begin{table}
  \caption{Quantitative metrics for material generation, computed over 16 views for each of the 16 objects in our test set. We compare against TRELLIS.2~\shortcite{xiang2025trellis2}: 
  $\textrm{Our}_{\textrm{trellis}}$ represents our uplifted materials with the base color generated by TRELLIS.2; 
  $\textrm{Our}_{\textrm{ref}}$ uses the reference base color.}
  \label{tab:matgen}
  \centering
  \setlength{\tabcolsep}{2pt}
  \renewcommand{\arraystretch}{0.9}
  \begin{small}
  \begin{tabular}{@{}lcccc@{}}
    \toprule
    & CLIP-FID~($\downarrow$) & CMMD~($\downarrow$) & LPIPS~($\downarrow$) & PSNR~($\uparrow$) \\
    \midrule
    TRELLIS.2 & 8.642 & 0.132 & 0.0510 & 24.54\\
    $\textrm{Our}_{\textrm{trellis}}$ & 6.527 & 0.058 & 0.0444 & 28.28\\
    $\textrm{Our}_{\textrm{ref}}$ & \sota{3.907} & \sota{0.020} & \sota{0.0215} & \sota{32.53} \\
    \bottomrule
  \end{tabular}
  \end{small}
\end{table}
\subsection{Additional Microscale Enhancement}
\label{sec:wave}

Besides uplifting non-diffuse BRDFs, materials can be enriched via other procedural enhancements. For instance, microscale effects such as wave-optics-induced glints can introduce fine visual breakups that improve realism for some materials. However, these small details are difficult to represent in textures and synthesize with diffusion models. Many microscale effects primarily associate with specular reflection, and as we intentionally exclude the diffuse term from our neural material representation, we can model these fine-scale effects ad-hoc, by modulating our non-diffuse BRDF. 

\figGlint

As an example, we introduce colorful glints on some objects by deriving a procedural noise model using Simplex noise, informed by wave optics surface scattering as inspired by \cite{yu2024appearance,yu2025realistic}.
As shown in Fig. \ref{fig:wave} and our supplemental video, a glinty appearance can be added to our haze enhancement by modulating our neural BRDF with distinct noise instances $\mathcal{N}^{(i)}$ at different surface points:
\begin{equation}
    f_{\text{neu}}^{(i)}(\lambda, \omega_i, \omega_o) = f_{\text{neu}}(\omega_i, \omega_o) \cdot \mathcal{N}^{(i)}(\lambda, \omega_i, \omega_o)
\end{equation}
This shows that procedural enhancements at different scales can be composed, and fine-scale enhancement can reintroduce detailed structures difficult for generative models to produce.
\section{Discussion and Future Work}

In this last section, we discuss some directions for future work.

\subsubsection*{\textbf{Visual Breakup Texture Selection}}
To enhance spatial variations in our uplifted materials, we modulated some decorator lobe weights using pre-defined masks (recall Fig. \ref{fig:masks}). This can lead to appropriate looking imperfections (e.g. scratches, fingerprints) as on hand-crafted materials, but the resulting appearance still depends on how mask frequency relates to object scale. In practice, meshes may not reflect real-world sizes, and a more robust solution needs to incorporate object size estimation. A preferable alternative is to automate mask creation and placement using a video-to-video material transfer approach similar to \cite{Hadadan2025GenerativeDetail}.

\subsubsection*{\textbf{Resolution Limit}} 
The quality of our generated materials is limited by the video resolution supported in the diffusion stage. While our neural material data can capture the aforementioned high-frequency spatial variations, the generative pipeline may smooth out these fine-scale details, especially for dust variations (see Fig.~\ref{fig:monkey}). This constraint, owing to the current resolution limit in video models, will likely be reduced as higher resolution models become available.

\subsubsection*{\textbf{Multi-Lobe Model Extension}}
Our multi-lobe material model was designed to produce the visual effects explored in this work. The BRDF lobes are lightweight and easy to adjust, but they only cover our selected subset of phenomena. Effects such as iridescence or layered absorption would require additional decorator lobes, which are easy to add thanks to our BRDF's modular structure. However, increasing the number of appearance modes will likely amplify limitations related to our neural representation, requiring higher dimensional latent spaces and improved latent space regularization.



\begin{acks}
We thank everyone who worked on the neural appearance project, in particular, Tizian Zeltner, Fabrice Rousselle, and Craig Kolb. The material test object in Fig.~\ref{fig:types} and ~\ref{fig:trainingdata} was created by Robin Marin and released under CC (https://creativecommons.org/licenses/by/3.0/).
\end{acks}

\bibliographystyle{ACM-Reference-Format}
\bibliography{bibliography}

@String{Computing = "Computing" }

@String{Computer = "{IEEE} Computer" }

@inproceedings{blattmann2023videoldm,
    title={Align your Latents: High-Resolution Video Synthesis with Latent Diffusion Models},
    author={Blattmann, Andreas and Rombach, Robin and Ling, Huan and Dockhorn, Tim and Kim, Seung Wook and Fidler, Sanja and Kreis, Karsten},
    booktitle={IEEE Conference on Computer Vision and Pattern Recognition ({CVPR})},
    year={2023}
}

@article{blattmann2023svd,
	title={Stable Video Diffusion: Scaling Latent Video Diffusion Models to Large Datasets}, 
	author={Andreas Blattmann and Tim Dockhorn and Sumith Kulal and Daniel Mendelevitch and Maciej Kilian and Dominik Lorenz and Yam Levi and Zion English and Vikram Voleti and Adam Letts and Varun Jampani and Robin Rombach},
	year={2023},
	journal={arXiv:2311.15127}
}

@article{cosmos_short,
  title={{Cosmos World Foundation Model Platform for Physical AI}},
  author={NVIDIA},
  journal={arXiv preprint arXiv:2501.03575},
  year={2025}
}

@inproceedings{hong2023cogvideo,
	title={{CogVideo: Large-scale Pretraining for Text-to-Video Generation via Transformers}},
	author={Wenyi Hong and Ming Ding and Wendi Zheng and Xinghan Liu and Jie Tang},
	booktitle={The Eleventh International Conference on Learning Representations },
	year={2023}
}

@inproceedings{litman2025materialfusion,
  author    = {Yehonathan Litman and Or Patashnik and Kangle Deng and Aviral Agrawal and Rushikesh Zawar and Fernando De la Torre and Shubham Tulsiani},
  title     = {{MaterialFusion: Enhancing Inverse Rendering with Material Diffusion Priors}},
  booktitle = {3DV},
  year      = {2025}
}

@inproceedings{munkberg2025videomat,
    author = {Jacob Munkberg and Zian Wang and Ruofan Liang and Tianchang Shen and Jon Hasselgren},
    title = {{VideoMat: Extracting PBR Materials from Video Diffusion Models}},
    booktitle = {Eurographics Symposium on Rendering - CGF Track},
    year = {2025}
}

@inproceedings{objaverse,
  title="{Objaverse: A Universe of Annotated {3D} Objects}",
  author={Matt Deitke and Dustin Schwenk and Jordi Salvador and Luca Weihs and
          Oscar Michel and Eli VanderBilt and Ludwig Schmidt and
          Kiana Ehsani and Aniruddha Kembhavi and Ali Farhadi},
  booktitle={Proceedings of the IEEE/CVF Conference on Computer Vision and Pattern Recognition},
  pages={13142--13153},
  year={2023}
}

@Manual{polyhaven,
   author = {Greg Zaal and et al.},
   title = {Poly Haven - The Public 3D Asset Library},
   url = {https://polyhaven.com},
   year = 2024
}

@article{qwen25,
  title={Qwen2.5 technical report},
  author={Qwen},
  journal={arXiv preprint arXiv:2412.15115},
  year={2024}
}

@misc{raffel2023t5,
	title={Exploring the Limits of Transfer Learning with a Unified Text-to-Text Transformer}, 
	author={Colin Raffel and Noam Shazeer and Adam Roberts and Katherine Lee and Sharan Narang and Michael Matena and Yanqi Zhou and Wei Li and Peter J. Liu},
	year={2023},
	eprint={1910.10683},
	archivePrefix={arXiv},
	primaryClass={cs.LG},
	url={https://arxiv.org/abs/1910.10683}
}

@inproceedings{sohl2015deep,
  title = {Deep Unsupervised Learning Using Nonequilibrium Thermodynamics},
  booktitle = {International Conference on Machine Learning},
  author = {{Sohl-Dickstein}, Jascha and Weiss, Eric and Maheswaranathan, Niru and Ganguli, Surya},
  year = {2015}
}

@article{ho2020denoising,
  title={Denoising diffusion probabilistic models},
  author={Ho, Jonathan and Jain, Ajay and Abbeel, Pieter},
  journal={Advances in Neural Information Processing Systems},
  volume={33},
  pages={6840--6851},
  year={2020}
}

@inproceedings{dhariwal2021diffusion,
  title={Diffusion Models Beat {GAN}s on Image Synthesis},
  author={Prafulla Dhariwal and Alexander Quinn Nichol},
  booktitle={Advances in Neural Information Processing Systems},
  year={2021},
}

@article{xiang2025trellis2,
    title={Native and Compact Structured Latents for 3D Generation},
    author={Xiang, Jianfeng and Chen, Xiaoxue and Xu, Sicheng and Wang, Ruicheng and Lv, Zelong and Deng, Yu and Zhu, Hongyuan and Dong, Yue and Zhao, Hao and Yuan, Nicholas Jing and Yang, Jiaolong},
    journal={Tech report},
    year={2025}
}

@article{wan2025,
title={Wan: Open and Advanced Large-Scale Video Generative Models}, 
author={Team Wan},
journal = {arXiv preprint arXiv:2503.20314},
year={2025}
}

@misc{yang2024cogvideox,
	title={{CogVideoX: Text-to-Video Diffusion Models with An Expert Transformer}}, 
	author={Zhuoyi Yang and Jiayan Teng and Wendi Zheng and Ming Ding and Shiyu Huang and Jiazheng Xu and Yuanming Yang and Wenyi Hong and Xiaohan Zhang and Guanyu Feng and Da Yin and Xiaotao Gu and Yuxuan Zhang and Weihan Wang and Yean Cheng and Ting Liu and Bin Xu and Yuxiao Dong and Jie Tang},
	year={2024},
	journal={arXiv:2408.06072}
}

@inproceedings{richardson2023texture,
  title={{TEXTure: Text-guided texturing of 3d shapes}},
  author={Richardson, Elad and Metzer, Gal and Alaluf, Yuval and Giryes, Raja and Cohen-Or, Daniel},
  booktitle={ACM SIGGRAPH 2023 conference proceedings},
  pages={1--11},
  year={2023}
}

@inproceedings{chen2023text2tex,
  title={Text2tex: Text-driven texture synthesis via diffusion models},
  author={Chen, Dave Zhenyu and Siddiqui, Yawar and Lee, Hsin-Ying and Tulyakov, Sergey and Nie{\ss}ner, Matthias},
  booktitle={Proceedings of the IEEE/CVF international conference on computer vision},
  pages={18558--18568},
  year={2023}
}

@inproceedings{zeng2024paint3d,
  title={Paint3d: Paint anything 3d with lighting-less texture diffusion models},
  author={Zeng, Xianfang and Chen, Xin and Qi, Zhongqi and Liu, Wen and Zhao, Zibo and Wang, Zhibin and Fu, Bin and Liu, Yong and Yu, Gang},
  booktitle={Proceedings of the IEEE/CVF conference on computer vision and pattern recognition},
  pages={4252--4262},
  year={2024}
}

@article{yeh2024texturedreamer,
      title={{TextureDreamer: Image-guided Texture Synthesis through Geometry-aware Diffusion}},
      author={Yeh, Yu-Ying and Huang, Jia-Bin and Kim, Changil and Xiao, Lei and Nguyen-Phuoc, Thu and Khan, Numair and Zhang, Cheng and Chandraker, Manmohan and Marshall, Carl S and Dong, Zhao and others},
      journal={arXiv preprint arXiv:2401.09416},
      year={2024}
}

@article{Zhang2024dreammat,
  author = {Zhang, Yuqing and Liu, Yuan and Xie, Zhiyu and Yang, Lei and Liu, Zhongyuan and Yang, Mengzhou and Zhang, Runze and Kou, Qilong and Lin, Cheng and Wang, Wenping and Jin, Xiaogang},
  title = {{DreamMat: High-quality PBR Material Generation with Geometry- and Light-aware Diffusion Models}},
  year = {2024},
  volume = {43},
  number = {4},
  journal = {ACM Trans. Graph.}, 
  articleno = {39}, 
}

@inproceedings{deng2024flashtex,
  title={{FlashTex:} Fast Relightable Mesh Texturing with {LightControlNet}},
  author={Deng, Kangle and Omernick, Timothy and Weiss, Alexander and Ramanan, Deva and Zhu, Jun-Yan and Zhou, Tinghui and Agrawala, Maneesh},
  booktitle={European Conference on Computer Vision (ECCV)},
  year={2024},
}

@inproceedings{zhang2024mapa,
  author = {Zhang, Shangzhan and Peng, Sida and Xu, Tao and Yang, Yuanbo and Chen, Tianrun and Xue, Nan and Shen, Yujun and Bao, Hujun and Hu, Ruizhen and Zhou, Xiaowei},
  title = {{MaPa: Text-driven Photorealistic Material Painting for 3D Shapes}},
  year = {2024},
  booktitle = {ACM SIGGRAPH 2024 Conference Papers},
  articleno = {4},
  numpages = {12},
}

@misc{ceylan2024matatlas,
      title={{MatAtlas: Text-driven Consistent Geometry Texturing and Material Assignment}}, 
      author={Duygu Ceylan and Valentin Deschaintre and Thibault Groueix and Rosalie Martin and Chun-Hao Huang and Romain Rouffet and Vladimir Kim and Gaëtan Lassagne},
      year={2024},
      eprint={2404.02899},
      archivePrefix={arXiv},
      url={https://arxiv.org/abs/2404.02899}, 
}

@misc{fang2024makeitreal,
    title="{Make-it-Real: Unleashing Large Multimodal Model for Painting 3D Objects with Realistic Materials}", 
    author={Ye Fang and Zeyi Sun and Tong Wu and Jiaqi Wang and Ziwei Liu and Gordon Wetzstein and Dahua Lin},
    year={2024},
    eprint={2404.16829},
    archivePrefix={arXiv},
    primaryClass={cs.CV}
}

@inproceedings{voleti2024sv3d,
  author    = {Voleti, Vikram and Yao, Chun-Han and Boss, Mark and Letts, Adam and Pankratz, David and Tochilkin,  Dmitrii and Laforte, Christian and Rombach, Robin and Jampani, Varun},
  title     = {{SV3D}: Novel Multi-view Synthesis and {3D} Generation from a Single Image using Latent Video Diffusion},
  booktitle = {European Conference on Computer Vision (ECCV)},
  year      = {2024},
}

@inproceedings{yang2024hi3d,
  title={{Hi3D: Pursuing High-Resolution Image-to-3D Generation with Video Diffusion Models}},
  author={Haibo Yang and Yang Chen and Yingwei Pan and Ting Yao and Zhineng Chen and Chong-Wah Ngo and Tao Mei},
  booktitle={ACM MM},
  year={2024}
}

@article{xiang2024structured,
    title   = {{Structured 3D Latents for Scalable and Versatile 3D Generation}},
    author  = {Xiang, Jianfeng and Lv, Zelong and Xu, Sicheng and Deng, Yu and Wang, Ruicheng and Zhang, Bowen and Chen, Dong and Tong, Xin and Yang, Jiaolong},
    journal = {arXiv preprint arXiv:2412.01506},
    year    = {2024}
}

@article{yu2024texgen,
    author = {Yu, Xin and Yuan, Ze and Guo, Yuan-Chen and Liu, Ying-Tian and Liu, Jianhui and Li, Yangguang and Cao, Yan-Pei and Liang, Ding and Qi, Xiaojuan},
    title = {{TEXGen: a Generative Diffusion Model for Mesh Textures}},
    journal = {ACM Trans. Graph.},
    volume = {43},
    number = {6},
    year = {2024},
    articleno = {213},
}

@article{shi2023MVDream,
  author = {Shi, Yichun and Wang, Peng and Ye, Jianglong and Mai, Long and Li, Kejie and Yang, Xiao},
  title = {{MVDream: Multi-view Diffusion for 3D Generation}},
  journal = {arXiv:2308.16512},
  year = {2023},
}

@article{zhang2024clay,
  title={{CLAY: A Controllable Large-scale Generative Model for Creating High-quality 3D Assets}},
  author={Zhang, Longwen and Wang, Ziyu and Zhang, Qixuan and Qiu, Qiwei and Pang, Anqi and Jiang, Haoran and Yang, Wei and Xu, Lan and Yu, Jingyi},
  journal={ACM Transactions on Graphics (TOG)},
  volume={43},
  number={4},
  pages={1--20},
  year={2024},
}

@article{boss2025sf3d,
  author={Boss, Mark and Huang, Zixuan and Vasishta, Aaryaman and Jampani, Varun},
  title={{SF3D}: Stable Fast {3D} Mesh Reconstruction with UV-unwrapping and Illumination Disentanglement},
  journal={Conference on Computer Vision and Pattern Recognition ({CVPR})},
  year={2025},
}

@article{feng2025romantex,
  title={{RomanTex: Decoupling 3D-aware Rotary Positional Embedded Multi-Attention Network for Texture Synthesis}},
  author={Feng, Yifei and Yang, Mingxin and Yang, Shuhui and Zhang, Sheng and Yu, Jiaao and Zhao, Zibo and Liu, Yuhong and Jiang, Jie and Guo, Chunchao},
  journal={arXiv preprint arXiv:2503.19011},
  year={2025}
}

@article{he2025materialmvp,
  title={{MaterialMVP: Illumination-Invariant Material Generation via Multi-view PBR Diffusion}},
  author={He, Zebin and Yang, Mingxin and Yang, Shuhui and Tang, Yixuan and Wang, Tao and Zhang, Kaihao and Chen, Guanying and Liu, Yuhong and Jiang, Jie and Guo, Chunchao and Luo, Wenhan},
  journal={arXiv preprint arXiv:2503.10289},
  year={2025}
}

@article{shao2025mvpainter,
    title={{MVPainter: Accurate and Detailed 3D Texture Generation via Multi-View Diffusion with Geometric Control}},
    author={Shao, Mingqi and Xiong, Feng and Sun, Zhaoxu and Xu, Mu},
    journal={arXiv preprint arXiv:2505.12635},
    year={2025},
    url={https://arxiv.org/abs/2505.12635}
}

@article{yang2025pandora3d,
    title={{Pandora3D: A Comprehensive Framework for High-Quality 3D Shape and Texture Generation}},
    author={Yang, Jiayu and Shang, Taizhang and Sun, Weixuan and Song, Xibin and Chen, Ziang and Wang, Senbo and Chen, Shenzhou and Liu, Weizhe and Li, Hongdong and Ji, Pan},
    journal={arXiv preprint arXiv:2502.14247},
    year={2025}
}

@article{engelhardt2025svim3d,
  title={SViM3D: Stable Video Material Diffusion for Single Image 3D Generation},
  author={Engelhardt, Andreas and Boss, Mark and Voleti, Vikram and Yao, Chun-Han and Lensch, Hendrik P. and Jampani, Varun},
  journal={International Conference on Computer Vision},
  year={2025}
}

@Manual{seed3d,
  title={Seed3D 1.0: From Images to High-Fidelity Simulation-Ready 3D Assets},
  author={Seed, ByteDance},
  url={https://seed3d.github.io/Seed3D/report.pdf},
  year = 2025
}

@article{chen2024primx,
    title="{3DTopia-XL: High-Quality 3D PBR Asset Generation via Primitive Diffusion}",
    author={Chen, Zhaoxi and Tang, Jiaxiang and Dong, Yuhao and Cao, Ziang and Hong, Fangzhou and Lan, Yushi and Wang, Tengfei and Xie, Haozhe and Wu, Tong and Saito, Shunsuke and Pan, Liang and Lin, Dahua and Liu, Ziwei},
    journal={arXiv preprint arXiv:2409.12957},
    year={2024}
}

@article{kuznetsov2021neumip,
    title = {{NeuMIP: Multi-Resolution Neural Materials}},
    author = {Kuznetsov, Alexandr and Mullia, Krishna and Xu, Zexiang
                  and Ha\v{s}an, Milo\v{s} and Ramamoorthi, Ravi},
    journal = {Transactions on Graphics (Proceedings of SIGGRAPH)},
    year = {2021},
    month = jul,
    volume = {40},
    number = {4},
    articleno = {175},
}

@article{zeltner2024nm,
    author = {Zeltner, Tizian and Rousselle, Fabrice and Weidlich, Andrea and Clarberg, Petrik and Nov\'{a}k, Jan and Bitterli, Benedikt and Evans, Alex and Davidovi\v{c}, Tom\'{a}\v{s} and Kallweit, Simon and Lefohn, Aaron},
    title = {{Real-time Neural Appearance Models}},
    year = {2024},
    volume = {43},
    number = {3},
    journal = {ACM Trans. Graph.},
    month = jun,
    articleno = {33},
}

@article{raghavanmullia2025genneumat,
  author    = {Raghavan, Nithin and Mullia, Krishna and Trevithick, Alexander and Luan, Fujun and Ha\v{s}an, Milo\v{s} and Ramamoorthi, Ravi},
  title     = {Generative Neural Materials},
  year      = {2025},
  isbn      = {9798400715402},
  series    = {SIGGRAPH Conference Papers '25},
  volume    = {43},
  articleno = {162},
  pages     = {11},
  doi       = {10.1145/3721238.3730746}
}

@inproceedings{10.1145/3744199.3744632,
author = {Andersson, Zap and Edmondson, Paul and Guertault, Julien and Herubel, Adrien and Kutz, Peter and Machizaud, Andr\'{e}a and Portsmouth, Jamie and Servant, Fr\'{e}d\'{e}ric and Stone, Jonathan},
title = {OpenPBR Surface: An open shading model for physically based materials},
year = {2025},
isbn = {9798400720086},
publisher = {Association for Computing Machinery},
address = {New York, NY, USA},
url = {https://doi.org/10.1145/3744199.3744632},
doi = {10.1145/3744199.3744632},
booktitle = {Proceedings of the Digital Production Symposium},
articleno = {1},
numpages = {12},
location = {
},
series = {DigiPro '25}
}

@inproceedings{Burley2012,
  author={Burley, Brent},
  title={Physically Based Shading at Disney},
  booktitle={SIGGRAPH Courses: Practical Physically Based Shading in Film and Game Production},
  year={2012}
}

@inproceedings{Kulla2017,
  author={Christopher Kulla and Alejandro Conty},
  title={Revisiting Physically Based Shading at Imageworks},
  booktitle={SIGGRAPH 2017 Course: Physically Based Shading in Theory and Practice },
  year={2017}
}

@misc{MaterialXSiggraph,
  title        = {{MaterialX}: An Open Standard for Network-Based {CG} Object Looks},
  author       = {Doug Smythe and Jonathan Stone},
  howpublished = {SIGGRAPH 2017 Birds of a Feather},
  year         = {2017},
  month        = {July},
  organization = {Industrial Light & Magic}
}

@article{gauthier2024matup,
    journal   = {Computer Graphics Forum},
    title     = {MatUp: Repurposing Image Upsamplers for SVBRDFs},
    author    = {Gauthier, Alban and Kerbl, Bernhard and Levallois, Jérémy 
                and Faury, Robin and Thiery, Jean-Marc and Boubekeur, Tamy},
    year      = {2024},
    volume    = {43},
    number    = {4},
  }

@inproceedings{Hadadan2025GenerativeDetail,
    author = {Hadadan, Saeed and Bitterli, Benedikt and Zeltner, Tizian and Nov\'{a}k, Jan and Rousselle, Fabrice and Munkberg, Jacob and Hasselgren, Jon and Wronski, Bartlomiej and Zwicker, Matthias},
    title = {Generative detail enhancement for physically based materials},
    year = {2025},
    publisher = {Association for Computing Machinery},
    address = {New York, NY, USA},
    doi = {10.1145/3721238.3730751},
    booktitle = {Proceedings of the Special Interest Group on Computer Graphics and Interactive Techniques Conference Conference Papers},
    articleno = {164},
    numpages = {11},
    series = {SIGGRAPH Conference Papers '25}
}

@article{Gulbrandsen2014Fresnel,
   author  = {Ole Gulbrandsen},
   title   = {Artist Friendly Metallic Fresnel},
   year    = {2014},
   month   = {December},
   day     = {9},
   journal = {Journal of Computer Graphics Techniques (JCGT)},
   volume  = {3},
   number  = {4},
   pages   = {64--72},
   url     = {http://jcgt.org/published/0003/04/03/},
   issn    = {2331-7418}
}

@inproceedings{10.5555/2383847.2383874,
author = {Walter, Bruce and Marschner, Stephen R. and Li, Hongsong and Torrance, Kenneth E.},
title = {Microfacet models for refraction through rough surfaces},
year = {2007},
isbn = {9783905673524},
publisher = {Eurographics Association},
address = {Goslar, DEU},
booktitle = {Proceedings of the 18th Eurographics Conference on Rendering Techniques},
pages = {195–206},
numpages = {12},
location = {Grenoble, France},
series = {EGSR'07}
}

@article{https://doi.org/10.1111/cgf.13475,
author = {Barla, P. and Pacanowski, R. and Vangorp, P.},
title = {A Composite BRDF Model for Hazy Gloss},
journal = {Computer Graphics Forum},
volume = {37},
number = {4},
pages = {55-66},
doi = {https://doi.org/10.1111/cgf.13475},
year = {2018}
}

@inproceedings{10.1145/1321261.1321292,
author = {Weidlich, Andrea and Wilkie, Alexander},
title = {Arbitrarily layered micro-facet surfaces},
year = {2007},
isbn = {9781595939128},
publisher = {Association for Computing Machinery},
address = {New York, NY, USA},
url = {https://doi.org/10.1145/1321261.1321292},
doi = {10.1145/1321261.1321292},
booktitle = {Proceedings of the 5th International Conference on Computer Graphics and Interactive Techniques in Australia and Southeast Asia},
pages = {171–178},
numpages = {8},
location = {Perth, Australia},
series = {GRAPHITE '07}
}

@article{10.1145/3658224,
author = {Lucas, Simon and Ribardi\`{e}re, Micka\"{e}l and Pacanowski, Romain and Barla, Pascal},
title = {A Fully-correlated Anisotropic Micrograin BSDF Model},
year = {2024},
issue_date = {July 2024},
publisher = {Association for Computing Machinery},
address = {New York, NY, USA},
volume = {43},
number = {4},
issn = {0730-0301},
url = {https://doi.org/10.1145/3658224},
doi = {10.1145/3658224},
journal = {ACM Trans. Graph.},
month = jul,
articleno = {111},
numpages = {14}
}

@article{10.1145/3130800.3130840,
author = {Werner, Sebastian and Velinov, Zdravko and Jakob, Wenzel and Hullin, Matthias B.},
title = {Scratch iridescence: wave-optical rendering of diffractive surface structure},
year = {2017},
issue_date = {December 2017},
publisher = {Association for Computing Machinery},
address = {New York, NY, USA},
volume = {36},
number = {6},
issn = {0730-0301},
url = {https://doi.org/10.1145/3130800.3130840},
doi = {10.1145/3130800.3130840},
journal = {ACM Trans. Graph.},
month = nov,
articleno = {207},
numpages = {14}
}

@article{10.1145/2897824.2925945,
author = {Raymond, Boris and Guennebaud, Ga\"{e}l and Barla, Pascal},
title = {Multi-scale rendering of scratched materials using a structured SV-BRDF model},
year = {2016},
issue_date = {July 2016},
publisher = {Association for Computing Machinery},
address = {New York, NY, USA},
volume = {35},
number = {4},
issn = {0730-0301},
url = {https://doi.org/10.1145/2897824.2925945},
doi = {10.1145/2897824.2925945},
journal = {ACM Trans. Graph.},
month = jul,
articleno = {57},
numpages = {11}
}

@article{10.1145/2815618,
author = {Dong, Zhao and Walter, Bruce and Marschner, Steve and Greenberg, Donald P.},
title = {Predicting Appearance from Measured Microgeometry of Metal Surfaces},
year = {2016},
issue_date = {December 2015},
publisher = {Association for Computing Machinery},
address = {New York, NY, USA},
volume = {35},
number = {1},
issn = {0730-0301},
url = {https://doi.org/10.1145/2815618},
doi = {10.1145/2815618},
journal = {ACM Trans. Graph.},
month = dec,
articleno = {9},
numpages = {13}
}

@article{10.1145/3355089.3356525,
author = {Kuznetsov, Alexandr and Ha\v{s}an, Milo\v{s} and Xu, Zexiang and Yan, Ling-Qi and Walter, Bruce and Kalantari, Nima Khademi and Marschner, Steve and Ramamoorthi, Ravi},
title = {Learning generative models for rendering specular microgeometry},
year = {2019},
issue_date = {December 2019},
publisher = {Association for Computing Machinery},
address = {New York, NY, USA},
volume = {38},
number = {6},
issn = {0730-0301},
url = {https://doi.org/10.1145/3355089.3356525},
doi = {10.1145/3355089.3356525},
journal = {ACM Trans. Graph.},
month = nov,
articleno = {225},
numpages = {14}
}

@inproceedings{10.1145/2897839.2927391,
author = {Atanasov, Asen and Koylazov, Vladimir},
title = {A practical stochastic algorithm for rendering mirror-like flakes},
year = {2016},
isbn = {9781450342827},
publisher = {Association for Computing Machinery},
address = {New York, NY, USA},
url = {https://doi.org/10.1145/2897839.2927391},
doi = {10.1145/2897839.2927391},
booktitle = {ACM SIGGRAPH 2016 Talks},
articleno = {67},
numpages = {2},
location = {Anaheim, California},
series = {SIGGRAPH '16}
}

@article{10.1007/s11390-024-4123-3,
author = {Xing, You-Xin and Tan, Hao-Wen and Xu, Yan-Ning and Wang, Lu},
title = {A Tiny Example Based Procedural Model for Real-Time Glinty Appearance Rendering},
year = {2024},
issue_date = {Jul 2024},
publisher = {Springer-Verlag},
address = {Berlin, Heidelberg},
volume = {39},
number = {4},
issn = {1000-9000},
url = {https://doi.org/10.1007/s11390-024-4123-3},
doi = {10.1007/s11390-024-4123-3},
journal = {J. Comput. Sci. Technol.},
month = sep,
pages = {771–784},
numpages = {14}
}

@article{Chermain2020Procedural,
author = {Chermain, Xavier and Sauvage, Basile and Dischler, Jean-Michel and Dachsbacher, Carsten},
title = {{Procedural Physically-based BRDF for Real-Time Rendering of Glints}},
journal = {Computer Graphics Forum (Proceedings of Pacific Graphics)},
volume = {39},
number = {7},
year = {2020},
pages = {243--253},
doi = {10.1111/cgf.14141}
}

@article{karis2013real,
  title={Real shading in {Unreal} {Engine} 4},
  author={Karis, Brian},
  journal={ACM SIGGRAPH Course on Physically Based Shading Theory and Practice},
  volume={4},
  number={3},
  pages={1},
  year={2013}
}

@inproceedings{chen2025pbrsr,
  author    = {Chen, Yujin and Nie, Yinyu and Ummenhofer, Benjamin and Birkl, Reiner and Paulitsch, Michael and Nie{\ss}ner, Matthias},
  title={PBR-SR: Mesh PBR Texture Super Resolution from 2D Image Priors},
  booktitle = {NeurIPS},
  year      = {2025}
}

@inproceedings{
    M3ashy2026, 
    author = {Chenliang Zhou and Zheyuan Hu and Alejandro Sztrajman and Yancheng Cai and Yaru Liu and Cengiz Oztireli}, 
    title = {M$^{3}$ashy: Multi-Modal Material Synthesis via Hyperdiffusion}, 
    year = {2026}, 
    booktitle = {Proceedings of the 40th AAAI Conference on Artificial Intelligence}, 
    location = {Singapore}, 
    series = {AAAI'26} 
}

@inproceedings{10.1145/3532836.3536240,
author = {Zeltner, Tizian and Burley, Brent and Chiang, Matt Jen-Yuan},
title = {Practical Multiple-Scattering Sheen Using Linearly Transformed Cosines},
year = {2022},
isbn = {9781450393713},
publisher = {Association for Computing Machinery},
address = {New York, NY, USA},
url = {https://doi.org/10.1145/3532836.3536240},
doi = {10.1145/3532836.3536240},
booktitle = {ACM SIGGRAPH 2022 Talks},
articleno = {7},
numpages = {2},
location = {Vancouver, BC, Canada},
series = {SIGGRAPH '22}
}

@inproceedings{10.1145/3528233.3530732,
author = {Fan, Jiahui and Wang, Beibei and Hasan, Milos and Yang, Jian and Yan, Ling-Qi},
title = {Neural Layered BRDFs},
year = {2022},
isbn = {9781450393379},
publisher = {Association for Computing Machinery},
address = {New York, NY, USA},
url = {https://doi.org/10.1145/3528233.3530732},
doi = {10.1145/3528233.3530732},
booktitle = {ACM SIGGRAPH 2022 Conference Proceedings},
articleno = {4},
numpages = {8},
location = {Vancouver, BC, Canada},
series = {SIGGRAPH '22}
}

@article {Matusik:2003,
	author = "Wojciech Matusik and Hanspeter Pfister and Matt Brand and Leonard McMillan",
	title = "A Data-Driven Reflectance Model",
	journal = "ACM Transactions on Graphics",
	year = "2003",
	month = jul,
	volume = "22",
	number = "3",
	pages = "759-769"
}

@inproceedings{10.5555/2383654.2383671,
author = {Ngan, Addy and Durand, Fr\'{e}do and Matusik, Wojciech},
title = {Experimental analysis of BRDF models},
year = {2005},
isbn = {3905673231},
publisher = {Eurographics Association},
address = {Goslar, DEU},
booktitle = {Proceedings of the Sixteenth Eurographics Conference on Rendering Techniques},
pages = {117–126},
numpages = {10},
location = {Konstanz, Germany},
series = {EGSR '05}
}

@inproceedings{10.1145/3543664.3543675,
author = {Emrose, Luke and Black, Curtis and Schrade, Emanuel},
title = {ASH - A Case For Layered Shading},
year = {2022},
isbn = {9781450394185},
publisher = {Association for Computing Machinery},
address = {New York, NY, USA},
url = {https://doi.org/10.1145/3543664.3543675},
doi = {10.1145/3543664.3543675},
booktitle = {Proceedings of the 2022 Digital Production Symposium},
articleno = {7},
numpages = {15},
location = {Vancouver, BC, Canada},
series = {DigiPro '22}
}

@online{iors,
  author  = {Michail N. Polyanskiy},
  title   = {Refractive index database},
  url     = {https://refractiveindex.info},
  urldate = {2026-01-14},
  year = 2025
}

@online{gltf2,
  author  = {Khronos},
  title   = {glTF 2.0 Specification},
  url     = {https://registry.khronos.org/glTF/specs/2.0/glTF-2.0.html#appendix-b-brdf-implementation},
  urldate = {2026-01-14},
  year = 2025
}

@inproceedings{Kynkaanniemi2022,
  author    = {Tuomas Kynkäänniemi and
               Tero Karras and
               Miika Aittala and
               Timo Aila and
               Jaakko Lehtinen},
  title     = {{The Role of ImageNet Classes in Fréchet Inception Distance}},
  booktitle = {Proc. ICLR},
  year      = {2023},
}

@inproceedings{zhang2018perceptual,
  title={The Unreasonable Effectiveness of Deep Features as a Perceptual Metric},
  author={Zhang, Richard and Isola, Phillip and Efros, Alexei A and Shechtman, Eli and Wang, Oliver},
  booktitle={CVPR},
  year={2018}
}

@misc{jayasumana2024cmmd,
      title={Rethinking FID: Towards a Better Evaluation Metric for Image Generation}, 
      author={Sadeep Jayasumana and Srikumar Ramalingam and Andreas Veit and Daniel Glasner and Ayan Chakrabarti and Sanjiv Kumar},
      year={2024},
      eprint={2401.09603},
      archivePrefix={arXiv},
}

@article{yu2024appearance,
  title={Appearance Modeling of Iridescent Feathers with Diverse Nanostructures},
  author={Yu, Yunchen and Weidlich, Andrea and Walter, Bruce and d'Eon, Eugene and Marschner, Steve},
  journal={ACM Transactions on Graphics (TOG)},
  volume={43},
  number={6},
  pages={1--18},
  year={2024},
  publisher={ACM New York, NY, USA}
}

@article{yu2025realistic,
  title={Realistic Cloth Rendering with a Ray-Wave Hybrid Shading Model},
  author={Yu, Yunchen and Walter, Bruce and Marschner, Steve and Weidlich, Andrea},
  journal={ACM Transactions on Graphics (TOG)},
  volume={44},
  number={6},
  pages={1--17},
  year={2025},
  publisher={ACM New York, NY, USA}
}

@article{Andersson2020,
  author    = {Pontus Andersson and
               Jim Nilsson and
               Tomas Akenine{-}M{\"{o}}ller and
               Magnus Oskarsson and
               Kalle {\AA}str{\"{o}}m and
               Mark D. Fairchild},
  title     = "{{\FLIP:} {A} Difference Evaluator for Alternating Images}",
  journal   = {Proceedings of the ACM on Computer Graphics and Interactive Techniques},
  volume    = {3},
  number    = {2},
  pages     = {15:1--15:23},
  year      = {2020},
  doi={10.1145/3406183}
}

@article{10.1145/3197517.3201321,
author = {Zeltner, Tizian and Jakob, Wenzel},
title = {The layer laboratory: a calculus for additive and subtractive composition of anisotropic surface reflectance},
year = {2018},
issue_date = {August 2018},
publisher = {Association for Computing Machinery},
address = {New York, NY, USA},
volume = {37},
number = {4},
issn = {0730-0301},
url = {https://doi.org/10.1145/3197517.3201321},
doi = {10.1145/3197517.3201321},
journal = {ACM Trans. Graph.},
month = jul,
articleno = {74},
numpages = {14}
}

@article{10.1145/2601097.2601139,
author = {Jakob, Wenzel and d'Eon, Eugene and Jakob, Otto and Marschner, Steve},
title = {A comprehensive framework for rendering layered materials},
year = {2014},
issue_date = {July 2014},
publisher = {Association for Computing Machinery},
address = {New York, NY, USA},
volume = {33},
number = {4},
issn = {0730-0301},
url = {https://doi.org/10.1145/2601097.2601139},
doi = {10.1145/2601097.2601139},
journal = {ACM Trans. Graph.},
month = jul,
articleno = {118},
numpages = {14}
}

@article{10.1145/3272127.3275053,
author = {Guo, Yu and Ha\v{s}an, Milo\v{s} and Zhao, Shuang},
title = {Position-free monte carlo simulation for arbitrary layered BSDFs},
year = {2018},
issue_date = {December 2018},
publisher = {Association for Computing Machinery},
address = {New York, NY, USA},
volume = {37},
number = {6},
issn = {0730-0301},
url = {https://doi.org/10.1145/3272127.3275053},
doi = {10.1145/3272127.3275053},
journal = {ACM Trans. Graph.},
month = dec,
articleno = {279},
numpages = {14}
}

@article{10.1145/3618365,
author = {Guo, Jie and Li, Zeru and He, Xueyan and Wang, Beibei and Li, Wenbin and Guo, Yanwen and Yan, Ling-Qi},
title = {MetaLayer: A Meta-Learned BSDF Model for Layered Materials},
year = {2023},
issue_date = {December 2023},
publisher = {Association for Computing Machinery},
address = {New York, NY, USA},
volume = {42},
number = {6},
issn = {0730-0301},
url = {https://doi.org/10.1145/3618365},
doi = {10.1145/3618365},
month = dec,
articleno = {222},
numpages = {15}
}

@article{10.1145/3546940,
author = {Wang, Beibei and Jin, Wenhua and Ha\v{s}an, Milo\v{s} and Yan, Ling-Qi},
title = {SpongeCake: A Layered Microflake Surface Appearance Model},
year = {2022},
issue_date = {February 2023},
publisher = {Association for Computing Machinery},
address = {New York, NY, USA},
volume = {42},
number = {1},
issn = {0730-0301},
url = {https://doi.org/10.1145/3546940},
doi = {10.1145/3546940},
journal = {ACM Trans. Graph.},
month = sep,
articleno = {8},
numpages = {16}
}

@article{hasselgren2026videomatgen,
  title={{VideoMatGen: PBR Materials through Joint Generative Modeling}},
  author={Jon Hasselgren and Zheng Zeng and Milos Hasan and Jacob Munkberg},
  year={2026},
  journal={arXiv preprint arXiv:2603.16566},
}

\end{document}